\documentclass[journal=jpcafh,manuscript=preprint,layout=onecolumn]{achemso}

\usepackage{graphicx} 
\usepackage{bm}
\usepackage{amsmath}
\usepackage{booktabs} 
\usepackage{float}
\usepackage{fullpage}
\usepackage{longtable}
\usepackage{amssymb, booktabs, subcaption}
\usepackage{parskip}
\SectionNumbersOn
\renewcommand{\arraystretch}{1.2}

\title{Electron-Transfer and Exchange-Interaction Model of the Ligand Hyperfine Structure of
Alkylated Iron-Sulfur Clusters}

\author{William C. Robinson}
\affiliation{Department of Chemistry and Biochemistry, Montana State University, Bozeman, MT 59717
USA}
\author{Victoria Pascutti}
\affiliation{Department of Chemistry and Biochemistry, Montana State University, Bozeman, MT 59717
USA}
\author{David A. Hall}
\affiliation{Department of Chemistry and Biochemistry, Montana State University, Bozeman, MT 59717
USA}
\author{Mart\'in A. Mosquera}
\email{martinmosquera@montana.edu}
\affiliation{Department of Chemistry and Biochemistry, Montana State University, Bozeman, MT 59717
USA}
\phone{+1-406-994-7750}

\begin{document}
\newtheorem{thm}{Theorem}
\newtheorem{cor}{Corollary}

\def\bea{\begin{eqnarray}}
\def\eea{\end{eqnarray}}
\def\ben{\begin{equation}}
\def\een{\end{equation}}
\def\benu{\begin{enumerate}}
\def\enu{\end{enumerate}}

\newcommand{\mr}[1]{\mathrm{#1}}
\newcommand{\mc}[1]{\mathcal{#1}}
\newcommand{\mb}[1]{\mathbf{#1}}
\newcommand{\lket}[1]{\langle #1|}
\newcommand{\rket}[1]{| #1\rangle}
\newcommand{\ud}{\mr{d}}
\newcommand{\ui}{\mr{i}}
\newcommand{\intdr}{\int \ud^3\mb{r}~}
\newcommand{\XC}{\mr{XC}}
\newcommand{\sA}{_\mr{A}}
\newcommand{\sB}{_\mr{B}}
\newcommand{\XX}{\mr{X}}
\newcommand{\cc}{\mr{c}}
\newcommand{\inL}{i_\mr{L}}
\newcommand{\jnL}{j_\mr{L}}
\newcommand{\HXC}{\mr{HXC}}
\newcommand{\GS}{\mr{GS}}
\newcommand{\LDA}{\mr{LDA}}
\newcommand{\GGA}{\mr{GGA}}
\newcommand{\LSDA}{\mr{LSDA}}
\newcommand{\ps}{\mr{ps}}
\newcommand{\h}{_{\mr{h}}}
\newcommand{\Ha}{\mr{H}}
\newcommand{\upa}{\uparrow}
\newcommand{\dwna}{\downarrow}
\newcommand{\s}{\mr{s}}
\newcommand{\dernr}[1]{\frac{\delta {#1} }{\delta n(\mb{r})}}
\newcommand{\dernrs}[1]{\frac{\delta {#1}}{\delta n_{\sigma}(\mb{x})}}
\newcommand{\derN}[1]{\frac{\partial {#1}}{\partial N}}
\newcommand{\eel}{$e^{-}$}





\begin{abstract}  
Iron-sulfur clusters conduct a wide variety of biochemical reactions that are conserved across all domains of life. 
The hyperfine structure of reactive ligands of these clusters can be
studied experimentally and theoretically by means of hyperfine spectroscopy, which can reveal
catalytic intermediates in these biochemical processes. Their theoretical prediction, however,
requires either advanced methods that describe strongly correlated systems, or Hamiltonian modeling
based on symmetry-broken electronic structure methods. This work shows that the addition of
electron-transfer interactions to the Heisenberg-Dirac-van Vleck Hamiltonian model leads to the
quantitative explanation of hyperfine coupling constants at active organic ligand sites. Comparison
with experimentally available results confirms our extended approach can be used in
calculations aimed at describing cutting-edge systems.
\end{abstract}

\maketitle
\section{Introduction}

Iron-sulfur (FeS) clusters are crucial components of biochemical processes in all living systems.
\cite{booker2022twenty,hoffman2023mechanism,backman2017new,mclaughlin2021overall,frey2008radical,magnusson2001characterization,nicolet2020structure,broderick2019radical,zhu2011mechanistic}
For example, RSAM clusters (radical {\slshape S}-adenosyl-{\footnotesize L}-methionine) are found
to-date in about 700,000 sequences and catalyze a very large number of radical reactions. Not all of
these reactions are well-understood in detail;
\cite{crack2021biological,read2021mitochondrial,honarmand2022iron,boncella2022expanding} elucidating
them could have wide implications not only in their specific field of science, but also in the understanding of
related diseases. In order to examine these radical reactions, magnetic spectroscopy has proven
determinant in the identification of intermediates in natural and synthetic reactions.  Besides SAM,
other prominent related examples involve nitrogenase, hydrogenase, and Rieske-type clusters, where pulsed- and
continuous-wave magnetic resonance techniques are often utilized to measure fine and hyperfine
interactions at atomic sites that are crucial to understanding the reactivity.

Iron-sulfur chemistries offer exciting opportunities for theoretical work aimed at advancing the
discovery and support of functionalities in these systems. HFCCs (hyperfine coupling constants) are
usually computed using
multireference\cite{shiozaki2016hyperfine,birnoschi2022hyperion,wysocki2024relativistic,samanta2018first}
(MR) and/or spin symmetry-broken (SB)
theories.\cite{noodleman1981valence,noodleman1992density,noodleman1991local} Computational MR
electronic structure theory has the advantage that it can directly predict HFCCs without additional
work by the user, provided it is coupled to a robust relativistic approach for core electrons. MR
methods that include dynamical correlation are among the most promising for this task. However, due
to the limits in active space size, they are currently constrained in application to relatively
small FeS clusters. SB electronic structure methods, in contrast, use spin-polarized single-reference
simulations to predict the so-called ‘raw’ HFCCs, which must then be corrected using projection
factors. It can be used with density functional theory (DFT) or wavefunction theory. 

SB-DFT (also known as broken-symmetry DFT, BS-DFT) is commonly applied to elucidate properties and
functionalities of FeS clusters and polynuclear transition metal complexes of natural or synthetic
origin.  This includes FeS clusters of different molecularities, from dimers to nitrogenase
co-factors,\cite{han2009dft, shoji2007theory, joshi2020accuracy, hubner2002structure,
pelmenschikov2013redox,blachly2015broken,
niu2011density,benediktsson2020quantum,lovell2001femo,noodleman2002insights} oxygen evolving
complexes,\cite{beal2017comparison,beal2018comparison, guo2017open,lohmiller2014structure} but also
mononuclear complexes.\cite{aragoni2020diradical,wang2015electronic} SB-DFT is used to study the
(super) exchange coupling between metal atoms and their hyperfine values, redox free energies, bond
lengths, and M\"osbauer shifts. However, the approach is open to theoretical extensions to compute
fine/hyperfine structure of (non-metal) ligand atomic sites. Past work in the field has identified
semi-empirical formulas,\cite{han2005active, han2009dft,rapatskiy2015characterization} which have
been validated in specific cases, but fundamental theoretical formulations for this goal are lacking
in the field. 

Motivated by this gap in understanding, we previously proposed a two-configuration method that
estimates HFCCs at ligand sites.\cite{jodts2023computational} This theory assumes that an anionic
ligand is bound to the FeS cluster, but has a contribution from a triplet electron transfer
configuration, where the weight of such configuration comes from SB-DFT. The present work extends
this previous formulation as an approach that is applicable to both bound and unbound ligand
scenarios. The trapping of an organic radical in proximity to the RSAM [4Fe-4S] cluster has been
recently reported,\cite{yang2024endor} where the radical activates the nominally silent state of
this cluster. If the ligand is bound to an iron atom of the FeS cluster, it is expected that the
ligand oxidizes the cluster, which can be modeled in a quantum mechanical fashion, Section
\ref{bound}. Otherwise, for the unbound case this theory assumes that the free ligand is in radical
form with two possible FeS cluster spin states, Section \ref{unbound}.







\section{Theory}

Fine and hyperfine effects derive from relativistic quantum mechanics. Fine effects are associated
to spin-orbit couplings, while hyperfine ones to the interaction between electronic and
nuclear spins of the system. This work concerns hyperfine effects that are split into the
so-called `Fermi contact' and `spin-dipolar' contributions. For a given nuclear site with
non-zero spin, the Fermi contact arises from the electronic spin-density evaluated at the nuclear position,
while the spin-dipolar tensor is due to the coupling of the spin vectors of the electron and the
nucleus. The hyperfine tensor of a site labeled $n$ is thus:
\ben
\mb{A}^{\mr{obs}}_n = a^{\mr{iso}}_n\mb{I} + \mb{T}_n^{\mr{dip}}
\een
where the $a^{\mr{iso}}_n$ is the isotropic Fermi contact (scalar) value, $\mb{I}$ the $3\times 3$ identity
matrix, and $\mb{T}^{\mr{dip}}_n$ the spin-dipolar tensor, which is anisotropic. These values can be produced by
quantum chemistry software, most commonly through a SB electronic structure method, offering
a practical pathway that requires spin-symmetrization through a Hamiltonian model.


The conventional approach to spin-symmetrize HFCCs is constrained to metal sites. It is based on a
model energy operator known as the Heisenberg-Dirac-Van Vleck (HDVV) Hamiltonian. In this
Hamiltonian, the metal sites are (super) exchange-coupled, so it would be expressed as
$\hat{H}_{\mr{HDVV}}=\sum_{i>j}J_{ij}\hat{\mb{S}}_i\cdot\hat{\mb{S}}_j$ (other related terms, like
double exchange, can be added if needed), where $\hat{\mb{S}}_i$ is the spin vector operator of the
atomic metal site $i$. Before solving such Hamiltonian, input from the SB approach is required, in which
a set of SB energies is generated by polarizing the spin of each site, usually in $z$-oriented
directions.
Four metal-sites yield, for example, yield six possible energies (more are possible if $d$-orbital
occupation can be controlled), which are used to determine the $\{J_{ij}\}$ exchange-coupling
constants. The Hamiltonian is then diagonalized, giving spin wavefunctions of a desired symmetry
(doublet, triplet, and so on), and the expectation value of the site spin. Knowledge of this value
and the intrinsic hyperfine tensor of the site lead to the estimation of experimentally observable
HFCCs for metals.

\begin{figure}[htp]
\centerline{\includegraphics[width=0.65\hsize]{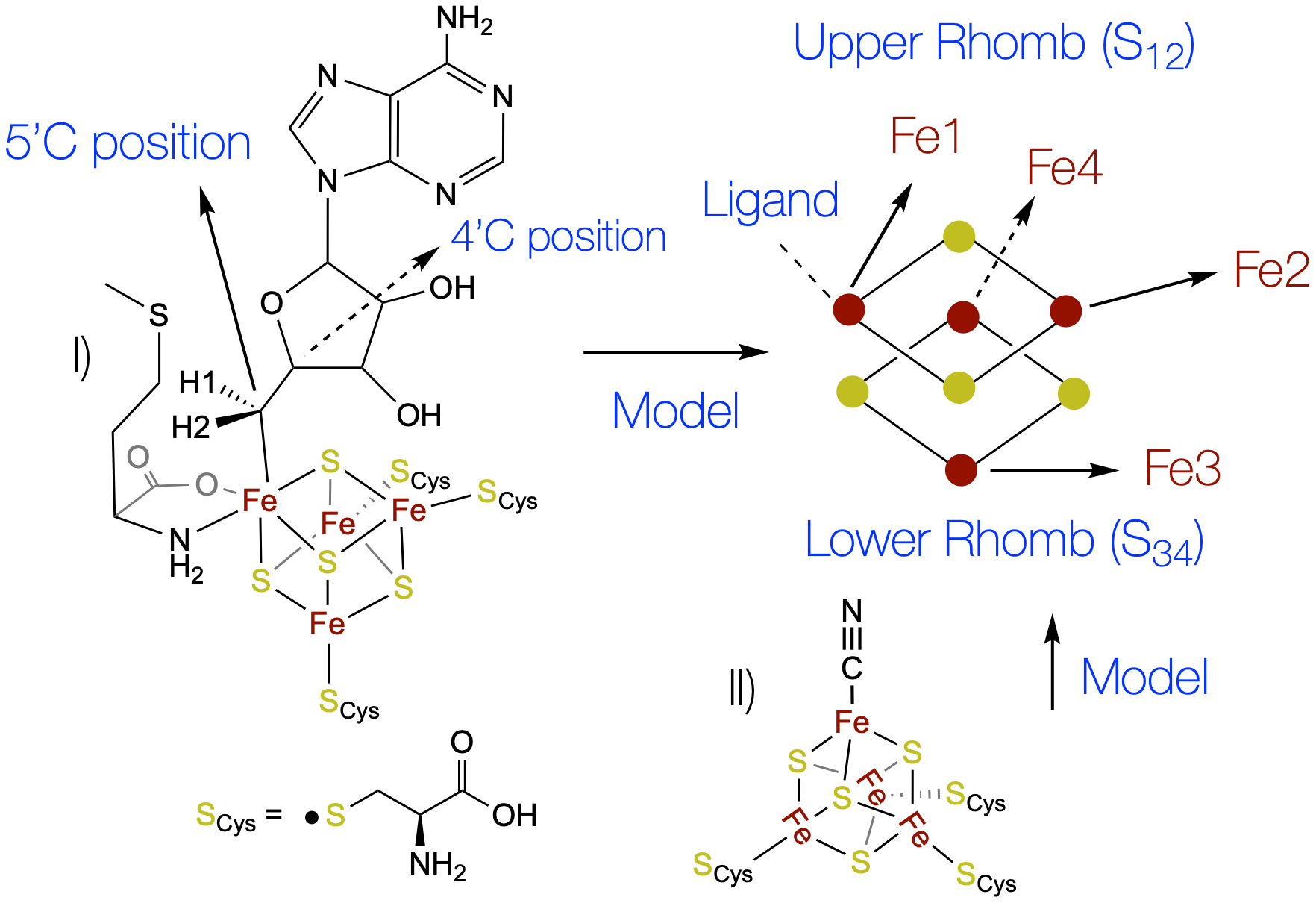}}
\caption{This work models the interaction between the ligand of interest and the clusters as a three
subsystem composite. Upper left: I) $\Omega$ intermediate, 5'-dAdo radical bound to a [4Fe-4S]${}^{3+}$
cluster; bottom right: II) [4Fe-4S]${}^{+}$ cluster featuring
a cyano ligand.}
\label{model_fig}
\end{figure}

For the study of ligand hyperfine structure, however, the above conventional HDVV model requires an
extension. When the overall system is electron-spin active, its ligands tend to produce hyperfine
signals from nuclear-spin-active positions, largely due to their non-zero spin distribution that
results from interaction with the FeS cluster. An alkyl ligand is expected to abstract an electron
from the FeS cluster that it binds to. From a quantum mechanical perspective, the ligand can thus be
modeled as the linear superposition of states where the ligand fully abstracts an electron and
states where it does not. In the latter states, the ligand would be a neutral radical. A radical
configuration clearly involves the presence of a delocalized spin-electron. Such a configuration
does give rise to the hyperfine signals, as opposed to configurations where a full electron is
abstracted. Our approach is an extended HDVV (E-HDVV) model for ligand spin properties based on SB
theory. The present formalism treats the unpaired ligand spin-electron (LSE) as a delocalized
object, as opposed to the spin electrons in transition metals, which are treated as localized
entities. Inclusion of the LSE in the model and spin-symmetrization of the resultant states, as we
show here, yields a quantitative explanation of experimental results. Herein we study the $\Omega$
intermediate\cite{horitani2016radical,byer2018paradigm,broderick2018mechanism} of SAM reductive
cleavage, which features 5'-dAdo as a ligand, Figure \ref{model_fig}. For the bound ligand case, a
cyano ligand is considered.  We then show our model is also applicable to ligand detachment process.

\subsection{Spin Wavefunctions and Hamiltonian}

For the sake of simplicity, and facilitated by the significant difference in energy scales, we study
the electronic and hyperfine Hamiltonians separately. A general form of electronic Hubbard-like
Hamiltonian is the following:
$\hat{H}=\sum_{i\sigma}\epsilon_{ij,\sigma}\hat{a}_{i\sigma}^{\dagger}\hat{a}_{j\sigma}+
\sum_{i,j}U_{ij}\hat{n}_i\hat{n}_j+\sum_{i>j}J_{ij}\hat{\mb{S}}_i\cdot\hat{\mb{S}}_j$, where
$\epsilon_{ij,\sigma}$ represents on-site and hopping energies, $U_{ij}$ intra- and inter-site
electrostatics, and $J_{ij}$ the exchange energies between different sites; $\hat{a}_i$ destroys an
electron at site $i$, and $\hat{n}_i$ is the occupation operator for that site. This Hamiltonian
includes an electron-hopping term between ligand and cluster, which is not part of the traditional
HDVV model Hamiltonian, but that is shown to be useful in our present study.

We model the cubic-like four-iron/four-sulfur cluster as two exchange-coupled rhombs, which we refer
to as upper and lower rhombs; UR and LR, respectively (Figure \ref{model_fig}), where spin operator of UR is
$\hat{\mb{S}}_{12}$, and for LR $\hat{\mb{S}}_{34}$. The upper rhomb is coupled to ligand through
exchange and hopping interactions. Therefore the Hamiltonian takes the form:
\ben
\hat{H}=\hat{H}_{0}-t\hat{E}_{\mr{L},1}+k\hat{\mb{S}}_{\mr{L}}\cdot
\hat{\mb{S}}_{12}+j\hat{\mb{S}}_{12}\cdot\hat{\mb{S}}_{34}
+\hat{H}_{\mr{iR}}
\een
where $t$ is the hopping parameter, $\hat{\mb{S}}_{\mr{L}}$ refers to the spin (vector) operator of the
ligand, $\hat{H}_{\mr{iR}}$ is the intra-rhomb exchange Hamiltonian, and
it has the form $\hat{H}_{\mr{iR}}=J\hat{\mb{S}}_1\cdot
\hat{\mb{S}}_2+J\hat{\mb{S}}_3\cdot\hat{\mb{S}}_4$; so $J$ is the exchange coupling between iron
atoms in their rhombs.  The operator $\hat{E}_{\mr{L},1}$ describes electron transfer between
ligand and unique iron site. Formally, it is written as
$\hat{E}_{\mr{L},1}=\hat{a}^{\dagger}_{\mr{1}}\hat{a}_{\mr{L}}+\hat{a}^{\dagger}_{\mr{L}}\hat{a}_{1}$.
The parameter $k$ refers to exchange coupling between ligand and upper rhomb, and $j$ to exchange
interaction between the upper and lower rhombs. The reference operator $\hat{H}_{0}$ accounts for
the remainder on-site and electrostatic interactions in the molecule. This term is implied in the
conventional HDVV approach as well, but in the present case, it is important to explicitly include
it due to the presence of spin configurations with charge redistribution. The above Hamiltonian is a
generalized form of the Hubbard
model,\cite{tchougreeff1992intra,tchougreeff1992heisenberg,tchougreeff1991ferromagnetism,schumann2010hubbard,de1995exact,montorsi1996rigorous,ozaki1992broken}
however, it has not been previously applied to ligand hyperfine properties

The spin-wavefunction of the system is denoted as $|\Psi\rangle$, and is taken as
the sum of spin-configurations $|\Psi\rangle=\sum_{\mu}C_{\mu}|\mr{SC}\mu\rangle$, where
$|\mr{SC}\mu\rangle$ is $\mu$-th spin-configuration wavefunction, and $C_{\mu}$ its linear superposition
coefficient (LSC). Once the space of SCs is selected, we apply first-order perturbation theory (PT)
to compute projection factors that quantify ligand HFCCs.

\begin{table}[htbp]
  \centering
  \caption{Model spin configurations for the $\Omega$ intermediate, ligand/cluster spins and
  charges. $Q_{\mr{L}}$ and $Q_{\mr{cl}}$ represent the charges of ligand and cluster,
  respectively.}
  \label{spin_table}
  \begin{small}
    \begin{tabular}{c|c|c|c|c|c|c}
          & $S_{\mr{L}}$ & $S_{\mr{cl}}$ & $Q_{\mr{L}}$ & $Q_{\mr{cl}}$ & $S_{12}$ & $S_{34}$ \\
    \hline
    $|\mr{SC0}\rangle$   & 1/2 & 0 & 0 & +2 & 9/2 & 9/2  \\
    \hline
    $|\mr{SC1}\rangle$   & 1/2 & 1 & 0 & +2 & 9/2 & 9/2 \\
    \hline
    $|\mr{SC2}\rangle$   & 0   & 1/2 & -1   & +3 & 4 & 9/2 \\
    \end{tabular}%
  \end{small}
\end{table}%

Focusing on the $\Omega$ intermediate, we introduce three spin configurations for the ligand +
cluster system, which are summarized in Table \ref{spin_table}; the rhomb spin values derive from
Ref. \cite{mouesca1994density}. In the first ET spin configuration, the ligand in radical form has
$S_{\mr{L}}=1/2$, while the cluster is of singlet spin symmetry ($S_{\mr{cl}}=0$). This
configuration is denoted as $|\mr{SC0}\rangle=|0,0\rangle_{\mr{cl}}|1/2,1/2\rangle_{\mr{L}}$, where
$|0,0\rangle_{\mr{cl}}$ is the singlet state of the cluster, and $|1/2,1/2\rangle_{\mr{L}}$ refers
to the ligand doublet state with secondary spin $m_{S_{\mr{L}}}=1/2$; we use the notation
$|S,m_S\rangle$, where the first number denotes the principal spin number of the subsystem. The
second spin configuration, $|\mr{SC}1\rangle$, describes the cluster in triplet state, and the
ligand in doublet state. This configuration, however, is spin-adapted as follows:
\ben
|\mr{SC}1\rangle=\sqrt{\frac{2}{3}}|1,1\rangle_{\mr{cl}}|1/2,-1/2\rangle_{\mr{L}}-
                     \sqrt{\frac{1}{3}}|1,0\rangle_{\mr{cl}}|1/2,1/2\rangle_{\mr{L}}
\een
\sloppy The third configuration involves the ligand in spin-inactive anionic form, so
$|\mr{SC}2\rangle=|1/2,1/2\rangle_{\mr{cl}}|0,0\rangle_{\mr{L}}$. 

Even though three configurations are used to model the quantum mechanical state of the
cluster+ligand system, each configuration is a sum of separate internal configurations. First, a
cluster state with principal and secondary spin numbers $S_{\mr{cl}}$ and $m_{\mr{cl}}$,
correspondingly, is expressed as:
\ben
|S_{\mr{cl}}, m_{\mr{cl}}\rangle_{\mr{cl}}=\sum_{m_{12}+
m_{34}=m_{\mr{cl}}}c(m_{12},m_{34})|S_{12},m_{12}\rangle_{\mr{UR}}
|S_{34},m_{34}\rangle_{\mr{LR}}
\een
Second, the state of a rhomb is in turn the linear combination of their iron states, so 
\ben
\begin{split}
|S_{12},m_{12}\rangle&=\sum_{m_1,m_2}d_{\mr{UR}}(m_1,m_2)|S_1,m_1\rangle_{\mr{Fe1}}|S_2,m_2\rangle_{\mr{Fe2}}~,\\
|S_{34},m_{34}\rangle&=\sum_{m_3,m_4}d_{\mr{LR}}(m_3,m_4)|S_3,m_3\rangle_{\mr{Fe3}}|S_4,m_4\rangle_{\mr{Fe4}}~.
\end{split}
\een
The coefficients $c(m_{12},m_{34})$, $d_{\mr{UR}}(m_1,m_2)$, and $d_{\mr{LR}}(m_3,m_4)$ are
Clebsch-Gordan (CG) coefficients. These are determined using the program from Ref. \cite{stone1980root}. 

After performing the CG-related algebra (Supporting Information), we arrive at the Hamiltonian of
the system in matrix form:
\begin{equation}\label{h_matrix}
\displaystyle
\mb{H} = \begin{bmatrix}
10J - \frac{99}{4}j +\Delta \epsilon_0 & -5/2k & 2/3 t\\
-5/2k & 10J-\frac{95}{4}j-\frac{1}{2}k+\Delta\epsilon_0 & 3/4 t\\
2/3t & 3/4 t & \frac{25}{4}J - 22j
\end{bmatrix}
\end{equation}
where $H_{\mu\nu}=\langle \mr{SC}\mu|\hat{H}|\mr{SC}\nu\rangle$, and $\Delta\epsilon_0>0$ is the
difference in electrostatic energy between the radical and anionic states:
\begin{equation}
\begin{split}
\Delta\epsilon_0 &= \langle\mr{SC}0|\hat{H}_0|\mr{SC0}\rangle - \langle\mr{SC}2|\hat{H}_0|\mr{SC2}\rangle\\
                 &= \langle\mr{SC}1|\hat{H}_0|\mr{SC1}\rangle - \langle\mr{SC}2|\hat{H}_0|\mr{SC2}\rangle
\end{split}
\end{equation}
The Hamiltonian matrix ($\mb{H}$) is separated into
its diagonal and off-diagonal forms, so $\mb{H}=\mb{H}_{\mr{diag}}+\bm{\Delta}$. This allows us to
apply PT: We consider a zero-th order spin wavefunction $|\Psi^{(0)}\rangle$, such
that the first-order correction takes the form $|\Psi^{(1)}\rangle=\sum_{\mu}C_{\mu}^{(1)}|\mr{SC}\mu\rangle$,
where coefficients are given by the expression:
\ben
C_{\mu}^{(1)}=-\frac{\langle \mr{SC}\mu|\hat{\Delta}|\Psi^{(0)}\rangle}{
\langle \mr{SC}\mu|\hat{H}_{\mr{diag}}|\mr{SC}\mu\rangle - \langle
\Psi^{(0)}|\hat{H}_{\mr{diag}}|\Psi^{(0)}\rangle}
\een
In what follows, we study two scenarios: In the first case, the spin configuration $|\mr{SC}2\rangle$
is the dominant state, which corresponds to ligand bound to cluster, and, in the second one,
$|\mr{SC}0\rangle$ is the main configuration (which describes ligand being partially detached from
the cluster).

\subsection{Bound Ligand Case}\label{bound}

For the bound case we express the spin wavefunction as
$|\Psi\rangle=|\mr{SC}2\rangle+C_0|\mr{SC}0\rangle+C_1|\mr{SC1}\rangle$, so
$|\Psi^{(0)}\rangle=|\mr{SC}2\rangle$. The state $|\mr{SC}2\rangle$ only couples via the ET operator
to the states $|\mr{SC}0\rangle$ and $|\mr{SC}1\rangle$. $|\mr{SC}2\rangle$ is a conventional state
where the ligand is {\slshape spin-inactive}. In contrast, 
the ligand is spin active in $|\mr{SC}0\rangle$
and $|\mr{SC}1\rangle$, where these configurations originate an
observable hyperfine signal at sites with non-negligible nuclear spin.

Starting from Eq. (\ref{h_matrix}), the 
application of PT gives the following SC coefficients:
\ben
C_0=\frac{-a_0t}{a_1J - a_2 j + \Delta \epsilon_{0}};~~C_1=\frac{-b_0 t}{b_1J - b_2j - b_3 k
+\Delta \epsilon_{0}}
\een
where $a_0=2/3$, $a_1=15/4$, $b_0=3/4$, $b_1=15/4$, $b_2=7/4$, and $b_3=1/2$.
The Hartree energy term (kinetic energy + electrostatics), $\Delta \epsilon_0$, which is greater than
2.5 eV, is 
naturally large in comparison to the exchange couplings, as these are much lower than 100 meV. 
This implies, that $(\langle\mr{SC}0|\hat{H}|\mr{SC}0\rangle -
\langle\mr{SC}2|\hat{H}|\mr{SC}2\rangle)
\approx (\langle\mr{SC}1|\hat{H}|\mr{SC}1\rangle -
\langle\mr{SC}2|\hat{H}|\mr{SC}2\rangle)$. Therefore, we can approximate the coefficients as
$C_0\approx -(2/3)(t/\Delta \epsilon_0)$ and $C_1\approx -(3/4)(t/\Delta \epsilon_0)$, 
and notice that both coefficients are directly proportional to the hopping parameter, $t$. 

Now, we introduce three variables: First, the `probability', or net weight, of the (minority)
radical ET state as $p_{\mr{ET}}=|C_0|^2+|C_1|^2$: This is the weight of simply transferring an
electron regardless of spin configuration (provided the system remains a doublet overall). Then,
we define the relative weights of the ligand spin-active configurations with respect to $p_{\mr{ET}}$ as
$\alpha = |C_0|^2/p_{\mr{ET}}$, and $\beta = |C_1|^2/p_{\mr{ET}}$. Considering that $C_0$ and $C_1$
share almost the same proportionality factor with regards to the hopping parameter, we have that the
relative weights are both approximately half the unit, $\alpha=\beta\approx 1/2$.

The hyperfine Hamiltonian reads
\ben
\hat{H}_{\mr{F}}=\sum_{\mr{all}~n}
\hat{\mb{S}}{\cdot}\mb{A}_n^{\mr{obs}}{\cdot}\hat{\mb{I}}_n=\sum_l\hat{\mb{S}}_{L}{\cdot}\mb{A}_l^{\mr{int}}{\cdot}\hat{\mb{I}}_l
+\sum_{i=1}^4\hat{\mb{S}}_i{\cdot}\mb{A}_i^{\mr{int}}{\cdot}\hat{\mb{I}}_i~, 
\een
where $\hat{\mb{S}}$ is the total spin operator; $l$ labels nuclear sites in the ligand, and $i$
metal sites. This energy operator is usually expressed in ‘observable’ (superscript obs) and local
forms (superscript `int', which stands for `intrinsic'). In the experimentally observable HFCC, the
net (true) spin of the whole system couples to the nuclear spins, as opposed to the local
expression, where each electronic spin site couples to its nuclear spin. For the ligand in local
expression, the unpaired electron spin ($\hat{\mb{S}}_{L}$) couples to the active nuclear sites. The
spin wavefunction, $|\Psi\rangle$, dictates how the observable hyperfine coupling is determined by
the local quantities, which derive from SB theory. The average $z$-spin at the ligand,
$\langle\Psi|\hat{S}_{\mr{L},z}|\Psi\rangle=\langle\hat{S}_{\mr{L},z}\rangle$, can be expressed as:
\ben
\langle \hat{S}_{\mr{L},z}\rangle = p_{\mr{ET}}[\alpha\langle \mr{SC}0|\hat{S}_{\mr{L},z}|\mr{SC}0\rangle+
\beta\langle \mr{SC}1|\hat{S}_{\mr{L},z}|\mr{SC}1\rangle]
\een
evaluation of this term gives $\langle \hat{S}_{\mr{L},z}\rangle= (1/6) \times p_{\mr{ET}}$.
The projection factor in this case reads:
\ben
K_{\mr{L}}=p_{\mr{ET}}^{-1}\frac{\langle \hat{S}_{\mr{L},z}\rangle}{\langle \hat{S}_z\rangle}
=\frac{1}{3}
\een
The $p_{\mr{ET}}$ factor cancels out as $|\mr{SC}0\rangle$ and $|\mr{SC}1\rangle$ are minority states in
comparison to $|\mr{SC2}\rangle$. The isotropic HFCC at a ligand site reads\footnote{Recall that isotropic
coupling is the trace of its hyperfine tensor, divided by 3: $a^{\mr{iso}}=\mr{tr}(\mb{A})/3$.}
$a^{\mr{iso}}_{l}=K_{\mr{L}}p_{\mr{ET}}a^{\mr{int}}_{l}$. As mentioned previously, electrostatics
dominates ET energetics and are the main driving force for electron hopping between ligand and
cluster. This remains the cases regardless of whether one is employing the symmetry-broken or true spin wavefunction.
Hence, we take the product $p_{\mr{ET}}a^{\mr{int}}_{\mr{L}}$ as the symmetry-broken theory ``raw''
HFCC value for the ligand site. Extending this hypothesis to the full hyperfine tensor, we have that the
observable tensor is given by a formula that operates similar to the metal-site HDVV
theory:
\ben
\mb{A}_l^{\mr{obs}}=K_{\mr{L}}\mb{A}_l^{\mr{SB}}
\een
where $\mb{A}_l^{\mr{SB}}$ is the tensor from the SB electronic structure simulation.

The magnitude of $K_{\mr{L}}$ was derived before in Ref.\cite{jodts2023computational} but with a
negative sign. In this case its positivity manifests from the contributions of spin configurations
$|\mr{SC}0\rangle$ and $|\mr{SC1}\rangle$. The observable signs of the ligand site HFCCs can be
determined by examination of their signs in the SB DFT configurations that are energetically
favorable. For example, if the unique iron site is preferentially in a spin-down state, then the
SB-DFT calculations would estimate the sign of the ligand site HFCCs relative to that preferred spin
orientation. We show this procedure in the Results and Discussion section.

The cyano ligand mathematics is not shown in this work, but it follows a similar procedure as the
$\Omega$ intermediate, with a few differences in the basis: The dominant state, $|\mr{SC2}\rangle$
has oxidation state $+1$, where $S_{12}=4$ and $S_{34}=9/2$. For the perturbative configurations,
the rhombs would be in $S_{12}=4$ and $S_{34}=4$ states. These spin numbers do not deviate
significantly from
the $\Omega$ intermediate case, and the same applies to the hopping off-diagonal Hamiltonian
elements. Similar to the above, electrostatics outweigh exchange energies in the calculation of PT
LSCs, so the model also leads to the projection factor $K_{\mr{L}}=1/3$. As shown in the
Supporting Information section, $K_{\mr{L}}=1/3$ also holds for a two-iron model with $S=1/2$,
suggesting this is a robust projection factor for alkylated systems of net doublet multiplicity.

\subsection{Radical-Cluster Exchange Interaction}\label{unbound}

In recent experiments \cite{yang2024endor}, supported by theoretical modeling, the 5'-dAdo radical has been
trapped in close proximity to the cluster, where the cluster is dominantly in its 2+ oxidation state. The
presence of the radical, however, induces its exchange coupling to the cluster, caused by the mixing of
the radical doublet state with the triplet state of the cluster. The relevant spin wavefunctions are
$|\Psi^{(0)}\rangle=|\mr{SC}0\rangle$ and $|\Psi^{(1)}\rangle=C_1|\mr{SC}1\rangle$. The $C_1$
coefficient is then determined through PT, where the Hamiltonian is the same used
in the previous case, Section \ref{bound}. This gives the result:
\ben
\begin{split}
C_1&=-\frac{H_{10}}{H_{11}-H_{00}}\\
   &=\frac{5}{2}\times \frac{k}{j}
\end{split}
\een
This assumes that $k\ll j$, as verified in Section \ref{results}.
The next step involves evaluation of the $z$-spin expectation values of the ${}^{57}$Fe sites, and
comparison with its total spin counterpart. This is similar to the previous case, but here it
gives the following expressions connecting SB theory with their observable site HFCCs:
\ben
\label{eq_set}
\begin{split}
\mb{A}_1^{\mr{obs}}&=-\frac{12}{\sqrt{3}}\frac{k}{j}\mb{A}_1^{\mr{int}}~,~~\mb{A}_2^{\mr{obs}}=-\frac{17}{\sqrt{3}}\frac{k}{j}\mb{A}_2^{\mr{int}}~,\\
\mb{A}_3^{\mr{obs}}&=+\frac{16}{\sqrt{3}}\frac{k}{j}\mb{A}_3^{\mr{int}}~,~~\mb{A}_4^{\mr{obs}}=+\frac{13}{\sqrt{3}}\frac{k}{j}\mb{A}_4^{\mr{int}}~.
\end{split}
\een
We then obtain symmetry factors for nominally inactive transition metal sites
(${}^{57}$Fe sites in this case). These relations imply that in this model, the ratio of
ligand-cluster and inter-rhomb exchange couplings is determinant for the estimation of the
metal-site HFCCs. We previously found a similar result for a low-spin two iron system coupled to
a methyl radical, which is now generalized for the [4Fe-4S] cluster.

\section{Computational Methodology}
SB-DFT simulations were performed with the ORCA program, version 6,\cite{neese2022software} and
based on the BP86 exchange-correlation functional for geometry optimization and non-metal hyperfine
tensors. This choice of functional derives from previous work.\cite{jodts2023computational} For
${}^{57}$Fe intrinsic HFCCs only we use the BHandHLYP exchange-correlation functional, which
involves 50 \% Hartree-Fock exchange; this improves said raw HFCCs over the BP86 functional by
around 30 \%. For SB-BP86 molecular optimizations, we used the def2-TZVP basis set for most C, S, N,
Fe, and O atoms, except for the atoms of interest, for which we applied def2-TZVPD, hydrogen atoms
were treated at the def2-SVP level; these were performed with the ``NormalSCF''and ``NormalOpt'' set
of SCF and optimization thresholds, as well as the def2/J auxiliary density-fitting basis set. The
basis set selection is changed for hyperfine calculations, all C, N, O, and H atoms were modeled
with the EPR-III,\cite{epriii} a triple-zeta + polarization basis set, fine-tuned for EPR
simulations; IGLO-III\cite{igloiii} is employed for sulfur atoms, and CP(PPP) for Fe,\cite{cpppp}
for this last basis set the Fe atomic grid was modified for high accuracy (sample input script is
shown in the SI), otherwise, the ``DefGrid2'' level grid was used.  EPR calculations were
performed under the ``TightSCF'' convergence thresholds, and without density fitting.

\section{Results and Discussion}\label{results}

\begin{figure}[htp]
\centerline{\includegraphics[width=0.65\hsize]{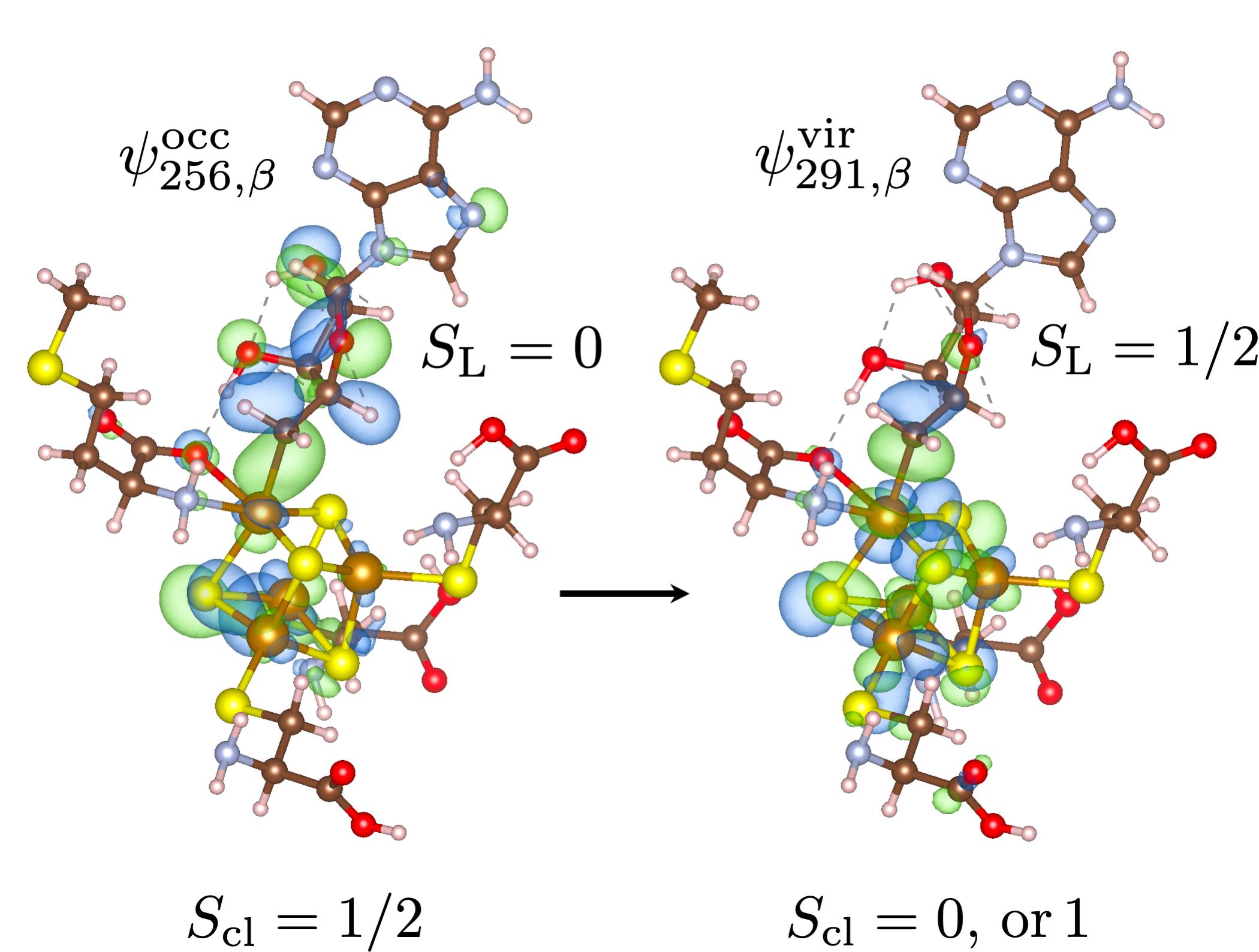}}
\caption{Example of a single-orbital transition with ligand-to-metal character. The occupied SB-DFT (BP86) canonical orbital,
left, shows $d\sigma$ character, whereas the virtual concentrates electron density on the cluster (a
contribution from $d\sigma^*$ can also be observed). The difference between orbital energies here is
about 4 eV, motivating the difference in energy scales used in this work. Isovalue level: 0.03.}
\label{fig_orbs}
\end{figure}

We begin this discussion by considering the $\Omega$ intermediate of reductive cleavage of
SAM,\cite{horitani2016radical,byer2018paradigm,broderick2018mechanism} in which the cluster is oxidized by
the 5'-dAdo ligand, [4Fe-4S]${}^{3+}$. First, we geometry-optimize all the possible spin
configurations of the system, and select the geometry with the lowest energy. Table
\ref{dAdo_numbers} shows the results of the SB-DFT calculations for the six possible spin flips,
calculated at the lowest energy spin configuration. In this context $\alpha$ or $\beta$ refer to
iron atom being fully spin polarized in the positive $z$ direction, or negative. These simulations
imply that the SB configuration $\beta\alpha\alpha\beta$ has the lowest energy and therefore that
the unique iron spin points in the negative direction (atoms Fe2 and Fe3, Figure \ref{model_fig},
would be positively spin-polarized, and Fe4 negatively spin-polarized). As a consequence, the signs
of the 5'-${}^{13}$C and 4'-${}^{13}$C sites (Figure \ref{model_fig}) are also positive, and the
signs for the protons adjacent to the 5' position are negative.  These four observations agree with
experiment.

\begin{table}[htbp]
  \centering
  \caption{5'-dAdo ligand HFCCs (SB-BP86, $K_{\mr{L}}$ corrected) at selected sites (carbon-13 and
  hydrogen-1 implied), values in MHz units, spin flips shown in sequential order with respect to
  Figure 1 labeling. Experimental proton HFCCs are reported negative, carbon-13 ones are positive.}
  \label{dAdo_numbers}
    \begin{tabular}{c|c|c|c|c|c}
    Configuration & Energy (meV) & 5'C    & 4'C    & H1    & H2 \\
    \hline
    $\beta\alpha\alpha\beta$  & 0     & 16    & 1.4   & -4    & -1 \\
    \hline
    $\beta\alpha\beta\alpha$  & 298 & 16    & 1.6   & -2    & -2 \\
    \hline
    $\beta\beta\alpha\alpha$  & 491 & 13    & 1     & -2    & -3 \\
    \hline
    $\alpha\alpha\beta\beta$  & 451 & -19   & -0.7  & -3    & 5 \\
    \hline
    $\alpha\beta\beta\alpha$  & 306 & -16   & -0.5  & 2     & 5 \\
    \hline
    $\alpha\beta\alpha\beta$  & 494 & -17   & -0.3  & 3     & 2 \\
    \hline
    Expt. Obs.                &     & 9     & 0.7   & $7-8$ & $7-8$ \\
    \end{tabular}
\end{table}

\begin{table}[htbp]
  \centering
  \caption{CN${}^-$ ligand diagonalized hyperfine tensors, SB-BP86 + $K_{\mr{L}}$, values in MHz units.}
  \label{CN_numbers}
    \begin{tabular}{c|c|c|c|c|c}
    Configuration & Energy (meV) & ${}^{13}$C    & ${}^{15}$N  \\
    \hline
    $\beta\alpha\beta\alpha$  & 0     & [$+0.7,~+7.3,~+7.4$] & [$-0.4,~-0.5,~-2.2$]  \\
    \hline
    $\beta\alpha\alpha\beta$  & 402   & [$+2.2,~+8.2,~+8.4$] & [$-0.2,~+0.7,~-1.8$]  \\
    \hline
    $\beta\beta\alpha\alpha$  & 439   & [$+0.3,~+7.1,~+7.2$] & [$-0.5,~-0.9,~-1.9$]  \\
    \hline
    $\alpha\beta\alpha\beta$  & 143   & [$+3.8,~-4.0,~-4.3$] & [$+0.1,~+1.3,~+2.1$]  \\
    \hline
    $\alpha\alpha\beta\beta$  & 469   & [$-4.4,~+3.5,~-4.7$] & [$+0.3,~+1.1,~+2.5$]  \\
    \hline
    $\alpha\beta\beta\alpha$  & 561   & [$+2.5,~-6.1,~-6.5$] & [$+0.0,~+1.0,~+2.3$]  \\
    \hline
    Expt. Obs.                &       & [$+0.9,~-4.7,~-5.0$] & [$+1.1,~+1.1,~+2.3$]   \\
    \end{tabular}
\end{table}

At the BP86 and chosen basis set levels [EPR-III/IGLO-III/CP(PPP)], the closest value computed for
the 5'C position is 13 MHz, compared to experimental estimation of 9 MHz. For the 4'C position we
find values that closely agree with experiment, 0.7 MHz. The proton values show variability, with
values as low as 1 MHz and as high as 5 MHz. In experimental work, the ENDOR spectrum of these
protons is not well resolved, as ENDOR did not produce the two peaks that are used to extract the
tensor. It is possible that the observed computational variability serves as an explanation for the
difficulty resolving these values via ENDOR. There is possibly an spectral overlap as these protons
bound to 5'C are in different spatial locations and are in proximity to different
atomic neighbors, so their hyperfine values should be different. In a synthetic analogue, our
methodology performed better than in the RSAM enzyme, the protons in this molecule are in a more
homogeneous environment.

\begin{figure}[htp]
\centerline{\includegraphics[width=0.75\hsize]{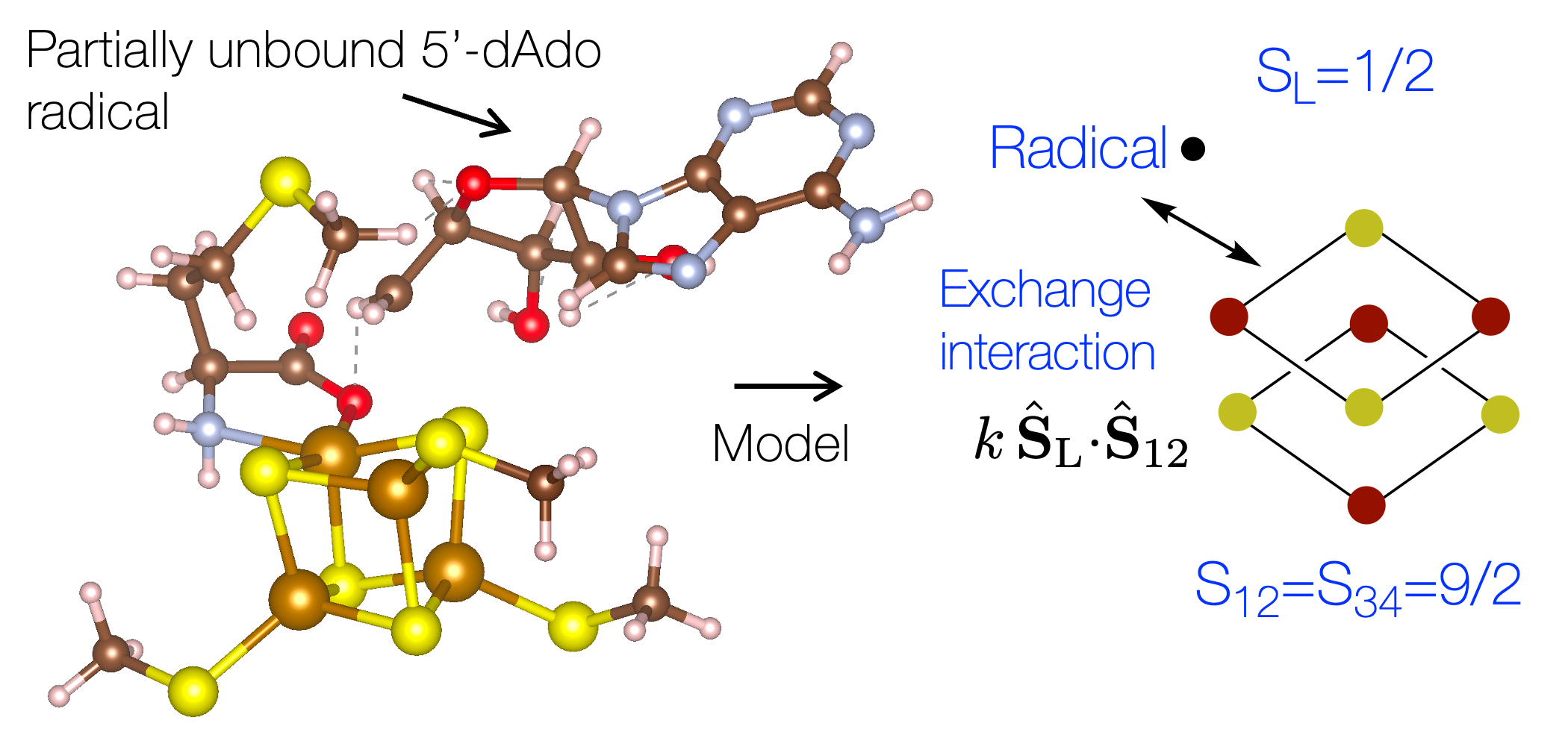}}
\caption{Structure studied in this work for the case of a partially unbound ligand in radical form.}
\label{unbound_fig}
\end{figure}

Table \ref{CN_numbers} presents SB-DFT results projected by $K_{\mr{L}}=1/3$ for the cyano ligand
bound to a reduced cluster, [4Fe-4S]${}^{+}$. In this case, BP86 gives $\beta\alpha\beta\alpha$ as
the most energetically stable, followed by $\alpha\beta\alpha\beta$. The former configuration gives
an axial hyperfine tensor for the ${}^{13}$C site. The magnitudes are somewhat overestimated with
respect to the experimental value reported by Suess et al.\cite{suess2015cysteine} for an auxiliary
cluster in the HydG enzyme. For the ${}^{15}$N position SB-DFT yields an axial tensor as well, in
agreement with experimental observation\cite{dowd2017spectroscopic} (IspH). For the [4Fe-4S] cluster
in {\slshape pyrococus furiosus}, Tesler et al.\cite{tesler1994cyanide} also report similar
experimental tensors: [0.1 -4.5 -4.5] and [1.8 1.0 -2.4] for ${}^{13}$C and ${}^{15}$N,
respectively. In this case, however, BP86 yields oppositve signs to experiment. For ligand
antiferromagnetically coupled to unique iron, this would indicate that iron atom is $\alpha$
spin-polarized. On the other hand, it can be noted that BP86 gives some variability in the first
entry of the ${}^{13}$C tensor, while it shows are more consistent behavior for the nitrogen
isotope. 

We now turn our attention to partially unbound ligand, Figure \ref{unbound_fig}. From a geometrical
perspective, the ligand in trapped radical form is challenging. What is known so far about this
system has been deduced from 5'C and methionine hyperfine tensors, giving a distance between 5'C
position and unique iron of 2.9 \AA. Here we employ such inferred geometry, except we only perform a
constrained optimization of the sulfur and hydrogen atoms that were too close to 5'C (these
positions are not resolved from experiment). In previous work, we used a two-iron modeled coupled to
a methyl radical to illustrate the activation of the nominally silent cluster by the radical. Now,
based on Eq. 15, we compute the isotropic HFCCs of the four iron atoms of the system. First, we
determine the exchange coupling constants, $j$, $k$, and $J$. This procedure gives the ratio
$k/j$ is 0.053, which lies in the regime of PT. 

In Table \ref{Fe_couplings} we show the computed values for the possible spin flip configurations of the system
$\alpha'$ refers to the $z$-spin of the organic radical being positive. In this case, the configuration
$\alpha'\alpha\beta\alpha\beta$ has the lowest energy. Employing these values and the high-spin
energy, we obtain $k=3.6~\mr{meV}$, $j=55~\mr{meV}$, and $J=21~\mr{meV}$. These lead us to find that
the ${}^{57}$Fe couplings are in the range of 6 to 9 MHz in magnitude. These are in close agreement
with experimental values that were found to be 5, 7, and 12 MHz. Our calculations, as well as
experiment, suggest that two iron atoms have overlapping HFCCs. 

\begin{table}[htbp]
  \centering
  \caption{${}^{57}$Fe HFCCs of nominally silent cluster, activated by ligand in radical state,
  hyperfine couplings in MHz units. These were computed with BHandHLYP and Eq. (\ref{eq_set}).}
\label{Fe_couplings}
    \begin{tabular}{c|c|c|c|c|c}
    Configuration & Energy (meV)    & Fe(1) & Fe(2) & Fe(3) & Fe(4) \\
    \hline
    $\alpha'\alpha\beta\alpha\beta$ & 0     & -7    & 8     & -9    & 8 \\
    \hline
    $\alpha'\beta\alpha\beta\alpha$ & 70    & 7     & -8    & 9     & -7 \\
    \hline
    $\alpha'\alpha\alpha\beta\beta$ & 281   & -6    & -9    & 8     & 7 \\
    \hline
    $\alpha'\beta\beta\alpha\alpha$ & 301   & 6     & 9     & -8    & -7 \\
    \end{tabular}%
\end{table}%

The experimental diagonalized hyperfine tensor of the 5'C position is [10, 10, 230] MHz, whereas
BP86 gives [30, 36, 219] MHz. Therefore, it can be inferred that BP86 tends to overestimate the spin
density and couplings at this position over the experiment, which also explains why the isotropic
values for the $\Omega$ intermediate are above the experimental value by a few MHz.

The hyperfine symmetry factors from this work are applicable to an arbitrary number of ligands as
long as the system has a dominant `majority' configuration (like $|\mr{SC0}\rangle$ in this case, or
$|\mr{SC}2\rangle$ in the previous one), where a projection factor is applied to data from SB-DFT.
As typical of treatments of hyperfine properties, we expect PT to be an excellent approach for
ligand HFCCs: Ligands in organometallic systems tend to have a dominant electronic configuration,
where the hyperfine signals originated from the perturbative configurations.  

The numerical value of a ligand projection factor is dependent upon type of organometallic bond and
total spin symmetry; future work will explore these generalizations, as well as the exploration of
basis sets and density functionals of higher orders. Another subject of future interest is the
potential unification of SB properties: For a given spin symmetry there should be a single set of
intrinsic hyperfine values, g-tensors, and bond lengths/angles. SB-DFT, for instance, does not yield
to a single optimized geometry, as a multireference method would.

\section{Conclusion}

This work presented a theoretical model to derive projection factors for ligand hyperfine
properties. We herein demonstrated its application to the reductive cleavage of SAM as well as a
bound cyano ligand, where this model was able to quantitatively explain experimentally observed
hyperfine coupling constants. The spin-wavefunction approach extends the conventional Heisenberg
model by introducing an additional pair of exchange and hopping parameters, however, these values
are not needed for the bound ligand case due to the application of PT and the consideration that
electrostatics outweigh exchange coupling. The present theoretical model is flexible, rigorous, and
extendable to other chemical scenarios. As an example, we applied it to a trapped 5'-dAdo radical,
where the nominally silent cluster is activated by this radical. Through the application of PT, we
were able to quantitatively explain ENDOR observations, and offer tools for investigating future
scenarios of interest.

\begin{acknowledgement}
M.A.M. and D.A.H. thank Prof. Brian Hoffman and Madeline Ho (Northwestern University) for helpful
discussions. M.A.M. thanks Dr. Hao Yang (Northwestern University) for sharing coordinates of the
trapped dAdo radical. Computational efforts were performed on the Tempest High Performance Computing
System, operated and supported by University Information Technology Research Cyberinfrastructure
(RRID:SCR\_026229) at Montana State University.
\end{acknowledgement}

\bibliography{refs}

\providecommand{\latin}[1]{#1}
\makeatletter
\providecommand{\doi}
  {\begingroup\let\do\@makeother\dospecials
  \catcode`\{=1 \catcode`\}=2 \doi@aux}
\providecommand{\doi@aux}[1]{\endgroup\texttt{#1}}
\makeatother
\providecommand*\mcitethebibliography{\thebibliography}
\csname @ifundefined\endcsname{endmcitethebibliography}
  {\let\endmcitethebibliography\endthebibliography}{}
\begin{mcitethebibliography}{60}
\providecommand*\natexlab[1]{#1}
\providecommand*\mciteSetBstSublistMode[1]{}
\providecommand*\mciteSetBstMaxWidthForm[2]{}
\providecommand*\mciteBstWouldAddEndPuncttrue
  {\def\EndOfBibitem{\unskip.}}
\providecommand*\mciteBstWouldAddEndPunctfalse
  {\let\EndOfBibitem\relax}
\providecommand*\mciteSetBstMidEndSepPunct[3]{}
\providecommand*\mciteSetBstSublistLabelBeginEnd[3]{}
\providecommand*\EndOfBibitem{}
\mciteSetBstSublistMode{f}
\mciteSetBstMaxWidthForm{subitem}{(\alph{mcitesubitemcount})}
\mciteSetBstSublistLabelBeginEnd
  {\mcitemaxwidthsubitemform\space}
  {\relax}
  {\relax}

\bibitem[Booker and Lloyd(2022)Booker, and Lloyd]{booker2022twenty}
Booker,~S.~J.; Lloyd,~C.~T. Twenty years of radical {SAM}! The genesis of the
  superfamily. 2022\relax
\mciteBstWouldAddEndPuncttrue
\mciteSetBstMidEndSepPunct{\mcitedefaultmidpunct}
{\mcitedefaultendpunct}{\mcitedefaultseppunct}\relax
\EndOfBibitem
\bibitem[Hoffman \latin{et~al.}(2023)Hoffman, Broderick, and
  Broderick]{hoffman2023mechanism}
Hoffman,~B.~M.; Broderick,~W.~E.; Broderick,~J.~B. Mechanism of radical
  initiation in the radical {SAM} enzyme superfamily. \emph{Ann. Rev. Biochem.}
  \textbf{2023}, \emph{92}, 333--349\relax
\mciteBstWouldAddEndPuncttrue
\mciteSetBstMidEndSepPunct{\mcitedefaultmidpunct}
{\mcitedefaultendpunct}{\mcitedefaultseppunct}\relax
\EndOfBibitem
\bibitem[Backman \latin{et~al.}(2017)Backman, Funk, Dawson, and
  Drennan]{backman2017new}
Backman,~L.~R.; Funk,~M.~A.; Dawson,~C.~D.; Drennan,~C.~L. New tricks for the
  glycyl radical enzyme family. \emph{Crit. Rev. Biochem. Mol. Biol.}
  \textbf{2017}, \emph{52}, 674--695\relax
\mciteBstWouldAddEndPuncttrue
\mciteSetBstMidEndSepPunct{\mcitedefaultmidpunct}
{\mcitedefaultendpunct}{\mcitedefaultseppunct}\relax
\EndOfBibitem
\bibitem[McLaughlin \latin{et~al.}(2021)McLaughlin, Pallitsch, Wallner, Van
  Der~Donk, and Hammerschmidt]{mclaughlin2021overall}
McLaughlin,~M.~I.; Pallitsch,~K.; Wallner,~G.; Van Der~Donk,~W.~A.;
  Hammerschmidt,~F. Overall retention of methyl stereochemistry during
  {B}12-dependent radical SAM methyl transfer in fosfomycin biosynthesis.
  \emph{Biochemistry} \textbf{2021}, \emph{60}, 1587--1596\relax
\mciteBstWouldAddEndPuncttrue
\mciteSetBstMidEndSepPunct{\mcitedefaultmidpunct}
{\mcitedefaultendpunct}{\mcitedefaultseppunct}\relax
\EndOfBibitem
\bibitem[Frey \latin{et~al.}(2008)Frey, Hegeman, and Ruzicka]{frey2008radical}
Frey,~P.~A.; Hegeman,~A.~D.; Ruzicka,~F.~J. The radical {SAM} superfamily.
  \emph{Crit. Rev. Biochem. Mol. Biol.} \textbf{2008}, \emph{43}, 63--88\relax
\mciteBstWouldAddEndPuncttrue
\mciteSetBstMidEndSepPunct{\mcitedefaultmidpunct}
{\mcitedefaultendpunct}{\mcitedefaultseppunct}\relax
\EndOfBibitem
\bibitem[Magnusson \latin{et~al.}(2001)Magnusson, Reed, and
  Frey]{magnusson2001characterization}
Magnusson,~O.~T.; Reed,~G.~H.; Frey,~P.~A. Characterization of an allylic
  analogue of the {$5'$}-deoxyadenosyl radical: {A}n intermediate in the
  reaction of lysine 2, 3-aminomutase. \emph{Biochemistry} \textbf{2001},
  \emph{40}, 7773--7782\relax
\mciteBstWouldAddEndPuncttrue
\mciteSetBstMidEndSepPunct{\mcitedefaultmidpunct}
{\mcitedefaultendpunct}{\mcitedefaultseppunct}\relax
\EndOfBibitem
\bibitem[Nicolet(2020)]{nicolet2020structure}
Nicolet,~Y. Structure--function relationships of radical {SAM} enzymes.
  \emph{Nat. Catal.} \textbf{2020}, \emph{3}, 337--350\relax
\mciteBstWouldAddEndPuncttrue
\mciteSetBstMidEndSepPunct{\mcitedefaultmidpunct}
{\mcitedefaultendpunct}{\mcitedefaultseppunct}\relax
\EndOfBibitem
\bibitem[Broderick and Broderick(2019)Broderick, and
  Broderick]{broderick2019radical}
Broderick,~W.~E.; Broderick,~J.~B. Radical {SAM} enzymes: surprises along the
  path to understanding mechanism. \emph{J. Biol. Inorg. Chem.} \textbf{2019},
  \emph{24}, 769--776\relax
\mciteBstWouldAddEndPuncttrue
\mciteSetBstMidEndSepPunct{\mcitedefaultmidpunct}
{\mcitedefaultendpunct}{\mcitedefaultseppunct}\relax
\EndOfBibitem
\bibitem[Zhu \latin{et~al.}(2011)Zhu, Dzikovski, Su, Torelli, Zhang, Ealick,
  Freed, and Lin]{zhu2011mechanistic}
Zhu,~X.; Dzikovski,~B.; Su,~X.; Torelli,~A.~T.; Zhang,~Y.; Ealick,~S.~E.;
  Freed,~J.~H.; Lin,~H. Mechanistic understanding of Pyrococcus horikoshii
  {Dph2}, a [4{F}e--4{S}] enzyme required for diphthamide biosynthesis.
  \emph{Mol. Biosyst.} \textbf{2011}, \emph{7}, 74--81\relax
\mciteBstWouldAddEndPuncttrue
\mciteSetBstMidEndSepPunct{\mcitedefaultmidpunct}
{\mcitedefaultendpunct}{\mcitedefaultseppunct}\relax
\EndOfBibitem
\bibitem[Crack and Le~Brun(2021)Crack, and Le~Brun]{crack2021biological}
Crack,~J.~C.; Le~Brun,~N.~E. Biological iron-sulfur clusters: {M}echanistic
  insights from mass spectrometry. \emph{Coord. Chem. Reviews} \textbf{2021},
  \emph{448}, 214171\relax
\mciteBstWouldAddEndPuncttrue
\mciteSetBstMidEndSepPunct{\mcitedefaultmidpunct}
{\mcitedefaultendpunct}{\mcitedefaultseppunct}\relax
\EndOfBibitem
\bibitem[Read \latin{et~al.}(2021)Read, Bentley, Archer, and
  Dunham-Snary]{read2021mitochondrial}
Read,~A.~D.; Bentley,~R.~E.; Archer,~S.~L.; Dunham-Snary,~K.~J. Mitochondrial
  iron--sulfur clusters: {S}tructure, function, and an emerging role in
  vascular biology. \emph{Redox Biol.} \textbf{2021}, \emph{47}, 102164\relax
\mciteBstWouldAddEndPuncttrue
\mciteSetBstMidEndSepPunct{\mcitedefaultmidpunct}
{\mcitedefaultendpunct}{\mcitedefaultseppunct}\relax
\EndOfBibitem
\bibitem[Honarmand~Ebrahimi \latin{et~al.}(2022)Honarmand~Ebrahimi,
  Ciofi-Baffoni, Hagedoorn, Nicolet, Le~Brun, Hagen, and
  Armstrong]{honarmand2022iron}
Honarmand~Ebrahimi,~K.; Ciofi-Baffoni,~S.; Hagedoorn,~P.-L.; Nicolet,~Y.;
  Le~Brun,~N.~E.; Hagen,~W.~R.; Armstrong,~F.~A. Iron--sulfur clusters as
  inhibitors and catalysts of viral replication. \emph{Nat. Chem.}
  \textbf{2022}, \emph{14}, 253--266\relax
\mciteBstWouldAddEndPuncttrue
\mciteSetBstMidEndSepPunct{\mcitedefaultmidpunct}
{\mcitedefaultendpunct}{\mcitedefaultseppunct}\relax
\EndOfBibitem
\bibitem[Boncella \latin{et~al.}(2022)Boncella, Sabo, Santore, Carter, Whalen,
  Hudspeth, and Morrison]{boncella2022expanding}
Boncella,~A.~E.; Sabo,~E.~T.; Santore,~R.~M.; Carter,~J.; Whalen,~J.;
  Hudspeth,~J.~D.; Morrison,~C.~N. The expanding utility of iron-sulfur
  clusters: {T}heir functional roles in biology, synthetic small molecules,
  maquettes and artificial proteins, biomimetic materials, and therapeutic
  strategies. \emph{Coord. Chem. Rev.} \textbf{2022}, \emph{453}, 214229\relax
\mciteBstWouldAddEndPuncttrue
\mciteSetBstMidEndSepPunct{\mcitedefaultmidpunct}
{\mcitedefaultendpunct}{\mcitedefaultseppunct}\relax
\EndOfBibitem
\bibitem[Shiozaki and Yanai(2016)Shiozaki, and Yanai]{shiozaki2016hyperfine}
Shiozaki,~T.; Yanai,~T. Hyperfine coupling constants from internally contracted
  multireference perturbation theory. \emph{J. Chem. Theory Comput.}
  \textbf{2016}, \emph{12}, 4347--4351\relax
\mciteBstWouldAddEndPuncttrue
\mciteSetBstMidEndSepPunct{\mcitedefaultmidpunct}
{\mcitedefaultendpunct}{\mcitedefaultseppunct}\relax
\EndOfBibitem
\bibitem[Birnoschi and Chilton(2022)Birnoschi, and
  Chilton]{birnoschi2022hyperion}
Birnoschi,~L.; Chilton,~N.~F. Hyperion: A new computational tool for
  relativistic ab initio hyperfine coupling. \emph{J. Chem. Theory Comput.}
  \textbf{2022}, \emph{18}, 4719--4732\relax
\mciteBstWouldAddEndPuncttrue
\mciteSetBstMidEndSepPunct{\mcitedefaultmidpunct}
{\mcitedefaultendpunct}{\mcitedefaultseppunct}\relax
\EndOfBibitem
\bibitem[Wysocki and Park(2024)Wysocki, and Park]{wysocki2024relativistic}
Wysocki,~A.~L.; Park,~K. Relativistic Douglas--Kroll--Hess calculations of
  hyperfine interactions within first-principles multireference methods.
  \emph{J. Chem. Phys.} \textbf{2024}, \emph{160}\relax
\mciteBstWouldAddEndPuncttrue
\mciteSetBstMidEndSepPunct{\mcitedefaultmidpunct}
{\mcitedefaultendpunct}{\mcitedefaultseppunct}\relax
\EndOfBibitem
\bibitem[Samanta and K{\"o}hn(2018)Samanta, and K{\"o}hn]{samanta2018first}
Samanta,~P.~K.; K{\"o}hn,~A. First-order properties from internally contracted
  multireference coupled-cluster theory with particular focus on hyperfine
  coupling tensors. \emph{J. Chem. Phys.} \textbf{2018}, \emph{149}\relax
\mciteBstWouldAddEndPuncttrue
\mciteSetBstMidEndSepPunct{\mcitedefaultmidpunct}
{\mcitedefaultendpunct}{\mcitedefaultseppunct}\relax
\EndOfBibitem
\bibitem[Noodleman(1981)]{noodleman1981valence}
Noodleman,~L. Valence bond description of antiferromagnetic coupling in
  transition metal dimers. \emph{J. Chem. Phys.} \textbf{1981}, \emph{74},
  5737--5743\relax
\mciteBstWouldAddEndPuncttrue
\mciteSetBstMidEndSepPunct{\mcitedefaultmidpunct}
{\mcitedefaultendpunct}{\mcitedefaultseppunct}\relax
\EndOfBibitem
\bibitem[Noodleman and Case(1992)Noodleman, and Case]{noodleman1992density}
Noodleman,~L.; Case,~D.~A. \emph{Adv. Inorg. Chem.}; Elsevier, 1992; Vol.~38;
  pp 423--470\relax
\mciteBstWouldAddEndPuncttrue
\mciteSetBstMidEndSepPunct{\mcitedefaultmidpunct}
{\mcitedefaultendpunct}{\mcitedefaultseppunct}\relax
\EndOfBibitem
\bibitem[Noodleman \latin{et~al.}(1991)Noodleman, Case, and
  Baerends]{noodleman1991local}
Noodleman,~L.; Case,~D.~A.; Baerends,~E.~J. \emph{Density Functional Methods in
  Chemistry}; Springer, 1991; pp 109--123\relax
\mciteBstWouldAddEndPuncttrue
\mciteSetBstMidEndSepPunct{\mcitedefaultmidpunct}
{\mcitedefaultendpunct}{\mcitedefaultseppunct}\relax
\EndOfBibitem
\bibitem[Han and Noodleman(2009)Han, and Noodleman]{han2009dft}
Han,~W.-G.; Noodleman,~L. {DFT} calculations of comparative energetics and
  {ENDOR}/M{\"o}ssbauer properties for two protonation states of the iron dimer
  cluster of ribonucleotide reductase intermediate X. \emph{Dalton Trans.}
  \textbf{2009}, 6045--6057\relax
\mciteBstWouldAddEndPuncttrue
\mciteSetBstMidEndSepPunct{\mcitedefaultmidpunct}
{\mcitedefaultendpunct}{\mcitedefaultseppunct}\relax
\EndOfBibitem
\bibitem[Shoji \latin{et~al.}(2007)Shoji, Koizumi, Kitagawa, Yamanaka, Okumura,
  and Yamaguchi]{shoji2007theory}
Shoji,~M.; Koizumi,~K.; Kitagawa,~Y.; Yamanaka,~S.; Okumura,~M.; Yamaguchi,~K.
  Theory of chemical bonds in metalloenzymes {IV}: {H}ybrid-DFT study of
  Rieske-type {[2Fe-2S]} clusters. \emph{Int. J. Quantum Chem.} \textbf{2007},
  \emph{107}, 609--627\relax
\mciteBstWouldAddEndPuncttrue
\mciteSetBstMidEndSepPunct{\mcitedefaultmidpunct}
{\mcitedefaultendpunct}{\mcitedefaultseppunct}\relax
\EndOfBibitem
\bibitem[Joshi \latin{et~al.}(2020)Joshi, Phillips, Mitchell, Christou,
  Jackson, and Peralta]{joshi2020accuracy}
Joshi,~R.~P.; Phillips,~J.~J.; Mitchell,~K.~J.; Christou,~G.; Jackson,~K.~A.;
  Peralta,~J.~E. Accuracy of density functional theory methods for the
  calculation of magnetic exchange couplings in binuclear iron ({III})
  complexes. \emph{Polyhedron} \textbf{2020}, \emph{176}, 114194\relax
\mciteBstWouldAddEndPuncttrue
\mciteSetBstMidEndSepPunct{\mcitedefaultmidpunct}
{\mcitedefaultendpunct}{\mcitedefaultseppunct}\relax
\EndOfBibitem
\bibitem[H{\"u}bner and Sauer(2002)H{\"u}bner, and Sauer]{hubner2002structure}
H{\"u}bner,~O.; Sauer,~J. Structure and thermochemistry of {Fe2S2}-/0/+ gas
  phase clusters and their fragments. B3LYP calculations. \emph{Phys. Chem.
  Chem. Phys.} \textbf{2002}, \emph{4}, 5234--5243\relax
\mciteBstWouldAddEndPuncttrue
\mciteSetBstMidEndSepPunct{\mcitedefaultmidpunct}
{\mcitedefaultendpunct}{\mcitedefaultseppunct}\relax
\EndOfBibitem
\bibitem[Pelmenschikov and Kaupp(2013)Pelmenschikov, and
  Kaupp]{pelmenschikov2013redox}
Pelmenschikov,~V.; Kaupp,~M. Redox-dependent structural transformations of the
  {[4Fe-3S]} proximal cluster in {O2}-tolerant membrane-bound
  {[NiFe]}-hydrogenase: a {DFT} study. \emph{J. Am. Chem. Soc.} \textbf{2013},
  \emph{135}, 11809--11823\relax
\mciteBstWouldAddEndPuncttrue
\mciteSetBstMidEndSepPunct{\mcitedefaultmidpunct}
{\mcitedefaultendpunct}{\mcitedefaultseppunct}\relax
\EndOfBibitem
\bibitem[Blachly \latin{et~al.}(2015)Blachly, Sandala, Giammona, Bashford,
  McCammon, and Noodleman]{blachly2015broken}
Blachly,~P.~G.; Sandala,~G.~M.; Giammona,~D.~A.; Bashford,~D.; McCammon,~J.~A.;
  Noodleman,~L. Broken-symmetry {DFT} computations for the reaction pathway of
  {IspH}, an iron--sulfur enzyme in pathogenic bacteria. \emph{Inorg. Chem.}
  \textbf{2015}, \emph{54}, 6439--6461\relax
\mciteBstWouldAddEndPuncttrue
\mciteSetBstMidEndSepPunct{\mcitedefaultmidpunct}
{\mcitedefaultendpunct}{\mcitedefaultseppunct}\relax
\EndOfBibitem
\bibitem[Niu and Ichiye(2011)Niu, and Ichiye]{niu2011density}
Niu,~S.; Ichiye,~T. Density functional theory calculations of redox properties
  of iron--sulphur protein analogues. \emph{Mol. Sim.} \textbf{2011},
  \emph{37}, 572--590\relax
\mciteBstWouldAddEndPuncttrue
\mciteSetBstMidEndSepPunct{\mcitedefaultmidpunct}
{\mcitedefaultendpunct}{\mcitedefaultseppunct}\relax
\EndOfBibitem
\bibitem[Benediktsson and Bjornsson(2020)Benediktsson, and
  Bjornsson]{benediktsson2020quantum}
Benediktsson,~B.; Bjornsson,~R. Quantum mechanics/molecular mechanics study of
  resting-state vanadium nitrogenase: molecular and electronic structure of the
  iron--vanadium cofactor. \emph{Inorg. Chem.} \textbf{2020}, \emph{59},
  11514--11527\relax
\mciteBstWouldAddEndPuncttrue
\mciteSetBstMidEndSepPunct{\mcitedefaultmidpunct}
{\mcitedefaultendpunct}{\mcitedefaultseppunct}\relax
\EndOfBibitem
\bibitem[Lovell \latin{et~al.}(2001)Lovell, Li, Liu, Case, and
  Noodleman]{lovell2001femo}
Lovell,~T.; Li,~J.; Liu,~T.; Case,~D.~A.; Noodleman,~L. FeMo cofactor of
  nitrogenase: A density functional study of states MN, MOX, MR, and MI.
  \emph{J. Am. Chem. Soc.} \textbf{2001}, \emph{123}, 12392--12410\relax
\mciteBstWouldAddEndPuncttrue
\mciteSetBstMidEndSepPunct{\mcitedefaultmidpunct}
{\mcitedefaultendpunct}{\mcitedefaultseppunct}\relax
\EndOfBibitem
\bibitem[Noodleman \latin{et~al.}(2002)Noodleman, Lovell, Liu, Himo, and
  Torres]{noodleman2002insights}
Noodleman,~L.; Lovell,~T.; Liu,~T.; Himo,~F.; Torres,~R.~A. Insights into
  properties and energetics of iron--sulfur proteins from simple clusters to
  nitrogenase. \emph{Curr. Opin. Chem. Biol.} \textbf{2002}, \emph{6},
  259--273\relax
\mciteBstWouldAddEndPuncttrue
\mciteSetBstMidEndSepPunct{\mcitedefaultmidpunct}
{\mcitedefaultendpunct}{\mcitedefaultseppunct}\relax
\EndOfBibitem
\bibitem[Beal \latin{et~al.}(2017)Beal, Corry, and
  O’Malley]{beal2017comparison}
Beal,~N.~J.; Corry,~T.~A.; O’Malley,~P.~J. Comparison between experimental
  and broken symmetry density functional theory ({BS-DFT}) calculated electron
  paramagnetic resonance ({EPR}) parameters of the {S2} state of the
  oxygen-evolving complex of photosystem {II} in Its native (calcium) and
  strontium-substituted form. \emph{J. Phys. Chem. B} \textbf{2017},
  \emph{121}, 11273--11283\relax
\mciteBstWouldAddEndPuncttrue
\mciteSetBstMidEndSepPunct{\mcitedefaultmidpunct}
{\mcitedefaultendpunct}{\mcitedefaultseppunct}\relax
\EndOfBibitem
\bibitem[Beal \latin{et~al.}(2018)Beal, Corry, and
  O’Malley]{beal2018comparison}
Beal,~N.~J.; Corry,~T.~A.; O’Malley,~P.~J. A comparison of experimental and
  broken symmetry density functional theory ({BS-DFT}) calculated electron
  paramagnetic resonance ({EPR}) parameters for intermediates involved in the
  {S2} to {S3} state transition of nature's oxygen evolving complex. \emph{J.
  Phys. Chem. B} \textbf{2018}, \emph{122}, 1394--1407\relax
\mciteBstWouldAddEndPuncttrue
\mciteSetBstMidEndSepPunct{\mcitedefaultmidpunct}
{\mcitedefaultendpunct}{\mcitedefaultseppunct}\relax
\EndOfBibitem
\bibitem[Guo \latin{et~al.}(2017)Guo, Li, He, Zhao, Gong, and
  Yang]{guo2017open}
Guo,~Y.; Li,~H.; He,~L.-L.; Zhao,~D.-X.; Gong,~L.-D.; Yang,~Z.-Z. The
  open-cubane oxo--oxyl coupling mechanism dominates photosynthetic oxygen
  evolution: a comprehensive {DFT} investigation on {O--O} bond formation in
  the {S4} state. \emph{Phys. Chem. Chem. Phys.} \textbf{2017}, \emph{19},
  13909--13923\relax
\mciteBstWouldAddEndPuncttrue
\mciteSetBstMidEndSepPunct{\mcitedefaultmidpunct}
{\mcitedefaultendpunct}{\mcitedefaultseppunct}\relax
\EndOfBibitem
\bibitem[Lohmiller \latin{et~al.}(2014)Lohmiller, Krewald, Navarro, Retegan,
  Rapatskiy, Nowaczyk, Boussac, Neese, Lubitz, Pantazis, and
  Cox]{lohmiller2014structure}
Lohmiller,~T.; Krewald,~V.; Navarro,~M.~P.; Retegan,~M.; Rapatskiy,~L.;
  Nowaczyk,~M.~M.; Boussac,~A.; Neese,~F.; Lubitz,~W.; Pantazis,~D.~A.
  \latin{et~al.}  Structure, ligands and substrate coordination of the
  oxygen-evolving complex of photosystem {II} in the {S2} state: a combined
  {EPR} and {DFT} study. \emph{Phys. Chem. Chem. Phys.} \textbf{2014},
  \emph{16}, 11877--11892\relax
\mciteBstWouldAddEndPuncttrue
\mciteSetBstMidEndSepPunct{\mcitedefaultmidpunct}
{\mcitedefaultendpunct}{\mcitedefaultseppunct}\relax
\EndOfBibitem
\bibitem[Aragoni \latin{et~al.}(2020)Aragoni, Caltagirone, Lippolis, Podda,
  Slawin, Woollins, Pintus, and Arca]{aragoni2020diradical}
Aragoni,~M.~C.; Caltagirone,~C.; Lippolis,~V.; Podda,~E.; Slawin,~A.~M.;
  Woollins,~J.~D.; Pintus,~A.; Arca,~M. Diradical character of neutral
  heteroleptic bis (1, 2-dithiolene) metal complexes: Case study of [{P}d
  ({M}e2timdt)(mnt)]({M}e2timdt= 1, 3-dimethyl-2, 4, 5-trithioxoimidazolidine;
  mnt2--= 1, 2-dicyano-1, 2-ethylenedithiolate). \emph{Inorg. Chem.}
  \textbf{2020}, \emph{59}, 17385--17401\relax
\mciteBstWouldAddEndPuncttrue
\mciteSetBstMidEndSepPunct{\mcitedefaultmidpunct}
{\mcitedefaultendpunct}{\mcitedefaultseppunct}\relax
\EndOfBibitem
\bibitem[Wang \latin{et~al.}(2015)Wang, England, Weyherm{\"u}ller, and
  Wieghardt]{wang2015electronic}
Wang,~M.; England,~J.; Weyherm{\"u}ller,~T.; Wieghardt,~K. Electronic
  Structures of ``Low-Valent'' Neutral Complexes [{NiL2}] 0 (S= 0; L= bpy,
  phen, tpy)--{A}n Experimental and DFT Computational Study. \emph{Eur. J.
  Inorg. Chem.} \textbf{2015}, \emph{2015}, 1511--1523\relax
\mciteBstWouldAddEndPuncttrue
\mciteSetBstMidEndSepPunct{\mcitedefaultmidpunct}
{\mcitedefaultendpunct}{\mcitedefaultseppunct}\relax
\EndOfBibitem
\bibitem[Han \latin{et~al.}(2005)Han, Liu, Lovell, and
  Noodleman]{han2005active}
Han,~W.-G.; Liu,~T.; Lovell,~T.; Noodleman,~L. Active site structure of class I
  ribonucleotide reductase intermediate X: A density functional theory analysis
  of structure, energetics, and spectroscopy. \emph{J. Am. Chem. Soc.}
  \textbf{2005}, \emph{127}, 15778--15790\relax
\mciteBstWouldAddEndPuncttrue
\mciteSetBstMidEndSepPunct{\mcitedefaultmidpunct}
{\mcitedefaultendpunct}{\mcitedefaultseppunct}\relax
\EndOfBibitem
\bibitem[Rapatskiy \latin{et~al.}(2015)Rapatskiy, Ames, Perez-Navarro,
  Savitsky, Griese, Weyherm\"uller, Shafaat, Hogbom, Neese, Pantazis, and
  Cox]{rapatskiy2015characterization}
Rapatskiy,~L.; Ames,~W.~M.; Perez-Navarro,~M.; Savitsky,~A.; Griese,~J.~J.;
  Weyherm\"uller,~T.; Shafaat,~H.~S.; Hogbom,~M.; Neese,~F.; Pantazis,~D.~A.
  \latin{et~al.}  Characterization of oxygen bridged manganese model complexes
  using multifrequency 17O-hyperfine {EPR} spectroscopies and density
  functional theory. \emph{J. Phys. Chem. B} \textbf{2015}, \emph{119},
  13904--13921\relax
\mciteBstWouldAddEndPuncttrue
\mciteSetBstMidEndSepPunct{\mcitedefaultmidpunct}
{\mcitedefaultendpunct}{\mcitedefaultseppunct}\relax
\EndOfBibitem
\bibitem[Jodts \latin{et~al.}(2023)Jodts, Wittkop, Ho, Broderick, Broderick,
  Hoffman, and Mosquera]{jodts2023computational}
Jodts,~R.~J.; Wittkop,~M.; Ho,~M.~B.; Broderick,~W.~E.; Broderick,~J.~B.;
  Hoffman,~B.~M.; Mosquera,~M.~A. Computational Description of Alkylated
  Iron--Sulfur Organometallic Clusters. \emph{J. Am. Chem. Soc.} \textbf{2023},
  \emph{145}, 13879--13887\relax
\mciteBstWouldAddEndPuncttrue
\mciteSetBstMidEndSepPunct{\mcitedefaultmidpunct}
{\mcitedefaultendpunct}{\mcitedefaultseppunct}\relax
\EndOfBibitem
\bibitem[Yang \latin{et~al.}(2024)Yang, Ho, Lundahl, Mosquera, Broderick,
  Broderick, and Hoffman]{yang2024endor}
Yang,~H.; Ho,~M.~B.; Lundahl,~M.~N.; Mosquera,~M.~A.; Broderick,~W.~E.;
  Broderick,~J.~B.; Hoffman,~B.~M. {ENDOR} Spectroscopy Reveals the ``Free''
  {$5'$}-Deoxyadenosyl Radical in a Radical {SAM} Enzyme Active Site Actually
  is Chaperoned by Close Interaction with the Methionine-Bound [4{Fe}--4{S}] 2+
  Cluster. \emph{J. Am. Chem. Soc.} \textbf{2024}, \emph{146}, 3710--3720\relax
\mciteBstWouldAddEndPuncttrue
\mciteSetBstMidEndSepPunct{\mcitedefaultmidpunct}
{\mcitedefaultendpunct}{\mcitedefaultseppunct}\relax
\EndOfBibitem
\bibitem[Horitani \latin{et~al.}(2016)Horitani, Shisler, Broderick, Hutcheson,
  Duschene, Marts, Hoffman, and Broderick]{horitani2016radical}
Horitani,~M.; Shisler,~K.; Broderick,~W.~E.; Hutcheson,~R.~U.; Duschene,~K.~S.;
  Marts,~A.~R.; Hoffman,~B.~M.; Broderick,~J.~B. Radical {SAM} catalysis via an
  organometallic intermediate with an Fe--[{$5'$}-C]-deoxyadenosyl bond.
  \emph{Science} \textbf{2016}, \emph{352}, 822--825\relax
\mciteBstWouldAddEndPuncttrue
\mciteSetBstMidEndSepPunct{\mcitedefaultmidpunct}
{\mcitedefaultendpunct}{\mcitedefaultseppunct}\relax
\EndOfBibitem
\bibitem[Byer \latin{et~al.}(2018)Byer, Yang, McDaniel, Kathiresan, Impano,
  Pagnier, Watts, Denler, Vagstad, Piel, Duschene, Shepard, Shields, Scott,
  Lilla, Yokoyama, Broderick, Hoffman, and Broderick]{byer2018paradigm}
Byer,~A.~S.; Yang,~H.; McDaniel,~E.~C.; Kathiresan,~V.; Impano,~S.;
  Pagnier,~A.; Watts,~H.; Denler,~C.; Vagstad,~A.~L.; Piel,~J. \latin{et~al.}
  Paradigm shift for radical {S}-adenosyl-{L}-methionine reactions: The
  organometallic intermediate {$\Omega$} is central to catalysis. \emph{J. Am.
  Chem. Soc.} \textbf{2018}, \emph{140}, 8634--8638\relax
\mciteBstWouldAddEndPuncttrue
\mciteSetBstMidEndSepPunct{\mcitedefaultmidpunct}
{\mcitedefaultendpunct}{\mcitedefaultseppunct}\relax
\EndOfBibitem
\bibitem[Broderick \latin{et~al.}(2018)Broderick, Hoffman, and
  Broderick]{broderick2018mechanism}
Broderick,~W.~E.; Hoffman,~B.~M.; Broderick,~J.~B. Mechanism of radical
  initiation in the radical {S}-adenosyl-{L}-methionine superfamily. \emph{Acc.
  Chem. Res.} \textbf{2018}, \emph{51}, 2611--2619\relax
\mciteBstWouldAddEndPuncttrue
\mciteSetBstMidEndSepPunct{\mcitedefaultmidpunct}
{\mcitedefaultendpunct}{\mcitedefaultseppunct}\relax
\EndOfBibitem
\bibitem[Tchougreeff and Misurkin(1992)Tchougreeff, and
  Misurkin]{tchougreeff1992intra}
Tchougreeff,~A.; Misurkin,~I. Intra-atomic exchange and ferromagnetic
  interaction in metallocene-based donor-acceptor stacked crystals. \emph{Phys.
  Rev. B} \textbf{1992}, \emph{46}, 5357\relax
\mciteBstWouldAddEndPuncttrue
\mciteSetBstMidEndSepPunct{\mcitedefaultmidpunct}
{\mcitedefaultendpunct}{\mcitedefaultseppunct}\relax
\EndOfBibitem
\bibitem[Tchougreeff(1992)]{tchougreeff1992heisenberg}
Tchougreeff,~A. Heisenberg Hamiltonian for charge-transfer organometallic
  ferromagnets. \emph{J. Chem. Phys.} \textbf{1992}, \emph{96},
  6026--6032\relax
\mciteBstWouldAddEndPuncttrue
\mciteSetBstMidEndSepPunct{\mcitedefaultmidpunct}
{\mcitedefaultendpunct}{\mcitedefaultseppunct}\relax
\EndOfBibitem
\bibitem[Tchougreeff and Misurkin(1991)Tchougreeff, and
  Misurkin]{tchougreeff1991ferromagnetism}
Tchougreeff,~A.; Misurkin,~I. Ferromagnetism of charge-transfer crystals.
  \emph{Chem. Phys.} \textbf{1991}, \emph{153}, 371--378\relax
\mciteBstWouldAddEndPuncttrue
\mciteSetBstMidEndSepPunct{\mcitedefaultmidpunct}
{\mcitedefaultendpunct}{\mcitedefaultseppunct}\relax
\EndOfBibitem
\bibitem[Schumann and Zwicker(2010)Schumann, and Zwicker]{schumann2010hubbard}
Schumann,~R.; Zwicker,~D. The {Hubbard} model extended by nearest-neighbor
  {C}oulomb and exchange interaction on a cubic cluster--rigorous and exact
  results. \emph{Ann. Phys.} \textbf{2010}, \emph{522}, 419--439\relax
\mciteBstWouldAddEndPuncttrue
\mciteSetBstMidEndSepPunct{\mcitedefaultmidpunct}
{\mcitedefaultendpunct}{\mcitedefaultseppunct}\relax
\EndOfBibitem
\bibitem[de~Boer and Schadschneider(1995)de~Boer, and
  Schadschneider]{de1995exact}
de~Boer,~J.; Schadschneider,~A. Exact ground states of generalized {H}ubbard
  models. \emph{Phys. Rev. Lett.} \textbf{1995}, \emph{75}, 4298\relax
\mciteBstWouldAddEndPuncttrue
\mciteSetBstMidEndSepPunct{\mcitedefaultmidpunct}
{\mcitedefaultendpunct}{\mcitedefaultseppunct}\relax
\EndOfBibitem
\bibitem[Montorsi and Campbell(1996)Montorsi, and
  Campbell]{montorsi1996rigorous}
Montorsi,~A.; Campbell,~D.~K. Rigorous results on superconducting ground states
  for attractive extended {H}ubbard models. \emph{Phys. Rev. B} \textbf{1996},
  \emph{53}, 5153\relax
\mciteBstWouldAddEndPuncttrue
\mciteSetBstMidEndSepPunct{\mcitedefaultmidpunct}
{\mcitedefaultendpunct}{\mcitedefaultseppunct}\relax
\EndOfBibitem
\bibitem[Ozaki(1992)]{ozaki1992broken}
Ozaki,~M.-A. Broken symmetry solutions of the extended {H}ubbard model.
  \emph{Int. J. Quantum Chem.} \textbf{1992}, \emph{42}, 55--85\relax
\mciteBstWouldAddEndPuncttrue
\mciteSetBstMidEndSepPunct{\mcitedefaultmidpunct}
{\mcitedefaultendpunct}{\mcitedefaultseppunct}\relax
\EndOfBibitem
\bibitem[Mouesca \latin{et~al.}(1994)Mouesca, Chen, Noodleman, Bashford, and
  Case]{mouesca1994density}
Mouesca,~J.-M.; Chen,~J.~L.; Noodleman,~L.; Bashford,~D.; Case,~D.~A. Density
  functional/Poisson-Boltzmann calculations of redox potentials for iron-sulfur
  clusters. \emph{J. Am. Chem. Soc.} \textbf{1994}, \emph{116},
  11898--11914\relax
\mciteBstWouldAddEndPuncttrue
\mciteSetBstMidEndSepPunct{\mcitedefaultmidpunct}
{\mcitedefaultendpunct}{\mcitedefaultseppunct}\relax
\EndOfBibitem
\bibitem[Stone and Wood(1980)Stone, and Wood]{stone1980root}
Stone,~A.; Wood,~C. Root-rational-fraction package for exact calculation of
  vector-coupling coefficients. \emph{Comput. Phys. Commun.} \textbf{1980},
  \emph{21}, 195--205\relax
\mciteBstWouldAddEndPuncttrue
\mciteSetBstMidEndSepPunct{\mcitedefaultmidpunct}
{\mcitedefaultendpunct}{\mcitedefaultseppunct}\relax
\EndOfBibitem
\bibitem[Neese(2022)]{neese2022software}
Neese,~F. Software update: {T}he {ORCA} program system—Version 5.0.
  \emph{WIRES: Comput. Mol. Sci.} \textbf{2022}, \emph{12}, e1606\relax
\mciteBstWouldAddEndPuncttrue
\mciteSetBstMidEndSepPunct{\mcitedefaultmidpunct}
{\mcitedefaultendpunct}{\mcitedefaultseppunct}\relax
\EndOfBibitem
\bibitem[Barone(1996)]{epriii}
Barone,~V. In \emph{Recent Advances in Density Functional Methods}; Chong,~P.,
  Ed.; World Scientific, 1996; pp 287--334\relax
\mciteBstWouldAddEndPuncttrue
\mciteSetBstMidEndSepPunct{\mcitedefaultmidpunct}
{\mcitedefaultendpunct}{\mcitedefaultseppunct}\relax
\EndOfBibitem
\bibitem[Schindler and Kutzelnigg(1982)Schindler, and Kutzelnigg]{igloiii}
Schindler,~M.; Kutzelnigg,~W. Theory of magnetic susceptibilities and NMR
  chemical shifts in terms of localized quantities. II. Application to some
  simple molecules. \emph{J. Chem. Phys.} \textbf{1982}, \emph{76},
  1919--1933\relax
\mciteBstWouldAddEndPuncttrue
\mciteSetBstMidEndSepPunct{\mcitedefaultmidpunct}
{\mcitedefaultendpunct}{\mcitedefaultseppunct}\relax
\EndOfBibitem
\bibitem[Neese(2002)]{cpppp}
Neese,~F. Prediction and interpretation of the 57Fe isomer shift in
  M{\"o}ssbauer spectra by density functional theory. \emph{Inorg. Chim. Acta}
  \textbf{2002}, \emph{337}, 181--192\relax
\mciteBstWouldAddEndPuncttrue
\mciteSetBstMidEndSepPunct{\mcitedefaultmidpunct}
{\mcitedefaultendpunct}{\mcitedefaultseppunct}\relax
\EndOfBibitem
\bibitem[Suess \latin{et~al.}(2015)Suess, B{\"u}rstel, De~La~Paz,
  Kuchenreuther, Pham, Cramer, Swartz, and Britt]{suess2015cysteine}
Suess,~D.~L.; B{\"u}rstel,~I.; De~La~Paz,~L.; Kuchenreuther,~J.~M.;
  Pham,~C.~C.; Cramer,~S.~P.; Swartz,~J.~R.; Britt,~R.~D. Cysteine as a ligand
  platform in the biosynthesis of the FeFe hydrogenase H cluster. \emph{PNAS}
  \textbf{2015}, \emph{112}, 11455--11460\relax
\mciteBstWouldAddEndPuncttrue
\mciteSetBstMidEndSepPunct{\mcitedefaultmidpunct}
{\mcitedefaultendpunct}{\mcitedefaultseppunct}\relax
\EndOfBibitem
\bibitem[O'Dowd \latin{et~al.}(2017)O'Dowd, Williams, Wang, No, Rao, Wang,
  McCammon, Cramer, and Oldfield]{dowd2017spectroscopic}
O'Dowd,~B.; Williams,~S.; Wang,~H.; No,~J.~H.; Rao,~G.; Wang,~W.;
  McCammon,~J.~A.; Cramer,~S.~P.; Oldfield,~E. Spectroscopic and Computational
  Investigations of Ligand Binding to IspH: Discovery of Non-diphosphate
  Inhibitors. \emph{ChemBioChem} \textbf{2017}, \emph{18}, 914--920\relax
\mciteBstWouldAddEndPuncttrue
\mciteSetBstMidEndSepPunct{\mcitedefaultmidpunct}
{\mcitedefaultendpunct}{\mcitedefaultseppunct}\relax
\EndOfBibitem
\bibitem[Tesler \latin{et~al.}(1995)Tesler, Smith, Adams, Conover, Johnson, and
  Hoffman]{tesler1994cyanide}
Tesler,~J.; Smith,~E.~T.; Adams,~M. W.~W.; Conover,~R.~C.; Johnson,~M.~K.;
  Hoffman,~B.~M. Cyanide Binding to the Novel 4Fe Ferredoxin from Pyrococcus
  furiosus: Investigation by EPR and ENDOR Spectroscopy. \emph{J. Am. Chem.
  Soc.} \textbf{1995}, \emph{117}, 5133--5140\relax
\mciteBstWouldAddEndPuncttrue
\mciteSetBstMidEndSepPunct{\mcitedefaultmidpunct}
{\mcitedefaultendpunct}{\mcitedefaultseppunct}\relax
\EndOfBibitem
\end{mcitethebibliography}

\newpage

\setcounter{equation}{0}
\setcounter{table}{0}
\setcounter{figure}{0}
\setcounter{section}{0}

\renewcommand{\theequation}{S\arabic{equation}}
\renewcommand{\thetable}{S\arabic{table}}
\renewcommand{\thefigure}{S\arabic{figure}}

\setlength{\parindent}{0pt}

\begin{center}
{\bfseries \Large Supporting Information: Electron-Transfer and Exchange-Interaction Model of the Ligand Hyperfine Structure of
Alkylated Iron-Sulfur Clusters}
\end{center}

\singlespacing

\section{Radical Coupled to Two-Iron Cluster}
Let us begin with a two state model, we will use \(|S,M\rangle\) notation. We define $|\mr{SC0}\rangle$ and
$|\mr{SC1}\rangle$:

\begin{equation}
    |\mathrm{SC0}\rangle = |0,0\rangle_{\mr{cluster}}|\frac{1}{2},\frac{1}{2}\rangle_{\mr{alkyl}}
\end{equation}

\begin{equation}
    |\mr{SC1}\rangle = \sqrt{\tfrac{2}{3}} |1,1\rangle_{\mr{cluster}} |\tfrac{1}{2}, -\tfrac{1}{2}\rangle_{\mr{alkyl}} -
    \sqrt{\tfrac{1}{3}}|1,0\rangle_{\mr{cluster}} |\tfrac{1}{2},\tfrac{1}{2}\rangle_{\mr{alkyl}}
\end{equation}

The Hamiltonian can be written as 

\begin{equation}
    \hat{H} = \hat{H}_0 + \hat{V}
\end{equation}

where $\hat{H}_0=J\hat{\mb{S}}_1\cdot\hat{\mb{S}}_2$ and $\hat{V}=K\hat{\mb{S}}_1\cdot\hat{\mb{S}}_3$. Thus,

\begin{equation}
    \hat{H}=J\hat{\mb{S}}_1\cdot\hat{\mb{S}}_2 + K\hat{\mb{S}}_1\cdot\hat{\mb{S}}_3
\end{equation}

\section{Perturbation Theory}

Let the peturbed system be described by 

\begin{equation}
    |\Psi\rangle = |\mathrm{SC0}\rangle + C_1|\mathrm{SC1}\rangle
\end{equation}

\begin{figure}
    \centering
    \includegraphics[width=0.7\textwidth]{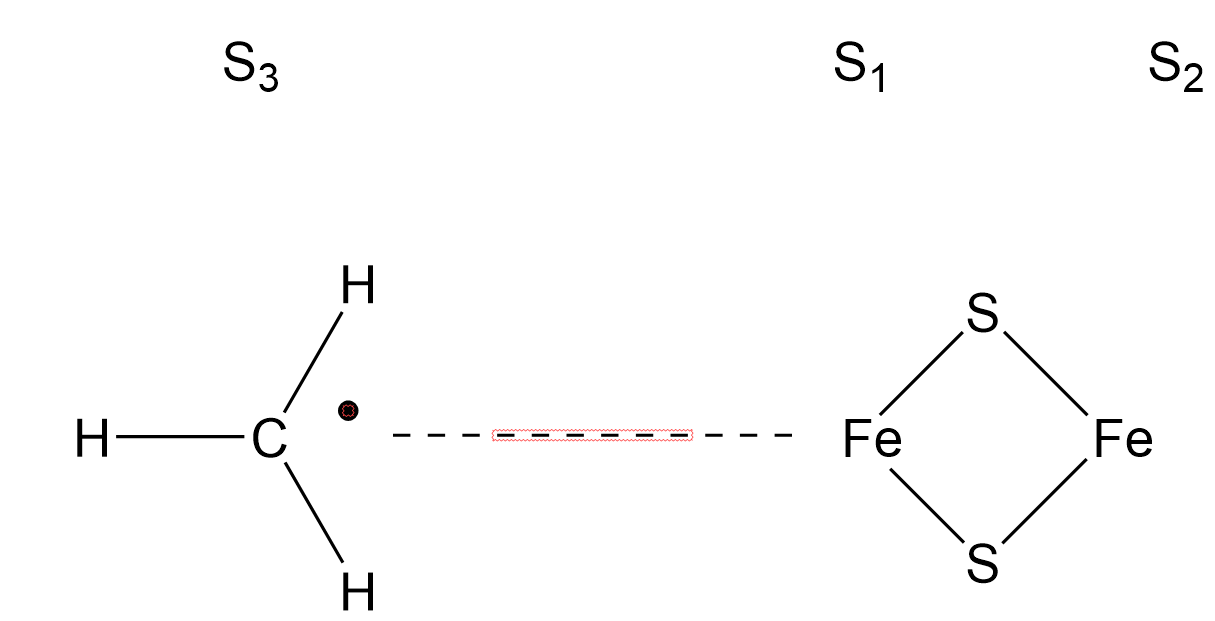}
    \caption{Two $S=2$ iron system with methyl radical (this can replaced by a general alkyl-like
    radical as they re covered by
    this model),
    \(S_{\mr{Total}} \equiv S_{\mr{T}} = \frac{1}{2}\).}
    \label{sfig_methyl}
\end{figure}

Then by perturbation theory we can obtain an expression for $C_1$:

\begin{equation}
    C_1 = \frac{\langle \mathrm{SC1}|\hat{V}|\mathrm{SC0}\rangle}{E_1 - E_0}
\end{equation}

\begin{equation}
    E_0 = \langle \mathrm{SC0} |J\hat{\mb{S}}_1\cdot\hat{\mb{S}}_2|\mathrm{SC0}\rangle = \tfrac{J}{2} \langle \mr{SC}0|
    \hat{\mb{S}}_{12}^2 - \hat{\mb{S}}_1^2 - \hat{\mb{S}}_1^2|\mr{SC}0 \rangle = -6J
\end{equation}

and

\begin{equation}
    E_1 = \langle \mathrm{SC1}|J \hat{\mb{S}}_1\cdot\hat{\mb{S}}_2 | \mathrm{SC1}\rangle = \tfrac{J}{2} [6-6-6] = -3J
\end{equation}

which means that 

\begin{equation}
    E_1 - E_0 = 3J
\end{equation}

For the calculation of exchange coupling between $S_3$ (the alkyl radical spin) and $S_1$, the so-called
unique Fe atom, we will need to obtain the Clebsch-Gordan coefficients (CGCs) a few times along the
way. What follows are the tables that are the result of Clebsch-Gordan (CG) Algebra. We do this now for
$|0,0\rangle_{\mr{cluster}}$ term in $\mr{SC}0$, $|1,1\rangle_{\mr{cluster}}$ term in $\mr{SC}1$,
and the $|1,0\rangle_{\mr{cluster}}$ term in $\mr{SC}1$.
\begin{table}[htp]
    \centering
    \begin{subtable}{0.32\textwidth}
        \centering
        \begin{tabular}{ccc}
            \toprule
            $m_1$ & $m_2$ & CGC \\
            \midrule
             2  & -2 & $\sqrt{\frac{1}{5}}$ \\
             1  & -1 & $-\sqrt{\frac{1}{5}}$ \\
             0  &  0 & $\sqrt{\frac{1}{5}}$ \\
            -1  &  1 & $-\sqrt{\frac{1}{5}}$ \\
            -2  &  2 & $\sqrt{\frac{1}{5}}$ \\
            \bottomrule
        \end{tabular}
        \caption{For $|0,0\rangle_{\mr{cluster}}$ in $\mr{SC}0$ with $S_{12}=0$.}
        \label{tab:cgc_constant}
    \end{subtable}
    \hfill
    \begin{subtable}{0.32\textwidth}
        \centering
        \begin{tabular}{ccc}
            \toprule
            $m_1$ & $m_2$ & CGC \\
            \midrule
             2  & -1 & $\sqrt{\frac{1}{5}}$ \\
             1  &  0 & $-\sqrt{\frac{3}{10}}$ \\
             0  &  1 & $\sqrt{\frac{3}{10}}$ \\
            -1  &  2 & $-\sqrt{\frac{1}{5}}$ \\
            \bottomrule
        \end{tabular}
        \caption{For $|1,1\rangle_{\mr{cluster}}$ in $\mr{SC}1$ with $S_{12}=1$.}
        \label{tab:cgc_asym_repeat}
    \end{subtable}
    \hfill
    \begin{subtable}{0.32\textwidth}
        \centering
        \begin{tabular}{ccc}
            \toprule
            $m_1$ & $m_2$ & CGC \\
            \midrule
            2   & -2 & $\sqrt{\frac{2}{5}}$ \\
            1   & -1 & $-\sqrt{\frac{1}{10}}$ \\
            0   &  0 & $0$ \\
            -1  &  1 & $\sqrt{\frac{1}{10}}$ \\
            -2  &  2 & $-\sqrt{\frac{2}{5}}$ \\
            \bottomrule
        \end{tabular}
        \caption{For $|1,0\rangle_{\mr{cluster}}$ in $\mr{SC}1$ with $S_{12}=1$.}
        \label{tab:cgc_zero_msum}
    \end{subtable}
    \caption{Coefficients for several cluster states}
\end{table}

Before we tabulate more calculated CGCs, it will be helpful to review some definitions regarding the
spin operator.

\begin{equation}
    \mb{S}_1   \cdot \mb{S}_3 = S_{x,1}S_{x,3} + S_{y,1}S_{y,3} + S_{z,1}S_{z,3}
\end{equation}
and 
\begin{equation}
    S_+ = S_x + iS_y \wedge S_-=S_x-iS_y
\end{equation}
\begin{equation}
   \Rightarrow S_x = \tfrac{1}{2}(S_++S_-) \wedge S_y=\tfrac{1}{2i}(S_+-S_-)
\end{equation}

From these relations, it is known that:

\begin{equation}
    \mb{S}_1\cdot\mb{S}_3 = \tfrac{1}{2}[S_{+,1}S_{-,3} + S_{+,3}S_{-,1}] +  S_{z,1}S_{z,3}
\end{equation}

 \noindent and with this we can return to the numerator for the coefficient of our first-order perturbation to make some substitutions

\begin{align}
\langle \mathrm{SC1} | \hat{V} | \mathrm{SC0}\rangle &= \sqrt{\tfrac{2}{3}} \langle1,1|\langle \tfrac{1}{2},-\tfrac{1}{2}|\hat{V} |0,0\rangle|\tfrac{1}{2},\tfrac{1}{2}\rangle \nonumber \\
&\quad - \sqrt{\tfrac{1}{3}} \langle1,0|\langle\tfrac{1}{2},\tfrac{1}{2}|\hat{V}| 0,0\rangle |\tfrac{1}{2},\tfrac{1}{2}\rangle
\end{align}

\noindent and knowing that
\begin{align}
    \hat{S}_+|S,m\rangle &= [{S(S+1) - m(m+1)}]^{\frac{1}{2}}\,|S,m+1\rangle \label{eq:Splus} \\
    \hat{S}_-|S,m\rangle &= [{S(S+1) - m(m-1)}]^{\frac{1}{2}}\,|S,m-1\rangle \label{eq:Sminus}
\end{align}

\noindent By substituting S15 and S16 into S14 and using equations S10 through S13 we obtain the following. 

\begin{align}
    \langle \mr{SC}1 | \hat{V} | \mr{SC}0\rangle &= \tfrac{K}{2}\sqrt{\tfrac{2}{3}}[\tfrac{3}{4}-\tfrac{1}{2}(-\tfrac{1}{2})]^{\tfrac{1}{2}}\langle1,1|S_{+,1}|0,0\rangle \nonumber \\
    &\quad -\tfrac{K}{\sqrt{3}}\tfrac{1}{2}\langle1,0|S_{z,1}|0,0\rangle
\end{align}

This leaves us with $\langle1,1|S_{+,1}|0,0\rangle$ and $\langle1,0|S_{z,1}|0,0\rangle$ to evaluate. The latter is straightforward and we obtain the following. 
\begin{equation}
    \langle1,0|\hat{S}_{z,1}|0,0\rangle = \sqrt{2}
\end{equation}
\noindent similarly $\langle1,0|\hat{S}_{z,2}|0,0\rangle = -\sqrt{2}$ because $\hat{S}_{z,\mr{T}}|0,0\rangle=0$. To
calculate $\langle1,1 |\hat{S}_{+,1}| 0,0\rangle$ calls for CG algebra again. For
$|0,0\rangle_{\mr{cluster}}$ with $S_1=2$, the CGCs obtained follow.
\begin{table}[htp]
    \centering
    \begin{tabular}{ccccc}
        \toprule
        $m_1$ & $m_2$ & CGC & $m_1(m_1+1)$ & $S_{+,1}$ prefactor \\
        \midrule
        2   & -2 & $\sqrt{\frac{1}{5}}$       &                 &                 \\
        1   & -1 & $-\sqrt{\frac{1}{5}}$      & 2               & 2               \\
        0   &  0 & $\sqrt{\frac{1}{5}}$       & 0 & $\sqrt{6}$ \\
        -1  &  1 & $-\sqrt{\frac{1}{5}}$      & 0               & $\sqrt{6}$      \\
        -2  &  2 & $\sqrt{\frac{1}{5}}$       & 2               & 2               \\
        \bottomrule
    \end{tabular}
    \caption{For $|0,0\rangle_{\mr{cluster}}$ with effect of $S_{+,1}$}
    \label{tab:cgc_extended}
\end{table}
\begin{table}[htp]
    \centering
    \begin{tabular}{ccc}
        \toprule
        $m_1$ & $m_2$ & CGC \\
        \midrule
        2   & -1 & $\sqrt{\frac{1}{5}}$ \\
        1   &  0 & $-\sqrt{\frac{3}{10}}$ \\
        0   &  1 & $\sqrt{\frac{3}{10}}$ \\
        -1  &  2 & $\sqrt{\frac{1}{5}}$ \\
        \bottomrule
    \end{tabular}
    \caption{For the $|1,1\rangle_{\mr{cluster}}$ the CGCs obtained follow.}
    \label{tab:cgc}
\end{table}

With these values we can arrive at the following

\begin{equation}
    \langle1,1 |\hat{S}_{+,1}| 0,0\rangle = -2
\end{equation}

Now we can substitute back into equation S14. 

\begin{equation}
    \langle \mr{SC}1 | \hat{V} | \mr{SC}0\rangle = -\sqrt{\frac{3}{2}}K
\end{equation}

And we can finally calculate the first order perturbation coeffienct, $C_1$

\begin{equation}
    C_1 = \sqrt{\frac{1}{6}}\frac{K}{J}
\end{equation}

\section{Calculation of Hyperfine Terms}

We now move on to the calculation of an observable isotropic hyperfine terms $a_i^{\mr{obs}}$ from
the calculated intrinsic value $a^{\mr{site}}_i$. 
\begin{align}
\langle \hat{S}_{z,1}\rangle &= C1 \langle \mr{SC}1|\hat{S}_{z,1}|\mr{SC}0\rangle + \mr{c.c.} \\
&\wedge \langle \mr{SC}1 |\hat{S}_{z,1}| \mr{SC}0\rangle = -\sqrt{\frac{1}{3}} \langle
1,0|\hat{S}_{z,1}| 0,0\rangle\\
&\Rightarrow \langle \hat{S}_{z,1}\rangle= -\sqrt{\frac{1}{3}} C_1[\langle 1,0|\hat{S}_{z,1}|
0,0\rangle +\mr{c.c.}]
\end{align}
Now recall from earlier that $\langle 1,0| S_{z,1} |0,0\rangle = \sqrt{2}$ and that $\langle
1,0|\hat{S}_{x/y,1}|0,0 \rangle$. This means we can begin to evaluate the following expression for
$\langle \Psi|a_i^{\mr{obs}}\hat{S}_{z,\mr{T}}|\Psi\rangle$.

\begin{align}
   \nonumber\langle \Psi|a_i^{\mr{obs}}\hat{S}_{z,\mr{T}}|\Psi\rangle &= a_i^{\mr{obs}} [\frac{1}{2} \langle \Psi | \Psi\rangle]\\
                                                            &= \frac{1}{2}a_i^{\mr{obs}}
\end{align}
Similarly for $\langle\Psi |a_i^{\mr{site}}| \Psi\rangle$ we obtain:

\begin{align}
    \langle\Psi |a_i^{\mr{site}} \hat{S}_{z,i}| \Psi\rangle = a_i^{\mr{site}} C_1 [\langle \mr{SC}1
    |\hat{S}_{z,i}| \mr{SC}0\rangle + \mr{c.c.}]
\end{align}

\noindent which we found earlier was $\langle \mr{SC}1 |\hat{S}_{z,1}| \mr{SC}0 \rangle =
-\sqrt{\frac{2}{3}}$. Using some algebra we can obtain a relationship between $a_1^{\mr{obs}}$ and
$a_1^{\mr{site}}$ (same for the second site): 

\begin{align}
    \nonumber&a_1^{\mr{obs}} = -\frac{4}{3} \frac{K}{J}a_1^{\mr{site}}\\
    &a_2^{\mr{obs}} = \frac{4}{3} \frac{K}{J}a_1^{\mr{site}}
\end{align}
This shows the dependency of the observable hyperfine property of a metal atom on the ratio of
exchange coupling constants (at infinite separation there are no signals, as expected).
In symmetry-broken density functional theory (SB-DFT) we locally set the spin configuration of the
three-site system and we follow the labelling laid out in Figure \ref{sfig_methyl}. For example,
\begin{align}
    \nonumber&|\alpha,\alpha,\alpha\rangle \rightarrow |m_{1}=2,m_{2}=2,m_{3}=\tfrac{1}{2}\rangle \\
    \nonumber&|\beta,\alpha,\alpha\rangle \rightarrow |m_{1}=-2,m_{2}=2,m_{3}=\tfrac{1}{2}\rangle \\
    \nonumber&|\alpha,\beta,\alpha\rangle \rightarrow |m_{1}=2,m_{2}=-2,m_{3}=\tfrac{1}{2}\rangle 
\end{align}

\noindent Locally the SB-DFT states describe states with full spin density and thus give us $a_i^{\mr{site}}$

\subsection{Anion Case}

Let us introduce a third spin configuration
$|\mr{SC}2\rangle=|1/2,1/2\rangle_{\mr{cluster}}|0,0\rangle_{\mr{alkyl}}$ that represents the ligand
as anion interacting with the $S_{\mr{cluster}} = \frac{1}{2}$ state. Our new wavefunction takes the
following form:
\begin{equation}
    |\Psi\rangle = C_0|\mathrm{SC0}\rangle + C_1|\mathrm{SC1}\rangle + C_2 |\mathrm{SC2}\rangle
\end{equation}

\noindent and with this we can introduce a new term to our Hamiltonian that accounts for hopping,
$t$. The new Hamiltonian is thus:

\begin{equation}
    \hat{H} = \hat{H}_0 + K\hat{\mb{S}}_3\cdot\hat{\mb{S}}_1 + J\hat{\mb{S}}_1\cdot\hat{\mb{S}}_2 - 
    t \hat{a}_1^{\dagger}\hat{a}_3 - \mr{H.c.}
\end{equation}

We will leave $\hat{H}_0$ (remainder energetic effects) for now and focus on calculating the CGCs
for $\mathrm{SC2}$ where we have $S_{\mr{cluster}}=1/2$ and $S_{\mr{L}}=0$. In this anionic case, we
have $\mr{Fe}^{3+}$ where $S =\frac{5}{2}$.

We also have a $\mr{Fe}^{2+}$ to deal with, and so we must find CGCs for adaptations related to
\begin{equation}
    |\phi\rangle=|\uparrow\downarrow\rangle_d|S_{1+}=2,m_{1+}\rangle
\end{equation}
where $|\rangle_d$ refers to the $d$-like orbital that the electron is abstracted from,
and we have to do so for each possible $m_1$ value, which is summarized in Tables
\ref{cgc_e_1}-\ref{cgc_e_5}.
We denote the secondary spin number for the $d$ electron as $m_d$, and the secondary spin of the
remaining electrons in Fe1 as $m_{1+}$.

\begin{table}[htp]
    \centering
    \begin{tabular}{ccc}
        \toprule
        $m_1$ & $m_2$ & CGC \\
        \midrule
        $\frac{5}{2}$   & -2 & $\sqrt{\frac{1}{3}}$ \\
        $\frac{3}{2}$  & -1 & $-\sqrt{\frac{4}{15}}$\\
        $\frac{1}{2}$  &  0 & $\sqrt{\frac{1}{5}}$                  \\
        $-\frac{1}{2}$  &  1 & $-\sqrt{\frac{2}{15}}$\\
        $-\frac{3}{2}$  &  2 & $\sqrt{\frac{1}{15}}$ \\
        \bottomrule
    \end{tabular}
    \caption{CGCs for the $|S_{\mr{cluster}} =
    \tfrac{1}{2},m_{\mr{cluster}}=\tfrac{1}{2}\rangle|S_{\mr{L}}=0,m_{\mr{L}}=0\rangle$}
    \label{tab:cgc_extended2}
\end{table}

\begin{table}[htp]
    \centering
    \begin{tabular}{ccc}
        \toprule
        & $m_1=\frac{5}{2}$ & \\
        \midrule
        $m_d$   & $m_{1+}$ & $C$ \\
        \midrule
        $\frac{1}{2}$  & 2 & $1$\\
        \bottomrule
    \end{tabular}
    \caption{CGCs for 30 with $m_1=\frac{5}{2}$}
    \label{cgc_e_1}
\end{table}
\begin{table}[htp]
    \centering
    \begin{tabular}{ccc}
        \toprule
        & $m_1=\frac{3}{2}$ & \\
        \midrule
        $m_d$   & $m_{1+}$ & $C$ \\
        \midrule
        $\frac{1}{2}$  & 1 & $\sqrt{\frac{1}{5}}$\\
        $-\frac{1}{2}$  & 2 & $\sqrt{\frac{4}{5}}$\\
        \bottomrule
    \end{tabular}
    \caption{CGCs for 30 with $m_1=\frac{3}{2}$}
    \label{cgc_e_2}
\end{table}
\begin{table}[htp]
    \centering
    \begin{tabular}{ccc}
        \toprule
        & $m_1=\frac{1}{2}$ & \\
        \midrule
        $m_d$   & $m_{1+}$ & $C$ \\
        \midrule
        $\frac{1}{2}$  & 0 & $\sqrt{\frac{3}{5}}$\\
        $-\frac{1}{2}$  & 1 & $\sqrt{\frac{2}{5}}$\\
        \bottomrule
    \end{tabular}
    \caption{CGCs for 30 with $m_1=\frac{1}{2}$}
    \label{cgc_e_3}
\end{table}
\begin{table}[htp]
    \centering
    \begin{tabular}{ccc}
        \toprule
        & $m_1=-\frac{1}{2}$ & \\
        \midrule
        $m_d$   & $m_{1+}$ & $C$ \\
        \midrule
        $\frac{1}{2}$  & -1 & $\sqrt{\frac{2}{5}}$\\
        $-\frac{1}{2}$  & 0 & $\sqrt{\frac{3}{5}}$\\
        \bottomrule
    \end{tabular}
    \caption{CGCs for 30 with $m_1={-\frac{1}{2}}$}
    \label{cgc_e_4}
\end{table}
\begin{table}[htp]
    \centering
    \begin{tabular}{ccc}
        \toprule
        & $m_1=-\frac{3}{2}$ & \\
        \midrule
        $m_d$   & $m_{1+}$ & $C$ \\
        \midrule
        $\frac{1}{2}$  & -2 & $\sqrt{\frac{1}{5}}$\\
        $-\frac{1}{2}$  & -1 & $\sqrt{\frac{4}{5}}$\\
        \bottomrule
    \end{tabular}
    \caption{CGCs for 30 with $m_1=-\frac{3}{2}$}
    \label{cgc_e_5}
\end{table}

Also, it is important to know that 
\begin{equation}
    \hat{a}_3|\phi\rangle_{\mr{Fe}^{2+}}= |\uparrow\rangle |2,m_{1+}\rangle + |\downarrow\rangle
    |2,m_{1+}\rangle
\end{equation}

This will come in handy later.

To summarize, so far we have the $\mr{SC}0$ state that corresponds to a case where we have a radical
and a low spin cluster, $S_{\mr{cluster}}=0$. Then we have $\mr{SC}1$, which corresponds to the case
where we have a radical and a high spin cluster, $S_{\mr{cluster}}=1$. Finally we have the latest
state, $\mr{SC}2$, which corresponds to the anion case with $S_{\mr{cluster}}=\frac{1}{2}$. With the
new Hamiltonian defined by equation S29 we have 4 terms: $\hat{H}_{\mr{ET}}$  for electron
transfer, $\hat{H}_k$ for iron-to-iron coupling, $\hat{H}_J$ for iron-to-carbon coupling, and
$\hat{H}_0$ which captures the electrostatic and kinetic effects. After some labor, we can summarize the results of
evaluating these parameters. 
\begin{equation}
    \langle \mr{SC}1 |\hat{H}_J| \mr{SC}0 \rangle = 0
\end{equation}
\begin{equation}
    \langle \mr{SC}2|\hat{H}_J| \mr{SC}2 \rangle = -7J
\end{equation}
\begin{equation}
    \langle \mr{SC}0 |\hat{H}_K| \mr{SC}0\rangle = 0
\end{equation}
\begin{equation}
    \langle \mr{SC}1|\hat{H}_K| \mr{SC}1\rangle = -\frac{5}{18} K
\end{equation}
\begin{equation}
    \langle \mr{SC}2 |\hat{H}_{\mr{ET}}| \mr{SC}1\rangle = \frac{7}{10}t
\end{equation}
\begin{equation}
    \langle \mr{SC}2|\hat{H}_{\mr{ET}}| \mr{SC}0\rangle = -t\frac{13}{5}\sqrt{\frac{1}{15}}
\end{equation}

which form the elements of our Hamiltonian in matrix form.

\begin{equation}
\hat{H} =
\begin{bmatrix}
E_0 - 6J                & -\sqrt{\frac{3}{2}}K     & -\frac{13}{5\sqrt{15}}t \\
-\sqrt{\frac{3}{2}}K    & E_0 - \frac{5}{18}K - 3J & \frac{7}{10}t \\
-\frac{13}{5\sqrt{15}}t & \frac{7}{10}t            & E_1 - 7J
\end{bmatrix}
\label{eq:H_matrix}
\end{equation}

With this matrix we can obtain very similar results as for the four-iron model, discussed next. For
instance, from this Hamiltonian we can notice that $K_{\mr{L}}$ is also $+1/3$. This results from
the similarities between this two-iron system and the two-rhomb model for [4Fe-4S].

\section{Two-Rhomb Model}

With the two iron case articulated we can turn our attention to the four iron case which can be
represented by two rhomboids interacting with a radical alkyl. As shown in Figure S2. 

\begin{figure}[H]
    \centering
    \includegraphics[width=0.33\textwidth]{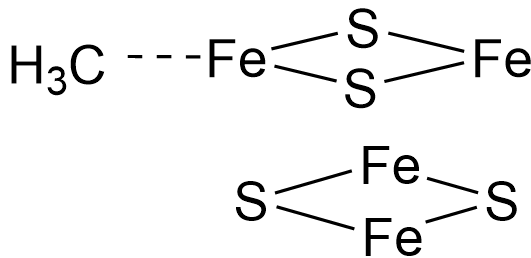}
    \caption{Two Rhomb system with methyl radical (model is applicable to alkyl-like ligands in
    general), \(S_{\mr{Total}} \equiv S_{\mr{T}} = \frac{1}{2}\).}
    \label{fig:two-rhomb}
\end{figure}

This calls for a new Hamiltonian and a review of the spin operators and what exactly they are referring to in this new setup. 
\begin{align}
    \nonumber&\hat{\mb{S}}_{12} = \hat{\mb{S}}_1 + \hat{\mb{S}}_2 \\
    &\hat{\mb{S}}_{34} = \hat{\mb{S}}_3 + \hat{\mb{S}}_4 
\end{align}
In the 3+ state of the cluster, Fe1, Fe2, and Fe3 are in 3+ ($S=5/2$), while Fe4 in 2+ state ($S=2$).
The Hamiltonian is as follows:
\begin{equation}
    \hat{H} =\hat{H}_0 -t\hat{a}^{\dagger}_1\hat{a}_{\mr{L}}-t\hat{a}^{\dagger}_{\mr{L}}\hat{a}_{1}
    +k \hat{\mb{S}}_{\mr{L}}\cdot\hat{\mb{S}}_{12} + j \hat{\mb{S}}_{12}\cdot \hat{\mb{S}}_{34} + J
    \hat{\mb{S}}_{1}\cdot \hat{\mb{S}}_{2} + J \hat{\mb{S}}_{3}\cdot \hat{\mb{S}}_{4} 
\end{equation}
For simplicity, we will ignore the effect of delocalization on the lower rhomb. We are primarily
interested in the way that the so-called unique iron atom in the upper rhomb interacts with the
alkyl. Now the states under consideration with this Hamiltonian remain the same as in the previous
case, but let us reiterate those for clarity. 

\begin{align}
    \nonumber&|\mr{SC}0\rangle= |0,0\rangle_{\mr{cluster}}
    |\tfrac{1}{2},\tfrac{1}{2}\rangle_{\mr{alkyl}} \\ \nonumber&|\mr{SC}1\rangle=
    \sqrt{\frac{2}{3}}|1,1\rangle_{\mr{cluster}} |\tfrac{1}{2},-\tfrac{1}{2}\rangle_{\mr{alkyl}} -
    \sqrt{\frac{1}{3}}|1,0\rangle_{\mr{cluster}}|\tfrac{1}{2},\tfrac{1}{2}\rangle_{\mr{alkyl}} \\
    &|\mr{SC}2\rangle= |\tfrac{1}{2},\tfrac{1}{2}\rangle_{\mr{cluster}} |0,0\rangle_{\mr{alkyl}}
\end{align}

However, because $|S_i-S_j| \leq|S_{ij}|\leq|S_i+S_j|$ the two rhombs have a large number of ways to
yield singlet and triplet states. For the 2+ cluster case, we will take $S_{12}=S_{34}=9/2$, and for
[4Fe-4S]${}^{3+}$ $S_{12}=4$ and $S_{34}=9/2$, Ref. 51, main document. Here we need to find CGCs
again. This time in two layers. 

We employ the generator mentioned before to make the Wigner $3j$ symbol to evaluate the
following expression:
\begin{equation}
\langle j_{1} m_{1} j_{2}m_2|j_{3}m_{3} \rangle = (-1)^{-j_{1}+j_{2}-m_{3}} \sqrt{2j_{3}+1} 
\begin{bmatrix}
    j_{1} & j_{2} & j_{3} \\
    m_{1} & m_{2} & -m_{3} \\
\end{bmatrix}
\end{equation}

The first two terms form what we will refer to as the `wigner prefactor' (WPF) while the last term
is our wigner 3j symbol together they form the CGCs we are looking for in this case. To tie this
directly to our convention the equation for the CGCs takes the following form:

\begin{equation}
\mr{CGC} = (-1)^{-S_{12}+S_{34}-m_{\mr{cluster}}} \sqrt{2S_{\mr{cluster}}+1} 
    \begin{bmatrix}
    S_{12} & S_{34} & S_{\mr{cluster}} \\
    m_{12} & m_{34} & -m_{\mr{cluster}} \\
    \end{bmatrix}
\end{equation}

The following tables summarize the fruit of laboring with the aforementioned Stone-Wood's generator.
In our notation, we show the rational number and its sign. For example, for $m_{12}=7/2$ and
$m_{34}=-7/2$ in Table S10.a, the CGC coefficient is $-\sqrt{1/10}$. In Table S10.b, for $m_{12}=7/2$
and $m_{34}=-5/2$, CGC is $-\sqrt{3}\times (-1)\sqrt{16/165}$.

\begin{table}[htp]
    \centering
    \renewcommand{\arraystretch}{1.3}

    \begin{subtable}[t]{0.31\textwidth}
        \centering
        \begin{tabular}{@{\hskip 6pt}c@{\hskip 6pt}c@{\hskip 6pt}c@{\hskip 6pt}}
            \toprule
            $m_{12}$ & $m_{34}$ & $\pm\sqrt{\text{coeff}}$ \\
            \midrule
            $\frac{9}{2}$  & $-\frac{9}{2}$ & $\frac{2}{10}$ \\
            $\frac{7}{2}$  & $-\frac{7}{2}$ & $-\frac{1}{10}$ \\
            $\frac{5}{2}$  & $-\frac{5}{2}$ & $\frac{1}{10}$ \\
            $\frac{3}{2}$  & $-\frac{3}{2}$ & $-\frac{1}{10}$ \\
            $\frac{1}{2}$  & $-\frac{1}{2}$ & $\frac{1}{10}$ \\
            $-\frac{1}{2}$ & $\frac{1}{2}$  & $\frac{1}{10}$ \\
            $-\frac{3}{2}$ & $\frac{3}{2}$  & $\frac{1}{10}$ \\
            $-\frac{5}{2}$ & $\frac{5}{2}$  & $-\frac{2}{10}$ \\
            $-\frac{7}{2}$ & $\frac{7}{2}$  & $\frac{1}{10}$ \\
            $-\frac{9}{2}$ & $\frac{9}{2}$  & $-\frac{1}{10}$ \\
            \bottomrule
        \end{tabular}
        \caption{3j coefficients corresponding to $m_{12}$ and $m_{34}$ pairings for the state with
        $S_{\mr{cluster}} = 0$, $m_{\mr{cluster}} = 0$, $S_{12} = \frac{9}{2}$, and $S_{34}
        = \frac{9}{2}$. WPF $= 1$.}
        \label{tab:a}
    \end{subtable}
    \hfill
    \begin{subtable}[t]{0.31\textwidth}
        \centering
        \begin{tabular}{@{\hskip 6pt}c@{\hskip 6pt}c@{\hskip 6pt}c@{\hskip 6pt}}
            \toprule
            $m_{12}$ & $m_{34}$ & $\pm\sqrt{\text{coeff}}$ \\
            \midrule
            $\frac{9}{2}$  & $-\frac{7}{2}$ & $\frac{3}{55}$ \\
            $\frac{7}{2}$  & $-\frac{5}{2}$ & $-\frac{16}{165}$ \\
            $\frac{5}{2}$  & $-\frac{3}{2}$ & $\frac{21}{165}$ \\
            $\frac{3}{2}$  & $-\frac{1}{2}$ & $-\frac{24}{165}$ \\
            $\frac{1}{2}$  & $\frac{1}{2}$  & $\frac{5}{33}$ \\
            $-\frac{1}{2}$ & $\frac{3}{2}$  & $-\frac{24}{165}$ \\
            $-\frac{3}{2}$ & $\frac{5}{2}$  & $\frac{21}{165}$ \\
            $-\frac{5}{2}$ & $\frac{7}{2}$  & $-\frac{16}{165}$ \\
            $-\frac{7}{2}$ & $\frac{9}{2}$  & $\frac{3}{55}$ \\
            \bottomrule
        \end{tabular}
        \caption{3j coefficients corresponding to $m_{12}$ and $m_{34}$ pairings for the state with
        $S_{\mr{cluster}} = 1$, $m_{\mr{cluster}} = 1$, $S_{12} = \frac{9}{2}$, and $S_{34} = \frac{9}{2}$. WPF $= -\sqrt{3}$.}
        \label{tab:b}
    \end{subtable}
    \hfill
    \begin{subtable}[t]{0.31\textwidth}
        \centering
        \begin{tabular}{@{\hskip 6pt}c@{\hskip 6pt}c@{\hskip 6pt}c@{\hskip 6pt}}
            \toprule
            $m_{12}$ & $m_{34}$ & $\pm\sqrt{\text{coeff}}$ \\
            \midrule
            $\frac{9}{2}$  & $-\frac{9}{2}$ & $\frac{27}{110}$ \\
            $\frac{7}{2}$  & $-\frac{7}{2}$ & $-\frac{49}{330}$ \\
            $\frac{5}{2}$  & $-\frac{5}{2}$ & $\frac{5}{66}$ \\
            $\frac{3}{2}$  & $-\frac{3}{2}$ & $-\frac{3}{110}$ \\
            $\frac{1}{2}$  & $-\frac{1}{2}$ & $\frac{1}{330}$ \\
            $-\frac{1}{2}$ & $\frac{1}{2}$  & $\frac{1}{330}$ \\
            $-\frac{3}{2}$ & $\frac{3}{2}$  & $-\frac{3}{110}$ \\
            $-\frac{5}{2}$ & $\frac{5}{2}$  & $\frac{5}{66}$ \\
            $-\frac{7}{2}$ & $\frac{7}{2}$  & $-\frac{49}{330}$ \\
            $-\frac{9}{2}$ & $\frac{9}{2}$  & $\frac{27}{110}$ \\
            \bottomrule
        \end{tabular}
        \caption{3j coefficients corresponding to $m_{12}$ and $m_{34}$ pairings for the state with
        $S_{\mr{cluster}} = 1$, $m_{\mr{cluster}} = 0$, $S_{12} = \frac{9}{2}$, and $S_{34} =
        \frac{9}{2}$. WPF $= \sqrt{3}$.}
        \label{tab:c}
    \end{subtable}

    \caption{(a)--(c) Coefficient tables for different $S_{\mr{cluster}}$, $m_{\mr{cluster}}$ values
    with $S_{12} = \frac{9}{2}$ and $S_{34} = \frac{9}{2}$.}
    \label{tab:all_cluster_tables}
\end{table}

For the next set of CGCs we need to recall that 

\begin{align}
\nonumber\hat{\mathbf{S}}_{AB} &= \hat{\mathbf{S}}_A + \hat{\mathbf{S}}_B \\
\hat{\mathbf{S}}_{AB}^2 &= \hat{\mathbf{S}}_A^2 + \hat{\mathbf{S}}_B^2 + 2 \, \hat{\mathbf{S}}_A \nonumber\cdot \hat{\mathbf{S}}_B \\
\Rightarrow \quad
\hat{\mathbf{S}}_A \cdot \hat{\mathbf{S}}_B &= \frac{1}{2} \left( \hat{\mathbf{S}}_{AB}^2 - \hat{\mathbf{S}}_A^2 - \hat{\mathbf{S}}_B^2 \right)
\end{align}

\begin{align}
\nonumber\left\langle S_{AB}', m_{AB}' \middle| 
\frac{1}{2} \left( \hat{\mathbf{S}}_{AB}^2 - \hat{\mathbf{S}}_A^2 - \hat{\mathbf{S}}_B^2 \right) 
\middle| S_{AB}, m_{AB} \right\rangle 
= \delta_{S_{AB}' S_{AB}} \, \delta_{m_{AB}' m_{AB}} \, \frac{1}{2} 
        [& S_{AB}(S_{AB}+1) \\
\nonumber&- S_A(S_A+1)\\
         &- S_B(S_B+1) ]
\end{align}

Then we obtain the next table of CGCs for the state described in Table S11.

\begin{table}[htp]
    \centering
    \renewcommand{\arraystretch}{1.3} 
    \begin{tabular}{@{\hskip 8pt}c@{\hskip 8pt}c@{\hskip 8pt}c@{\hskip 8pt}}
        \toprule
        $m_{12}$ & $m_{34}$ & $\pm\sqrt{\text{coeff}}$ \\
        \midrule
            $\frac{8}{2}$  & $-\frac{7}{2}$ & $\frac{1}{45}$ \\
            $\frac{6}{2}$  & $-\frac{5}{2}$ & $-\frac{2}{45}$ \\
            $\frac{4}{2}$  & $-\frac{3}{2}$ & $\frac{1}{15}$ \\
            $\frac{2}{2}$  & $-\frac{1}{2}$ & $-\frac{4}{45}$ \\
            $0$           & $\frac{1}{2}$  & $\frac{1}{9}$ \\
            $-\frac{2}{2}$ & $\frac{3}{2}$  & $-\frac{2}{15}$ \\
            $-\frac{4}{2}$ & $\frac{5}{2}$  & $\frac{7}{45}$ \\
            $-\frac{6}{2}$ & $\frac{7}{2}$  & $-\frac{8}{45}$ \\
            $-\frac{8}{2}$ & $\frac{9}{2}$  & $\frac{1}{5}$ \\
        \bottomrule
    \end{tabular}
    \caption{3j coefficients corresponding to $m_{12}$ and $m_{34}$ pairings for the state with $S_{\mr{cluster}}=\frac{1}{2}$, 
    $m_{\mr{cluster}}=\frac{1}{2}$, $S_{12}=\frac{8}{2}$, and $S_{34}=\frac{9}{2}$. WPF $= \sqrt{2}$}
    \label{tab:M_coeff}
\end{table}

As in the previous case we have to build the Hamiltonian matrix. First we have the following Hamiltonian elements

\begin{align*}
\hat{H} &= \hat{H}_0 + \hat{H}_{\mr{ET}} + \hat{H}_k + \hat{H}_J \\
\hat{H}_k &= k \, \hat{\mathbf{S}}_{\mr{L}} \cdot \hat{\mathbf{S}}_{12} \\
\hat{H}_J &=  j \, \hat{\mathbf{S}}_{12} \cdot \hat{\mathbf{S}}_{34} + J \, \hat{\mathbf{S}}_1 \cdot \hat{\mathbf{S}}_2 
          + J \, \hat{\mathbf{S}}_3 \cdot \hat{\mathbf{S}}_4  \qquad  \\
\hat{H}_{\mr{ET}} &= -t \left( a_{\mr{L}}^\dagger a_1 + a_{1}^\dagger a_{\mr{L}} \right)
\end{align*}

Some relatively straightforward elements:

\begin{align*}
\langle{\mr{SC}0}|{\hat{H}_k}|{\mr{SC}0}\rangle &= 0 
\qquad \text{(for same reason as in two-iron case)} \\
\langle{\mr{SC}0}|{\hat{H}_0}|{\mr{SC}0}\rangle &= \tilde{E}_0 \\
\langle{\mr{SC}1}|{\hat{H}_0}|{\mr{SC}1}\rangle &= \tilde{E}_0 \\
\langle{\mr{SC}2}|{\hat{H}_0}|{\mr{SC}2}\rangle &= \tilde{E}_1
\end{align*}

Evaluating $H_J$ gives:

\begin{align*}
\langle{\mr{SC}0}|{\hat{H}_J}|{\mr{SC}0}\rangle &= 10J - \frac{99}{4} j\\
\langle{\mr{SC}1}|{\hat{H}_J}|{\mr{SC}1}\rangle &= 10J - \frac{95}{2} j  \\  
\langle{\mr{SC}2}|{\hat{H}_J}|{\mr{SC}2}\rangle &= \frac{25}{4}J - 22 j    
\end{align*}

First let us consider the product of $\hat{\mb{S}}_{12}$ and $\hat{\mb{S}}_{34}$

\begin{align*}
\hat{\mb{S}}_{12} \cdot \hat{\mb{S}}_{34} 
&= \frac{1}{2} \left( \hat{\mb{S}}_{\mr{cluster}}^2 - \hat{\mb{S}}_{12}^2 - \hat{\mb{S}}_{34}^2 \right)
\end{align*}

The alkyl-to-unique iron couplings are described by the following term, which we will now simplify
to a numerical result using the CGC procedure we saw before:

\begin{align*}
\hat{\mathbf{S}}_{\mr{L}} \cdot \hat{\mathbf{S}}_{12}
&= \frac{1}{2} \left( \hat{S}_{+,{\mr{L}}} \hat{S}_{-,12} + \hat{S}_{-,{\mr{L}}} \hat{S}_{+,12} \right)
+ \hat{S}_{z,{\mr{L}}} \hat{S}_{z,12}
\end{align*}

The transition element for the first term in parenthesis is
\begin{equation}
    {\hat{S}_{+,{\mr{L}}}|{\mr{SC}0}\rangle = 0 \qquad \hat{S}_{-,\mr{L}} |{\tfrac{1}{2},-\tfrac{1}{2}}\rangle = 0}    
\end{equation}

\begin{equation}
\langle{\mr{SC}1}|{\hat{\mb{S}}_{\mr{L}} \cdot \hat{\mb{S}}_{12}}|{\mr{SC}0}\rangle 
= \langle{1,1}|{\frac{1}{2} \hat{S}_{+,12}}|{1,0}\rangle \frac{\sqrt{2}}{3}
+ \langle{\mr{SC}1}|{\hat{S}_{z,\mr{L}} \hat{S}_{z,12}}|{\mr{SC}0}\rangle
\end{equation}
 and recall that as stated before our pre-factor comes from:
 
\begin{align}
\nonumber\hat{S}_{+} |{S,m}\rangle &= \sqrt{[S(S+1) - m(m+1)]} |{S,m+1}\rangle \\
\nonumber\hat{S}_{-} |{S,m}\rangle &= \sqrt{[S(S+1) - m(m-1)]} |{S,m-1}\rangle.
\end{align}

Now we can plug in the results obtained earlier to obtain an expression for
$\hat{S}_{-,\mr{L}}\hat{S}_{+,12}$

\begin{align*}
\hat{S}_{-,\mr{L}} |{\tfrac{1}{2},\tfrac{1}{2}}\rangle &= \sqrt{\frac{3}{4} + \frac{1}{4}}
|{\tfrac{1}{2}, -\tfrac{1}{2}}\rangle\\ &=|{\tfrac{1}{2}, -\tfrac{1}{2}}\rangle.
\end{align*}

Continuing the evaluation with the next element is trivial:

\begin{align*}
\langle{\mr{SC}1}|{\hat{S}_{z,\mr{L}} \hat{S}_{z,12}}|{\mr{SC}0}\rangle
&= -\frac{1}{2} \cdot \sqrt{\frac{1}{3}} \langle{1,0}|{\hat{S}_{z,12}}|{0,0}\rangle_{\mr{cluster}}
\end{align*}

The next step will call for CGCs and so we return to these tables generated using Stone-Wood's
generator:

\begin{table}[htp!]
\centering
\renewcommand{\arraystretch}{1.3}
\begin{tabular}{c c c c c c}
\toprule
$m_{12}$ & $m_{34}$ & $\pm \sqrt{\text{coeff}}$ & $m_{12}(m_{12}+1)$ & $\left[(S_{12}(S_{12}+1)) -
m_{12}(m_{12}+1)\right]^{1/2}$ \\
\midrule
$\frac{9}{2}$  & $-\frac{9}{2}$  & $\frac{1}{10}$  & $\frac{99}{4}$ & $3$ \\
$\frac{7}{2}$  & $-\frac{7}{2}$  & $-\frac{1}{10}$ & $\frac{63}{4}$ & $4$ \\
$\frac{5}{2}$  & $-\frac{5}{2}$  & $\frac{1}{10}$  & $\frac{35}{4}$ & $\sqrt{21}$ \\
$\frac{3}{2}$  & $-\frac{3}{2}$  & $-\frac{1}{10}$ & $\frac{15}{4}$ & $\sqrt{24}$ \\
$\frac{1}{2}$  & $-\frac{1}{2}$  & $\frac{1}{10}$  & $\frac{3}{4}$  & $\sqrt{25}$ \\
$-\frac{1}{2}$ & $\frac{1}{2}$   & $-\frac{1}{10}$ & $-\frac{1}{4}$  & $\sqrt{25}$ \\
$-\frac{3}{2}$ & $\frac{3}{2}$   & $\frac{1}{10}$  & $\frac{3}{4}$ & $\sqrt{24}$ \\
$-\frac{5}{2}$ & $\frac{5}{2}$   & $-\frac{1}{10}$ & $\frac{15}{4}$ & $\sqrt{21}$ \\
$-\frac{7}{2}$ & $\frac{7}{2}$   & $\frac{1}{10}$  & $\frac{35}{4}$ & $4$ \\
$-\frac{9}{2}$ & $\frac{9}{2}$   & $-\frac{1}{10}$ & $\frac{63}{4}$ & $3$ \\
\bottomrule
\end{tabular}
\caption{CG coefficients, multiplicities, prefactors, and angular momentum quantities for the state
where $S_{\mr{cluster}}=0$, $m_{\mr{cluster}}=0$, $S_{12}=\frac{9}{2}$, and
$S_{34}=\frac{9}{2}$.}
\label{tab:cg_coeffs}
\end{table}

\begin{table}[htp!]
\centering
\renewcommand{\arraystretch}{1.3}
\begin{tabular}{c c c}
\toprule
$m_{12}$ & $m_{34}$ & $\pm \sqrt{\text{coeff}}$ \\
\midrule
$\frac{9}{2}$  & $-\frac{7}{2}$  & $\frac{3}{55}$ \\
$\frac{7}{2}$  & $-\frac{5}{2}$  & $-\frac{16}{165}$ \\
$\frac{5}{2}$  & $-\frac{3}{2}$  & $\frac{21}{165}$ \\
$\frac{3}{2}$  & $-\frac{1}{2}$  & $-\frac{24}{165}$ \\
$\frac{1}{2}$  & $\frac{1}{2}$   & $\frac{5}{33}$ \\
$-\frac{1}{2}$ & $\frac{3}{2}$   & $-\frac{24}{165}$ \\
$-\frac{3}{2}$ & $\frac{5}{2}$   & $\frac{21}{165}$ \\
$-\frac{5}{2}$ & $\frac{7}{2}$   & $-\frac{16}{165}$ \\
$-\frac{7}{2}$ & $\frac{9}{2}$   & $\frac{3}{55}$ \\
\bottomrule
\end{tabular}
\caption{CG coefficients and angular momentum quantities for the state where $S_{\mr{cluster}}=1$, $m_{\mr{cluster}}=1$, $S_{12}=0$, and $S_{34}=0$. WPF in each case is $-\sqrt{3}$}
\label{tab:coeffs_2}
\end{table}

Next we want to evaluate the electron transfer element: $\langle \mr{SC}2|\hat{H}_{\mr{ET}}|\mr{SC}1\rangle$, which comes down to the following expression.
\begin{align*}
&-t\, \langle \tfrac{1}{2}, \tfrac{1}{2}|_{\mr{cluster}} \langle 0,0 | _{\mr{alkyl}} \, \hat{a}^\dagger_{\mr{L}} \hat{a}_{1} 
| 0,0 \rangle_{\mr{cluster}} |\tfrac{1}{2}, \tfrac{1}{2} \rangle_{\mr{alkyl}} 
= \langle \mr{SC}2 | \hat{H}_{\mr{ET}} \vert \mr{SC}0 \rangle \\
&\quad= +t  
\langle \tfrac{1}{2}, \tfrac{1}{2} \vert_{\mr{cluster}} \left[\vert \downarrow \rangle_d \otimes \vert 0,0
\rangle_{\mr{cluster-1}} 
- \vert \uparrow \rangle_d \otimes \vert 0,0 \rangle_{\mr{cluster}-1}
\right] \\
&\quad= -t\, 
\underbrace{\langle \tfrac{1}{2}, \tfrac{1}{2} \vert}_{\mr{cluster}} \underbrace{\vert\downarrow
\rangle_d
\otimes 
\vert 0,0 \rangle_{\mr{cluster-1}}}_{\mr{cluster}}
\qquad 
\end{align*}
$\vert 0,0\rangle_{\mr{cluster}-1}$ refers to the resulting cluster state afer the application of the
annihilation operator $\hat{a}_1$ and factorizing out $|\downarrow\rangle_d$, which refers to the $d$ orbital that lost an
electron. To evaluate the overall overlap element, we need to split the upper rhomb as the
symmetrized product of the unpaired $d$ electron and the remaining electrons: 
\begin{equation}
\left| S_{12}, m_{12} \right\rangle = \sum_{m_d, m_{\mr{Rh}}} 
C(m_d, m_{\mr{Rh}}) 
\left| S_d, m_d \right\rangle 
\left| S_{\mr{Rh}}, m_{\mr{Rh}} \right\rangle
\end{equation}
Here the $d$ electron in Fe1 is assigned spin variables $S_d$ and $m_d$, and the rest or
the rhomb the spin variables $S_{\mr{Rh}}$ and $m_{\mr{Rh}}$. These expansions, Table S14, are applied to the state
$\langle 1/2,1/2|_{\mr{cluster}}$.
\begin{table}[htp!]
\centering
\renewcommand{\arraystretch}{1.3}
\begin{tabular}{ccc}
\toprule
$m_d$ & $m_{\mr{Rh}}$ & CGC \\
\midrule
\multicolumn{3}{c}{$m_{12} = 4$} \\
$\phantom{-}\frac{1}{2}$ & $\phantom{-}\frac{7}{2}$ & $\phantom{-}0.3162$ \\
$-\frac{1}{2}$           & $\phantom{-}\frac{9}{2}$ & $-0.9487$ \\
\addlinespace
\multicolumn{3}{c}{$m_{12} = 3$} \\
$\phantom{-}\frac{1}{2}$ & $\phantom{-}\frac{5}{2}$ & $\phantom{-}0.4472$ \\
$-\frac{1}{2}$           & $\phantom{-}\frac{7}{2}$ & $-0.8944$ \\
\addlinespace
\multicolumn{3}{c}{$m_{12} = 2$} \\
$\phantom{-}\frac{1}{2}$ & $\phantom{-}\frac{3}{2}$ & $\phantom{-}0.5471$ \\
$-\frac{1}{2}$           & $\phantom{-}\frac{5}{2}$ & $-0.8367$ \\
\addlinespace
\multicolumn{3}{c}{$m_{12} = 1$} \\
$\phantom{-}\frac{1}{2}$ & $\phantom{-}\frac{1}{2}$ & $\phantom{-}0.6325$ \\
$-\frac{1}{2}$           & $\phantom{-}\frac{3}{2}$ & $-0.7746$ \\
\addlinespace
\multicolumn{3}{c}{$m_{12} = 0$} \\
$\phantom{-}\frac{1}{2}$ & $-\frac{1}{2}$           & $\phantom{-}0.7071$ \\
$-\frac{1}{2}$           & $\phantom{-}\frac{1}{2}$ & $-0.7071$ \\
\addlinespace
\multicolumn{3}{c}{$m_{12} = -1$} \\
$-\frac{1}{2}$           & $-\frac{1}{2}$           & $-0.6325$ \\
$\phantom{-}\frac{1}{2}$ & $-\frac{3}{2}$           & $\phantom{-}0.7746$ \\
\addlinespace
\multicolumn{3}{c}{$m_{12} = -2$} \\
$-\frac{1}{2}$           & $-\frac{3}{2}$           & $-0.5477$ \\
$\phantom{-}\frac{1}{2}$ & $-\frac{5}{2}$           & $\phantom{-}0.8367$ \\
\addlinespace
\multicolumn{3}{c}{$m_{12} = -3$} \\
$-\frac{1}{2}$           & $-\frac{5}{2}$           & $-0.4472$ \\
$\phantom{-}\frac{1}{2}$ & $-\frac{7}{2}$           & $\phantom{-}0.8944$ \\
\addlinespace
\multicolumn{3}{c}{$m_{12} = -4$} \\
$-\frac{1}{2}$           & $-\frac{7}{2}$           & $-0.3162$ \\
$\phantom{-}\frac{1}{2}$ & $-\frac{9}{2}$           & $\phantom{-}0.9487$ \\
\bottomrule
\end{tabular}
\caption{CG coefficients for $S_{12} = 4$, various $m_{12}$.}
\label{tab:CG_S12_4}
\end{table}

Now we have all the ingredients to evaluate the couplings of the electron transfer for each state.

For $\mr{SC}2$ to $\mr{SC}0$ we have:
\begin{align}
    \nonumber&(\langle \tfrac{1}{2},\tfrac{1}{2}| |\uparrow\rangle \otimes|0,0\rangle)_{\mr{cluster}} = -0.6708 \approx -\tfrac{2}{3}\\
    &\Rightarrow \langle \mr{SC}2|\hat{H}_{\mr{ET}}|\mr{SC}0\rangle \approx \tfrac{2}{3}t.
\end{align}
For $\mr{SC}2$ to $\mr{SC}1$ we have:
\begin{align}
    \nonumber&(\langle \tfrac{1}{2},\tfrac{1}{2}| |\uparrow\rangle \otimes|1,0\rangle)_{\mr{cluster}} = 0.4282 \approx \tfrac{17}{40} \\
    \nonumber&(\langle \tfrac{1}{2},\tfrac{1}{2}| |\downarrow\rangle \otimes|1,1\rangle)_{\mr{cluster}} = -0.6055 \approx -\tfrac{3}{5}\\
    &\Rightarrow \langle \mr{SC}2 |\hat{H}_{\mr{ET}}|\mr{SC}1\rangle \approx \tfrac{3}{4}t.
\end{align}

We now turn our attention back to $\hat{H}_k$ and the $\hat{\mb{S}}_{\mr{L}}\cdot\hat{\mb{S}}_{12}$ elements.

\begin{equation}
\langle \mr{SC}1 \vert \hat{\mb{S}}_{\mr{L}}\cdot \hat{\mb{S}}_{12} \vert \mr{SC}0 \rangle 
= \langle 1,1\vert \, \tfrac{1}{2} \, \hat{S}_{+,12} \, \vert 0,0 \rangle \cdot \sqrt{\tfrac{2}{3}}
+ \langle \mr{SC}1 \vert \hat{S}_{z,\mr{L}} \hat{S}_{z,12} \vert \mr{SC}0 \rangle
\end{equation}

\begin{equation}
\langle \mr{SC}1 \vert \hat{S}_{z,\mr{L}} \hat{S}_{z,12} \vert \mr{SC}0 \rangle 
= -\tfrac{1}{2} \cdot \sqrt{\tfrac{1}{3}} \cdot 
\langle 1,0 \vert S_{z,12}|0,0 \rangle_{\mr{cluster}}
\end{equation}

\begin{align}
\langle 1,1 \vert S_{+,12}|0,0 \rangle_{\mr{cluster}} = -4.639 \approx \frac{-116}{25}
\end{align}

\begin{align}
\langle 1, 0 \vert S_{z,12} \vert 0, 0 \rangle = 2.872 \approx \frac{359}{125}
\end{align}
and so we obtain 
\begin{equation}
\langle \mr{SC}1|\hat{H}_k|\mr{SC}0 \rangle \approx -2.512k \approx -\frac{5}{2}k    
\end{equation}

To determine $\langle \mr{SC}1 | \hat{H}_k | \mr{SC}1 \rangle$ we want to evaluate the matrix element in terms of spin operator contractions between subsystems.
\[
\langle \mr{SC}1 \vert \hat{H}_k \vert \mr{SC}1 \rangle
\]

The Hamiltonian involves a product of spin operators acting on two subsystems:
\begin{equation}
\langle \mr{SC}1 \vert \hat{H}_k \vert \mr{SC}1 \rangle 
= k \langle \mr{SC}1 \vert S_{z,12} S_{z,\mr{L}} \vert \mr{SC}1 \rangle
-\frac{k}{2} \sqrt{\frac{2}{9}} 
\langle 1,1 \vert \langle \tfrac{1}{2}, -\tfrac{1}{2} \vert 
\hat{S}_{+,12} \hat{S}_{-,\mr{L}} \vert 1,0 \rangle 
\vert \tfrac{1}{2}, \tfrac{1}{2} \rangle - \mr{c.c.}
\end{equation}

The above can be decomposed as:
\begin{equation}
\langle \mr{SC}1 \vert S_{z,12} S_{z,\mr{L}} \vert \mr{SC}1\rangle 
= \frac{2}{3} \langle 1,1 \vert S_{z,12} \vert 1,1 \rangle \left(-\frac{1}{2} \right)
+ \frac{1}{6} \langle 1,0 \vert S_{z,\mr{L}} \vert 1,0 \rangle
\end{equation}

\noindent next we need to evaluate $\langle 1,1 | S_{z,12} | 1,1 \rangle$, which yields

\begin{align}
\langle 1,1 \vert S_{z,12} \vert 1,1 \rangle = \frac{1}{2}
\end{align}

\noindent find known diagonal contribution

\begin{equation}
\langle \mr{SC}1 \vert S_{z,12} S_{z,\mr{L}} \vert \mr{SC}1 \rangle = -\frac{1}{6}
\end{equation}
\noindent next evaluate spin ladder bracket,

\begin{align}
\langle 1,1 \vert \langle \tfrac{1}{2}, -\tfrac{1}{2} \vert 
S_{-,\mr{L}} S_{+,12} \vert 1,0 \rangle \vert \tfrac{1}{2}, \tfrac{1}{2} \rangle
&= \frac{1}{\sqrt{2}}.
\end{align}

\noindent Our final result of putting all components together:
\begin{equation}
\langle \mr{SC}1 \vert \hat{H}_k \vert \mr{SC}1 \rangle 
= k \left[ -\frac{1}{6} - \frac{1}{3} \right] 
= -\frac{k}{2}
\end{equation}

Finally after all that labor we obtain:
\[
{\langle \mr{SC}1 \vert \hat{H}_k \vert \mr{SC}1 \rangle = -\frac{k}{2}}.
\]
With all of these elements obtained we now have our Hamiltonian matrix for the 2-rhomb model. 

\[
\hat{H} =
\left[
\begin{array}{ccc}
10J - \tfrac{99}{4} j + \tilde{E}_0 & -5/2 k & 2/3 t \\
-5/2 k & 10J - \tfrac{95}{4} j - \tfrac{1}{2} k + \tilde{E}_0 & 3/4 t \\
2/3 t & 3/4 t & \tfrac{25}{4} J - 22 j + \tilde{E}_1
\end{array}
\right]
\]

Next, PT is applied to it.

\subsection{PT Spin Averages, Partially Unbound Ligand}
For some $i^{th}$ site, we consider the average spin to be
\begin{equation}
\langle S_{z,i} \rangle = |c_2|^2 \langle S_{z,i} \rangle_{\mr{SC}2} + |c_1|^2 \langle S_{z,i}
\rangle_{\mr{SC}1} + |c_0|^2 \langle S_{z,i} \rangle_{\mr{SC}0} + 2 c_1 c_0 \langle \mr{SC}1 |
S_{z,i} | \mr{SC}0 \rangle
\end{equation}
The full case spin average where the signs come from the rhomb assignment is as follows
\begin{equation}
\langle S_{z,\mr{L}} \rangle = |c_1|^2 \left( \frac{1}{3} \cdot \frac{1}{2} - \frac{2}{3} \cdot \frac{2}{2} \right) + |c_0|^2 \cdot \frac{1}{2} 
= \left( |c_0|^2 - \frac{1}{3} |c_1|^2 \right) \cdot \frac{1}{2}
\end{equation}
In the radical cluster case it reads:
\begin{equation}
\langle S_{z,1} \rangle = |c_1|^2 \langle S_{z,1} \rangle_{\mr{SC}1} + |c_0|^2 \langle S_{z,1}
\rangle_{\mr{SC}0} + 2 c_1 c_0 \langle \mr{SC}1 | S_{z,1} | \mr{SC}0 \rangle
\end{equation}
\noindent and with some evaluation we obtain the following four statements:
\begin{align*}
    \langle 1,1 |S_{z,1}| 1,1\rangle = 0.2222 \approx \frac{11}{50}
\end{align*}
\begin{align*}
    \langle 1,0 | S_{z,1} | 1,0 \rangle = 0 \text{ due to symmetry}
\end{align*}
\begin{align*}
    \langle 0,0 | S_{z,1} |0,0 \rangle = 0 \text{ also due to symmetry}
\end{align*}
\begin{align*}
    \langle \mr{SC}1 | S_{z,1} | \mr{SC}0\rangle = - \sqrt{\frac{1}{3}} \langle 1,0 |S_{z,1}| 0,0\rangle.
\end{align*}
We can evaluate that last term as follows:
\begin{align*}
    \langle 1,0 |S_{z,1}| 0,0\rangle = 1.166 \approx \frac{6}{5}
\end{align*}
From PT we have that 
\begin{equation}
c_1=\frac{5}{2}\frac{k}{j}
\end{equation}
As before we are now ready to plug everything in and obtain formulae for our adjusted hyperfine
coupling constants according to this model.

For $\langle\hat{S}_{z,2}\rangle$ we obtain

\begin{align}
    \langle 1,1|{S}_{z,2}|1,1 \rangle &= \frac{7}{25} \\
    \langle 1,0|{S}_{z,2}|0,0 \rangle &= 1.706 \\
    \Rightarrow \langle\hat{S}_{z,2}\rangle &= \frac{1}{5}|c_1|^2 - \frac{17}{5\sqrt{3}}c_1c_0.
\end{align}

For $\langle\hat{S}_{z,3}\rangle$ we obtain

\begin{align}
    \langle 1,1|{S}_{z,3}|1,1 \rangle &= \frac{7}{25} \\
    \langle 1,0|{S}_{z,3}|0,0 \rangle &= -1.596 \\
    \Rightarrow \langle\hat{S}_{z,2}\rangle &= \frac{14}{75}|c_1|^2 + \frac{16}{5\sqrt{3}}c_1c_0.
\end{align}

For $\langle\hat{S}_{z,4}\rangle$ we obtain

\begin{align}
    \langle 1,1|{S}_{z,4}|1,1 \rangle &= \frac{11}{50} \\
    \langle 1,0|{S}_{z,4}|0,0 \rangle &= -1.276 \\
    \Rightarrow \langle\hat{S}_{z,4}\rangle &= \frac{22}{150}|c_1|^2 + \frac{13}{5\sqrt{3}}c_1c_0.
\end{align}

These results lead to the observable ${}^{57}$Fe HFCC expressions:

\begin{equation}
    a_1^{\mr{obs}} = -\frac{12}{\sqrt{13}} \frac{k}{j}a_1^{\mr{site}}
\end{equation}

\begin{equation}
    a_2^{\mr{obs}} = -\frac{17}{\sqrt{3}}\frac{k}{j} a_2^{\mr{site}}
\end{equation}

\begin{equation}
    a_3^{\mr{obs}} = \frac{16}{\sqrt{3}} \frac{k}{j} a_3^{\mr{site}}
\end{equation}

\begin{equation}
    a_4^{\mr{obs}} = \frac{13}{\sqrt{3}} \frac{k}{j} a_4^{\mr{site}}
\end{equation}

The same relation holds for the tensors shown in the main document.

\newpage

\section{Optimized XYZ Coordinate Tables}

{\bfseries Sample Optimization File, $\Omega$ intermediate}

\begin{verbatim}
!UKS BP86 Def2-TZVP def2/J NormalSCF NormalOpt SlowConv NoTrah

%pal nprocs 32 end

%maxcore 3900

%cpcm
  epsilon 4
end

%scf
  maxiter 3000
  DIISMaxEq 15
  directresetfreq 1
  guessmode cmatrix
  flipspin 1,3
  finalms 0.5
end

%basis
  NewGTO H "def2-SVP" end
end

*xyz  -2 20
Fe         1.79405       -4.43402        4.09734
Fe         1.99882       -5.84036        6.49346
Fe         0.41364       -3.70828        6.64011
Fe        -0.41244       -5.85232        5.27733
C          3.73874       -4.61736        3.37894 newGTO "def2-TZVPD" end
H          3.81176       -3.96093        2.49538 newGTO "def2-TZVPD" end
H          4.38133       -4.19709        4.16976 newGTO "def2-TZVPD" end
C          4.30501       -5.96930        2.99249 newGTO "def2-TZVPD" end
C          5.23107        0.37509       -0.43007
S         -0.51100       -3.76190        4.56657
.
.
.
*
\end{verbatim}

{\bfseries Constrained Optimization Script}

\begin{verbatim}
!UKS BP86 Def2-TZVP def2/J NormalSCF NormalOpt SlowConv NoTRAH

%pal nprocs 32 end

%maxcore 3900

%cpcm
  epsilon 4
end

%geom
  Constraints
    {C 0:50 C}
  end
end

*xyz  -2 20
 Fe    -4.520690  -23.509440   21.567610
 Fe    -6.322300  -24.038390   19.412030
 Fe    -6.151680  -21.561170   20.469740
 Fe    -4.125320  -22.578810   18.963700
 C     -3.313650  -26.115430   22.050490
 H     -4.358320  -25.950890   21.796740
 H     -2.837620  -25.475440   22.792840
 C     -2.534020  -27.356190   21.844890
 S     -6.910000  -23.469580   21.485240
 S     -6.210440  -22.007480   18.282180
 S     -3.886430  -21.420780   20.860780
 S     -4.170250  -24.786840   19.647830
 N     -4.626780  -23.012180   23.739840
 C     -3.632190  -23.680430   24.551460
 C     -2.461840  -24.005980   23.638140
 O     -2.586660  -23.949120   22.380900
 C     -4.197400  -24.947480   25.239320
 C     -5.193640  -25.735660   24.372430
 C     -5.797120  -27.317700   22.182400
 O     -1.377460  -24.329960   24.190590
 H     -4.360400  -22.205250   23.562800
 H     -5.378730  -22.980630   24.171900
 H     -3.338400  -23.085400   25.215390
 H     -4.630320  -24.689650   26.031540
 H     -3.480660  -25.512760   25.459740
 O     -3.392000  -28.419700   21.349390
 C     -1.350050  -27.240990   20.817200
 C     -2.039830  -27.745910   19.547470
 O     -1.119750  -28.080690   18.532380
 C     -2.879780  -28.914030   20.106870
 N     -1.969680  -31.721500   15.870730
 C     -1.391080  -31.720620   17.085630
 N     -1.669330  -30.948240   18.147270
 C     -2.670890  -30.099980   17.881410
 C     -3.372510  -29.979120   16.672730
 C     -2.970910  -30.848690   15.634810
 N     -3.541090  -30.824840   14.402480
 N     -4.354360  -29.003640   16.755070
 C     -4.244280  -28.540480   17.982900
 N     -3.238550  -29.150060   18.723100
 H     -3.295510  -31.555180   13.745590
 H     -4.376670  -30.275110   14.251330
 H     -1.615890  -28.422540   17.769980
 H     -2.253540  -29.810860   20.241500
 H     -2.736620  -26.949990   19.218200
 H     -2.089110  -27.698430   22.800850
 H     -4.848820  -27.794090   18.360360
 H     -0.631180  -32.406060   17.218900
 O     -0.804070  -25.957060   20.686100
 H     -0.566320  -27.995860   21.065870
 H     -0.333010  -25.772760   21.539550
 S     -5.012726  -27.459139   23.838776
 S     -7.493902  -19.648940   20.973159
 H     -5.376592  -25.193489   23.464536
 H     -6.148952  -25.699355   24.886483
 H     -5.140211  -27.603563   21.396907
 H     -6.196284  -26.324317   22.052073
 H     -6.650022  -27.989150   22.183217
 C     -7.301805  -19.503019   22.780907
 H     -7.867655  -18.648142   23.129528
 H     -6.264076  -19.357917   23.053989
 H     -7.673595  -20.386389   23.284971
 S     -2.472294  -22.099926   17.426758
 C     -2.888591  -23.197983   16.034539
 H     -2.175554  -23.021775   15.240023
 H     -3.882865  -22.990475   15.662723
 H     -2.828804  -24.235618   16.332876
 S     -7.771864  -25.571405   18.447795
 C     -7.142942  -25.695655   16.741818
 H     -7.725424  -26.442722   16.219050
 H     -6.104142  -25.996619   16.730571
 H     -7.243338  -24.750003   16.227162
*
\end{verbatim}

{\bfseries Sample EPR Single-Point Script}

\begin{verbatim}
!UKS BP86 EPR-III TightSCF SlowConv NoTrah

%pal nprocs 64 end

%maxcore 3900

%cpcm
  epsilon 4
end

%scf
  maxiter 3000
  DIISMaxEq 15
  directresetfreq 1
  guessmode cmatrix
	flipspin 2,3
	finalms 0.5
end

%basis
	NewGTO Fe "CP(PPP)" end 
	NewGTO S "IGLO-III" end  
end  

%method
 SpecialGridAtoms 26
 SpecialGridIntAcc 7
end

*xyz  -2 20
{XYZ Coordinates}
*

%eprnmr 
  gtensor true
  nuclei = all C {aiso, adip, rho} 
  nuclei = all H {aiso, adip, rho} 
  nuclei = all Fe {aiso, adip, rho} 
end

\end{verbatim}

\begin{center}
\begin{longtable}{c r r r}
\caption{Atomic coordinates, cyano bound cluster, $\beta\beta\alpha\alpha$ configuration} \label{tab:Fe4S4CynEPR BP86 0,1} \\
\toprule
Element & x (\AA) & y (\AA) & z (\AA) \\
\midrule
\endfirsthead
\toprule
Element & x (\AA) & y (\AA) & z (\AA) \\
\midrule
\endhead
Fe &  8.090030 &  4.458520 &  5.340110 \\
Fe &  7.042100 &  2.635880 &  7.086030 \\
Fe &  9.513960 &  2.431310 &  6.201030 \\
Fe &  8.783610 &  4.401990 &  7.895790 \\
S &  5.189180 &  1.938630 &  8.174700 \\
S &  11.396010 &  1.260390 &  5.672770 \\
S &  7.670620 &  2.235110 &  4.943190 \\
S &  6.791760 &  4.895170 &  7.126220 \\
S &  10.156040 &  4.643730 &  6.054850 \\
S &  8.779910 &  2.147030 &  8.334410 \\
S &  9.068510 &  5.336330 &  9.933900 \\
N &  7.657250 &  6.479910 &  3.024070 \\
C &  7.727960 &  5.671920 &  3.884340 \\
C &  5.629730 &  0.274170 &  8.832560 \\
C &  11.238890 & -0.553190 &  6.041850 \\
C &  6.467140 &  4.318560 &  11.271290 \\
H &  11.349640 & -0.695240 &  7.123110 \\
C &  8.955780 &  4.042450 &  11.228840 \\
H &  9.750090 &  3.302260 &  11.095330 \\
H &  9.094890 &  4.545350 &  12.198830 \\
C &  7.602100 &  3.292750 &  11.273280 \\
O &  6.190400 & -1.370320 &  4.877540 \\
H &  5.023930 & -1.805060 &  8.760160 \\
N &  7.552240 &  2.457300 &  12.489370 \\
H &  7.411570 &  3.077800 &  13.292860 \\
H &  7.570510 &  2.638080 &  10.401320 \\
C &  9.914750 & -1.201850 &  5.610880 \\
H &  12.077930 & -1.036210 &  5.527830 \\
H &  5.586330 &  3.468480 &  9.695920 \\
O &  5.497390 &  4.228190 &  10.352590 \\
O &  6.401510 &  5.179140 &  12.138210 \\
N &  8.854400 & -0.720640 &  6.479040 \\
H &  6.748160 &  1.832310 &  12.445670 \\
C &  4.877700 & -0.881120 &  8.182930 \\
H &  6.701160 &  0.126750 &  8.671310 \\
H &  5.435090 &  0.266030 &  9.913190 \\
N &  3.529170 & -0.610130 &  7.962140 \\
H &  3.052890 & -1.280270 &  7.360330 \\
C &  5.239320 & -0.945340 &  5.749860 \\
H &  3.377440 &  0.349660 &  7.627470 \\
H &  5.902730 & -1.121740 &  3.972370 \\
O &  5.705260 & -1.271320 &  6.919890 \\
H &  7.903250 & -0.881510 &  6.131800 \\
C &  9.786330 & -1.085380 &  4.079580 \\
H &  9.015680 &  0.272450 &  6.671520 \\
H &  10.035110 & -2.289460 &  5.747980 \\
O &  10.482080 & -1.792010 &  3.361380 \\
O &  8.910290 & -0.253740 &  3.509230 \\
H &  8.439380 &  0.376200 &  4.134380 \\
\bottomrule
\end{longtable}
\end{center}

\begin{center}
\begin{longtable}{c r r r}
\caption{Cyano-bound cluster, $\beta\alpha\beta\alpha$ conf.} \label{tab:Fe4S4CynEPR BP86 0,2} \\
\toprule
Element & x (\AA) & y (\AA) & z (\AA) \\
\midrule
\endfirsthead
\toprule
Element & x (\AA) & y (\AA) & z (\AA) \\
\midrule
\endhead
Fe &  8.068340 &  4.454020 &  5.411390 \\
Fe &  7.126830 &  2.653990 &  7.093330 \\
Fe &  9.612610 &  2.430290 &  6.277380 \\
Fe &  8.783030 &  4.351730 &  7.944180 \\
S &  5.285380 &  1.911060 &  8.217810 \\
S &  11.393300 &  1.232250 &  5.588790 \\
S &  7.688760 &  2.229350 &  4.995420 \\
S &  6.715260 &  4.931570 &  7.055950 \\
S &  10.168060 &  4.642990 &  6.255890 \\
S &  8.954400 &  2.079040 &  8.355340 \\
S &  9.041800 &  5.359440 &  9.949740 \\
N &  7.909200 &  6.462350 &  3.046200 \\
C &  7.943150 &  5.656710 &  3.910400 \\
C &  5.763080 &  0.256250 &  8.861670 \\
C &  11.220090 & -0.582570 &  5.955390 \\
C &  6.382800 &  4.372340 &  11.246120 \\
H &  11.330680 & -0.725980 &  7.036330 \\
C &  8.866370 &  4.079540 &  11.255170 \\
H &  9.657260 &  3.334590 &  11.146850 \\
H &  8.970750 &  4.591610 &  12.223920 \\
C &  7.510770 &  3.338800 &  11.248360 \\
O &  6.114630 & -1.382570 &  4.884270 \\
H &  5.154820 & -1.821150 &  8.815110 \\
N &  7.431210 &  2.479010 &  12.443870 \\
H &  7.279360 &  3.079160 &  13.261450 \\
H &  7.491360 &  2.700420 &  10.362460 \\
C &  9.888580 & -1.218400 &  5.527620 \\
H &  12.053980 & -1.071940 &  5.438490 \\
H &  5.503900 &  3.517290 &  9.681970 \\
O &  5.410370 &  4.283770 &  10.329800 \\
O &  6.323060 &  5.235670 &  12.110240 \\
N &  8.838240 & -0.732300 &  6.403110 \\
H &  6.630110 &  1.851010 &  12.381820 \\
C &  4.978690 & -0.899290 &  8.241810 \\
H &  6.826330 &  0.103030 &  8.657200 \\
H &  5.607200 &  0.244900 &  9.948720 \\
N &  3.612130 & -0.630110 &  8.109390 \\
H &  3.115620 & -1.296180 &  7.519020 \\
C &  5.207870 & -0.950190 &  5.796190 \\
H &  3.450080 &  0.327920 &  7.772670 \\
H &  5.799150 & -1.118340 &  3.992540 \\
O &  5.726940 & -1.279920 &  6.946270 \\
H &  7.883900 & -0.895650 &  6.068120 \\
C &  9.748720 & -1.085740 &  3.999800 \\
H &  9.008530 &  0.261870 &  6.582960 \\
H &  9.997010 & -2.309220 &  5.652710 \\
O &  10.436800 & -1.786090 &  3.268720 \\
O &  8.867380 & -0.246430 &  3.446150 \\
H &  8.399250 &  0.369410 &  4.085160 \\
\bottomrule
\end{longtable}
\end{center}

\begin{center}
\begin{longtable}{c r r r}
\caption{Cyano-bound cluster, $\beta\alpha\alpha\beta$ conf.} \label{tab:Fe4S4CynEPR BP86 0,3} \\
\toprule
Element & x (\AA) & y (\AA) & z (\AA) \\
\midrule
\endfirsthead
\toprule
Element & x (\AA) & y (\AA) & z (\AA) \\
\midrule
\endhead
Fe &  8.147010 &  4.469440 &  5.475290 \\
Fe &  7.133610 &  2.680220 &  7.109860 \\
Fe &  9.668770 &  2.466410 &  6.203600 \\
Fe &  8.978650 &  4.219180 &  8.071510 \\
S &  5.367400 &  1.787830 &  8.192450 \\
S &  11.404750 &  1.170740 &  5.549710 \\
S &  7.723380 &  2.368940 &  4.912610 \\
S &  6.917220 &  4.834700 &  7.308780 \\
S &  10.263090 &  4.571910 &  6.242940 \\
S &  8.960890 &  2.035240 &  8.333410 \\
S &  9.478310 &  5.228600 &  10.005300 \\
N &  7.910760 &  6.691900 &  3.325630 \\
C &  7.960630 &  5.855830 &  4.160930 \\
C &  5.881300 &  0.162920 &  8.884320 \\
C &  11.190360 & -0.636980 &  5.913200 \\
C &  6.352770 &  4.521900 &  11.067620 \\
H &  11.297560 & -0.786820 &  6.993340 \\
C &  8.836130 &  4.109350 &  11.330460 \\
H &  9.530160 &  3.271050 &  11.455760 \\
H &  8.777840 &  4.714420 &  12.242870 \\
C &  7.455500 &  3.474620 &  11.101770 \\
O &  6.017320 & -1.383270 &  4.903570 \\
H &  5.328580 & -1.934590 &  8.885090 \\
N &  7.207250 &  2.568500 &  12.241320 \\
H &  6.956030 &  3.113360 &  13.075990 \\
H &  7.484530 &  2.861050 &  10.196740 \\
C &  9.842190 & -1.223390 &  5.478950 \\
H &  12.011260 & -1.146290 &  5.395170 \\
H &  5.452040 &  3.599080 &  9.562710 \\
O &  5.327820 &  4.350990 &  10.215880 \\
O &  6.310140 &  5.430740 &  11.884070 \\
N &  8.807000 & -0.713310 &  6.355970 \\
H &  6.424650 &  1.927530 &  12.082160 \\
C &  5.089270 & -1.020120 &  8.321660 \\
H &  6.941170 &  0.016150 &  8.659780 \\
H &  5.745780 &  0.183890 &  9.974460 \\
N &  3.700620 & -0.779330 &  8.313620 \\
H &  3.189120 & -1.442600 &  7.731830 \\
C &  5.147370 & -1.007450 &  5.873620 \\
H &  3.517070 &  0.175520 &  7.976970 \\
H &  5.653730 & -1.081900 &  4.043470 \\
O &  5.736610 & -1.363140 &  6.985760 \\
H &  7.849290 & -0.875530 &  6.032990 \\
C &  9.709500 & -1.065220 &  3.950970 \\
H &  8.985990 &  0.279420 &  6.529960 \\
H &  9.906190 & -2.320210 &  5.587550 \\
O &  10.394140 & -1.761460 &  3.213450 \\
O &  8.832170 & -0.214960 &  3.404650 \\
H &  8.374220 &  0.404460 &  4.043590 \\
\bottomrule
\end{longtable}
\end{center}

\begin{center}
\begin{longtable}{c r r r}
\caption{Cyano-bound cluster, $\alpha\beta\beta\alpha$ conf.} \label{tab:Fe4S4CynEPR BP86 1,2} \\
\toprule
Element & x (\AA) & y (\AA) & z (\AA) \\
\midrule
\endfirsthead
\toprule
Element & x (\AA) & y (\AA) & z (\AA) \\
\midrule
\endhead
Fe &  8.134420 &  4.513240 &  5.362610 \\
Fe &  6.999450 &  2.771930 &  6.982230 \\
Fe &  9.634690 &  2.521240 &  6.153270 \\
Fe &  8.839500 &  4.330560 &  7.927690 \\
S &  5.250230 &  1.867790 &  8.040850 \\
S &  11.349630 &  1.202880 &  5.583380 \\
S &  7.720910 &  2.396420 &  4.850630 \\
S &  6.762040 &  4.945620 &  7.125220 \\
S &  10.290970 &  4.629740 &  6.159580 \\
S &  8.800150 &  2.174300 &  8.206480 \\
S &  9.066740 &  5.342340 &  9.925830 \\
N &  7.954570 &  6.777420 &  3.176560 \\
C &  7.987860 &  5.943410 &  4.008070 \\
C &  5.754210 &  0.282710 &  8.833590 \\
C &  11.186930 & -0.586290 &  6.049890 \\
C &  6.398600 &  4.296790 &  11.155880 \\
H &  11.283060 & -0.694070 &  7.137950 \\
C &  8.892680 &  4.039630 &  11.212620 \\
H &  9.693410 &  3.304040 &  11.108720 \\
H &  8.976070 &  4.540660 &  12.189540 \\
C &  7.545600 &  3.282960 &  11.172000 \\
O &  6.086620 & -1.398450 &  4.923720 \\
H &  5.238020 & -1.826430 &  8.889200 \\
N &  7.455910 &  2.417160 &  12.362460 \\
H &  7.296910 &  3.012640 &  13.182620 \\
H &  7.545600 &  2.648240 &  10.284010 \\
C &  9.856530 & -1.219980 &  5.618590 \\
H &  12.024470 & -1.099440 &  5.562250 \\
H &  5.525650 &  3.426130 &  9.579680 \\
O &  5.422590 &  4.176550 &  10.246390 \\
O &  6.322250 &  5.166880 &  12.012280 \\
N &  8.797910 & -0.687800 &  6.450770 \\
H &  6.655680 &  1.788170 &  12.293990 \\
C &  5.006310 & -0.934480 &  8.288940 \\
H &  6.825780 &  0.145390 &  8.656910 \\
H &  5.573720 &  0.347750 &  9.914360 \\
N &  3.618290 & -0.719010 &  8.209130 \\
H &  3.133920 & -1.406000 &  7.632100 \\
C &  5.171620 & -0.998910 &  5.842780 \\
H &  3.425940 &  0.227310 &  7.853810 \\
H &  5.758640 & -1.143490 &  4.034280 \\
O &  5.712580 & -1.319780 &  6.987770 \\
H &  7.847630 & -0.858130 &  6.110350 \\
C &  9.739910 & -1.105390 &  4.085970 \\
H &  8.980110 &  0.305960 &  6.617820 \\
H &  9.942340 & -2.310390 &  5.764530 \\
O &  10.429890 & -1.811370 &  3.363150 \\
O &  8.857820 & -0.271630 &  3.520650 \\
H &  8.402790 &  0.354520 &  4.152210 \\
\bottomrule
\end{longtable}
\end{center}

\begin{center}
\begin{longtable}{c r r r}
\caption{Cyano-bound cluster, $\alpha\beta\alpha\beta$ conf.} \label{tab:Fe4S4CynEPR BP86 1,3} \\
\toprule
Element & x (\AA) & y (\AA) & z (\AA) \\
\midrule
\endfirsthead
\toprule
Element & x (\AA) & y (\AA) & z (\AA) \\
\midrule
\endhead
Fe &  8.153180 &  4.497460 &  5.368870 \\
Fe &  7.060890 &  2.713880 &  7.016000 \\
Fe &  9.568390 &  2.483930 &  6.216640 \\
Fe &  8.780860 &  4.399840 &  7.914980 \\
S &  5.279060 &  1.925030 &  8.120570 \\
S &  11.355440 &  1.260120 &  5.535880 \\
S &  7.623600 &  2.278620 &  4.899660 \\
S &  6.781130 &  4.981660 &  6.989050 \\
S &  10.259480 &  4.705510 &  6.279300 \\
S &  8.865930 &  2.163820 &  8.254800 \\
S &  8.969750 &  5.319760 &  9.941120 \\
N &  8.038900 &  6.514480 &  2.940030 \\
C &  8.061500 &  5.711270 &  3.801030 \\
C &  5.753170 &  0.304710 &  8.860090 \\
C &  11.205830 & -0.531720 &  5.989360 \\
C &  6.378730 &  4.270130 &  11.228350 \\
H &  11.315450 & -0.638650 &  7.075430 \\
C &  8.862280 &  4.007860 &  11.219440 \\
H &  9.666070 &  3.278050 &  11.082210 \\
H &  8.985610 &  4.499320 &  12.197330 \\
C &  7.517200 &  3.245850 &  11.232910 \\
O &  6.105570 & -1.381870 &  4.928190 \\
H &  5.190570 & -1.790020 &  8.875800 \\
N &  7.458020 &  2.393580 &  12.435570 \\
H &  7.312570 &  3.002520 &  13.247530 \\
H &  7.496960 &  2.602390 &  10.352040 \\
C &  9.878010 & -1.194060 &  5.583580 \\
H &  12.042630 & -1.033770 &  5.489760 \\
H &  5.515560 &  3.426600 &  9.638550 \\
O &  5.422760 &  4.190870 &  10.296000 \\
O &  6.305710 &  5.123990 &  12.101750 \\
N &  8.826660 & -0.704360 &  6.455990 \\
H &  6.654830 &  1.767530 &  12.383490 \\
C &  4.989400 & -0.886160 &  8.282790 \\
H &  6.821550 &  0.156730 &  8.677310 \\
H &  5.576930 &  0.340880 &  9.943470 \\
N &  3.611970 & -0.643440 &  8.169880 \\
H &  3.122780 & -1.320980 &  7.586100 \\
C &  5.194870 & -0.954910 &  5.838440 \\
H &  3.436100 &  0.308040 &  7.820830 \\
H &  5.789070 & -1.123690 &  4.035210 \\
O &  5.720110 & -1.277250 &  6.989070 \\
H &  7.873540 & -0.875040 &  6.121700 \\
C &  9.724690 & -1.082930 &  4.056380 \\
H &  8.988800 &  0.291250 &  6.635180 \\
H &  10.007030 & -2.279400 &  5.729440 \\
O &  10.409010 & -1.789870 &  3.327570 \\
O &  8.838330 & -0.250370 &  3.502710 \\
H &  8.380670 &  0.382120 &  4.135880 \\
\bottomrule
\end{longtable}
\end{center}

\begin{center}
\begin{longtable}{c r r r}
\caption{Cyano bound cluster, $\alpha\alpha\beta\beta$ conf.} \label{tab:Fe4S4CynEPR BP86 2,3} \\
\toprule
Element & x (\AA) & y (\AA) & z (\AA) \\
\midrule
\endfirsthead
\toprule
Element & x (\AA) & y (\AA) & z (\AA) \\
\midrule
\endhead
Fe &  8.079920 &  4.518860 &  5.409840 \\
Fe &  7.056040 &  2.670040 &  7.027530 \\
Fe &  9.546060 &  2.465670 &  6.157520 \\
Fe &  8.759350 &  4.412210 &  7.936840 \\
S &  5.188130 &  1.938040 &  8.109090 \\
S &  11.352830 &  1.255240 &  5.621960 \\
S &  7.708760 &  2.289010 &  4.854420 \\
S &  6.755130 &  4.956810 &  7.202120 \\
S &  10.126080 &  4.653350 &  6.159960 \\
S &  8.786050 &  2.181410 &  8.243600 \\
S &  9.150830 &  5.297190 &  9.935270 \\
N &  7.654050 &  6.546770 &  3.032860 \\
C &  7.717340 &  5.739860 &  3.888920 \\
C &  5.648780 &  0.304450 &  8.822950 \\
C &  11.231820 & -0.544750 &  6.063740 \\
C &  6.478900 &  4.311040 &  11.215270 \\
H &  11.338110 & -0.659460 &  7.149870 \\
C &  8.966780 &  4.000750 &  11.227900 \\
H &  9.751590 &  3.248540 &  11.111380 \\
H &  9.086180 &  4.496440 &  12.204710 \\
C &  7.602480 &  3.273990 &  11.220000 \\
O &  6.171100 & -1.380700 &  4.899080 \\
H &  5.080420 & -1.786150 &  8.807580 \\
N &  7.510140 &  2.416620 &  12.417000 \\
H &  7.355590 &  3.019710 &  13.232000 \\
H &  7.581600 &  2.634160 &  10.335030 \\
C &  9.912460 & -1.214420 &  5.643150 \\
H &  12.076470 & -1.033730 &  5.563870 \\
H &  5.593640 &  3.470900 &  9.638600 \\
O &  5.500650 &  4.219230 &  10.304970 \\
O &  6.420610 &  5.175630 &  12.078710 \\
N &  8.849070 & -0.712180 &  6.493260 \\
H &  6.706980 &  1.790920 &  12.352780 \\
C &  4.906710 & -0.879480 &  8.211460 \\
H &  6.721940 &  0.160250 &  8.667130 \\
H &  5.453130 &  0.329770 &  9.903000 \\
N &  3.544620 & -0.633550 &  8.021800 \\
H &  3.071350 & -1.315460 &  7.430680 \\
C &  5.227870 & -0.953270 &  5.777050 \\
H &  3.376960 &  0.319170 &  7.673630 \\
H &  5.878030 & -1.134980 &  3.994850 \\
O &  5.706290 & -1.276860 &  6.944730 \\
H &  7.899730 & -0.878610 &  6.145110 \\
C &  9.780900 & -1.115960 &  4.111960 \\
H &  9.020200 &  0.284410 &  6.663780 \\
H &  10.037830 & -2.299010 &  5.797910 \\
O &  10.473400 & -1.826480 &  3.394180 \\
O &  8.895860 & -0.293070 &  3.540750 \\
H &  8.438430 &  0.346980 &  4.163500 \\
\bottomrule
\end{longtable}
\end{center}

\begin{center}
\begin{longtable}{c r r r}
\caption{$\Omega$ intermediate, $\alpha\alpha\beta\beta$ conf.} \label{tab:aabb epr 3023101} \\
\toprule
Element & x (\AA) & y (\AA) & z (\AA) \\
\midrule
\endfirsthead
\toprule
Element & x (\AA) & y (\AA) & z (\AA) \\
\midrule
\endhead
Fe &  1.805720 & -4.542690 &  4.260280 \\
Fe &  2.039100 & -6.199600 &  6.585490 \\
Fe &  0.450710 & -4.068790 &  6.727040 \\
Fe & -0.257400 & -6.101930 &  5.277610 \\
C &  3.773750 & -4.697740 &  3.472320 \\
H &  3.765450 & -3.999430 &  2.624230 \\
H &  4.391270 & -4.290860 &  4.283840 \\
C &  4.356050 & -6.009340 &  3.012610 \\
C &  5.258540 &  0.304510 & -0.143580 \\
S & -0.420840 & -4.017860 &  4.693320 \\
C &  0.405170 & -2.435490 &  9.518920 \\
S & -2.130810 & -7.039800 &  4.644420 \\
N &  2.023450 & -2.468420 &  3.709550 \\
C &  1.247670 & -2.080530 &  2.507160 \\
O &  1.285130 & -4.449460 &  2.029840 \\
C &  1.736960 & -0.783750 &  1.839880 \\
S &  3.608690 &  0.721760 &  0.493660 \\
S &  0.014770 & -6.129710 &  7.459650 \\
S & -0.463470 & -2.414410 &  7.892860 \\
C &  4.923240 & -7.369320 &  1.109910 \\
C &  6.060770 & -6.353590 &  1.372160 \\
C & -2.225500 & -8.644650 &  5.553760 \\
O &  5.788500 & -5.773820 &  2.639760 \\
C &  9.709440 & -5.562720 & -0.958570 \\
S &  3.375040 & -7.478690 &  7.841400 \\
S &  1.633760 & -6.800680 &  4.447460 \\
C &  3.728870 & -6.664950 &  1.770150 \\
C &  1.118340 & -3.273540 &  1.520670 \\
N &  8.497820 & -5.637900 & -0.390320 \\
H &  5.146470 & -8.313980 &  1.630060 \\
S &  2.614170 & -4.032020 &  6.461590 \\
C &  3.134140 & -0.886540 &  1.227110 \\
H &  1.014660 & -0.522940 &  1.052660 \\
O &  4.717140 & -7.660030 & -0.260100 \\
C &  5.070240 & -7.375420 &  7.126830 \\
N &  7.400080 & -6.936440 &  1.383840 \\
O &  0.826450 & -3.038320 &  0.334050 \\
H &  2.946300 & -7.382080 &  2.068050 \\
H &  5.656530 &  1.214660 & -0.613170 \\
H &  5.930030 & -0.002880 &  0.671150 \\
N &  10.816770 & -6.279250 & -0.684070 \\
H &  5.198820 & -0.493190 & -0.897810 \\
N &  11.855320 & -7.957210 &  0.556880 \\
C &  8.462900 & -6.573270 &  0.574720 \\
C &  7.879190 & -7.928410 &  2.219740 \\
C &  9.529450 & -7.393820 &  0.976330 \\
O &  3.239600 & -5.749420 &  0.798410 \\
C &  10.752530 & -7.215800 &  0.291390 \\
N &  9.147350 & -8.234500 &  2.009860 \\
H &  2.415330 & -5.300490 &  1.175890 \\
H &  7.246930 & -8.379400 &  2.980910 \\
H &  4.372000 & -6.744860 &  3.834130 \\
H &  1.714450 &  0.025130 &  2.589150 \\
H &  9.819320 & -4.817980 & -1.754020 \\
H &  11.867100 & -8.565050 &  1.368850 \\
H &  6.079910 & -5.597310 &  0.568290 \\
H &  0.217350 & -1.905040 &  2.858710 \\
H &  1.746220 & -1.893570 &  4.511170 \\
H &  12.736300 & -7.693840 &  0.127200 \\
H &  3.024190 & -2.302920 &  3.562990 \\
H &  3.893140 & -1.144280 &  1.983710 \\
H &  4.099100 & -6.952190 & -0.561340 \\
H &  3.148480 & -1.652910 &  0.438440 \\
C &  6.167690 & -7.538750 &  8.197160 \\
H &  5.199950 & -6.406470 &  6.619570 \\
H &  5.187370 & -8.171450 &  6.379300 \\
N &  6.161080 & -6.423460 &  9.127700 \\
H &  6.052590 & -5.504420 &  8.698040 \\
H &  6.898770 & -6.438370 &  9.830400 \\
C &  7.503700 & -7.793470 &  7.462990 \\
H &  5.969090 & -8.469610 &  8.755090 \\
O &  7.780200 & -8.838110 &  6.902260 \\
O &  8.380840 & -6.759740 &  7.463200 \\
H &  7.964250 & -6.042210 &  8.004630 \\
C & -0.215980 & -1.435570 &  10.517550 \\
H &  1.468210 & -2.186780 &  9.369640 \\
H &  0.347160 & -3.449260 &  9.936920 \\
N & -0.038660 & -0.057550 &  10.075160 \\
H &  0.962380 &  0.161670 &  10.048270 \\
H & -0.447700 &  0.582320 &  10.761750 \\
C &  0.392120 & -1.712990 &  11.897950 \\
H & -1.294990 & -1.648920 &  10.587620 \\
O &  1.292640 & -1.073620 &  12.416350 \\
O & -0.184700 & -2.786480 &  12.499980 \\
H &  0.284050 & -2.931150 &  13.354280 \\
C & -3.200650 & -9.630710 &  4.877130 \\
H & -2.558370 & -8.459630 &  6.587720 \\
H & -1.218640 & -9.081910 &  5.597910 \\
N & -4.573990 & -9.151760 &  4.953030 \\
H & -4.855130 & -9.095410 &  5.937270 \\
H & -5.202770 & -9.835520 &  4.522560 \\
C & -3.004520 & -11.024540 &  5.506240 \\
H & -2.921340 & -9.709960 &  3.810780 \\
O & -3.768440 & -11.537110 &  6.301750 \\
O & -1.885680 & -11.706590 &  5.130730 \\
H & -1.372920 & -11.183480 &  4.476550 \\
\bottomrule
\end{longtable}
\end{center}

\begin{center}
\begin{longtable}{c r r r}
\caption{$\Omega$ intermediate, $\alpha\beta\alpha\beta$ conf.} \label{tab:abab epr 3025556} \\
\toprule
Element & x (\AA) & y (\AA) & z (\AA) \\
\midrule
\endfirsthead
\toprule
Element & x (\AA) & y (\AA) & z (\AA) \\
\midrule
\endhead
Fe &  1.849700 & -4.301240 &  4.205170 \\
Fe &  1.947590 & -5.701960 &  6.610570 \\
Fe &  0.431540 & -3.569990 &  6.609330 \\
Fe & -0.302120 & -5.809270 &  5.344380 \\
C &  3.820520 & -4.404930 &  3.478680 \\
H &  3.800150 & -3.765450 &  2.584930 \\
H &  4.426710 & -3.929130 &  4.260910 \\
C &  4.444530 & -5.728170 &  3.113850 \\
C &  5.103960 &  0.443860 & -0.482710 \\
S & -0.480050 & -3.770650 &  4.578070 \\
C &  0.288020 & -1.914400 &  9.390800 \\
S & -1.945920 & -7.239070 &  4.897220 \\
N &  1.964440 & -2.195300 &  3.514880 \\
C &  1.181180 & -1.947700 &  2.285620 \\
O &  1.349190 & -4.341890 &  2.020200 \\
C &  1.600820 & -0.685740 &  1.511540 \\
S &  3.421740 &  0.818070 &  0.093720 \\
S & -0.027450 & -5.525040 &  7.561110 \\
S & -0.313430 & -1.722990 &  7.659280 \\
C &  5.085950 & -7.184290 &  1.304780 \\
C &  6.180900 & -6.112810 &  1.516270 \\
C & -1.662140 & -8.695020 &  5.988650 \\
O &  5.878050 & -5.479650 &  2.751020 \\
C &  9.826160 & -5.241770 & -0.789970 \\
S &  3.037860 & -7.172430 &  7.813990 \\
S &  1.686680 & -6.508840 &  4.583160 \\
C &  3.858490 & -6.484700 &  1.908010 \\
C &  1.130310 & -3.224600 &  1.405940 \\
N &  8.610030 & -5.352730 & -0.237560 \\
H &  5.335320 & -8.087280 &  1.883750 \\
S &  2.674020 & -3.692190 &  6.353540 \\
C &  3.015270 & -0.755480 &  0.934940 \\
H &  0.882120 & -0.540820 &  0.691840 \\
O &  4.907000 & -7.564660 & -0.047070 \\
C &  4.399640 & -7.909490 &  6.812080 \\
N &  7.542710 & -6.638250 &  1.564380 \\
O &  0.849450 & -3.116730 &  0.198680 \\
H &  3.098950 & -7.211290 &  2.240060 \\
H &  5.461730 &  1.335030 & -1.016680 \\
H &  5.774830 &  0.238640 &  0.364070 \\
N &  10.960460 & -5.895960 & -0.473240 \\
H &  5.102050 & -0.412410 & -1.172350 \\
N &  12.054380 & -7.476030 &  0.844910 \\
C &  8.601990 & -6.253550 &  0.760350 \\
C &  8.052500 & -7.576680 &  2.442620 \\
C &  9.698340 & -7.008080 &  1.208210 \\
O &  3.349600 & -5.651620 &  0.873830 \\
C &  10.923200 & -6.798820 &  0.534950 \\
N &  9.337380 & -7.829290 &  2.265040 \\
H &  2.508390 & -5.207890 &  1.216520 \\
H &  7.427070 & -8.034620 &  3.205240 \\
H &  4.467120 & -6.407520 &  3.982220 \\
H &  1.512670 &  0.182440 &  2.185740 \\
H &  9.914980 & -4.521810 & -1.610470 \\
H &  12.077930 & -8.072750 &  1.664390 \\
H &  6.174020 & -5.398880 &  0.674180 \\
H &  0.138420 & -1.803730 &  2.614720 \\
H &  1.645620 & -1.583500 &  4.271900 \\
H &  12.926870 & -7.205710 &  0.402960 \\
H &  2.956040 & -1.987860 &  3.362430 \\
H &  3.767980 & -0.917310 &  1.723850 \\
H &  4.265220 & -6.901240 & -0.396330 \\
H &  3.088760 & -1.575210 &  0.205290 \\
C &  5.255660 & -8.887800 &  7.638210 \\
H &  5.042780 & -7.107810 &  6.413170 \\
H &  3.960930 & -8.451680 &  5.963660 \\
N &  5.933400 & -8.199230 &  8.725150 \\
H &  6.350220 & -7.303770 &  8.468690 \\
H &  6.573510 & -8.776960 &  9.268520 \\
C &  6.165600 & -9.658030 &  6.656590 \\
H &  4.585640 & -9.652440 &  8.066910 \\
O &  5.766980 & -10.523850 &  5.899250 \\
O &  7.476620 & -9.306130 &  6.663470 \\
H &  7.583030 & -8.621620 &  7.370640 \\
C & -0.311740 & -0.850170 &  10.334590 \\
H &  1.386430 & -1.833410 &  9.409950 \\
H &  0.013150 & -2.915570 &  9.748950 \\
N &  0.153010 &  0.490140 &  9.998320 \\
H &  1.166540 &  0.542300 &  10.143870 \\
H & -0.258490 &  1.171570 &  10.642070 \\
C &  0.020790 & -1.264090 &  11.774450 \\
H & -1.408460 & -0.881440 &  10.231690 \\
O &  0.925260 & -0.804430 &  12.451760 \\
O & -0.814070 & -2.235510 &  12.228520 \\
H & -0.513480 & -2.481750 &  13.133690 \\
C & -2.497420 & -9.914370 &  5.547980 \\
H & -1.924220 & -8.428300 &  7.024710 \\
H & -0.591480 & -8.940670 &  5.967810 \\
N & -3.922400 & -9.663430 &  5.709210 \\
H & -4.126250 & -9.510370 &  6.702070 \\
H & -4.452920 & -10.496780 &  5.440520 \\
C & -2.005290 & -11.160740 &  6.313670 \\
H & -2.297400 & -10.095370 &  4.476270 \\
O & -2.620840 & -11.708350 &  7.207940 \\
O & -0.797000 & -11.663720 &  5.937090 \\
H & -0.417780 & -11.127800 &  5.206810 \\
\bottomrule
\end{longtable}
\end{center}

\begin{center}
\begin{longtable}{c r r r}
\caption{$\Omega$ intermediate, $\alpha\beta\beta\alpha$ conf.} \label{tab:abba epr 3025558} \\
\toprule
Element & x (\AA) & y (\AA) & z (\AA) \\
\midrule
\endfirsthead
\toprule
Element & x (\AA) & y (\AA) & z (\AA) \\
\midrule
\endhead
Fe &  2.054370 & -4.504300 &  4.164350 \\
Fe &  2.495820 & -6.033640 &  6.389260 \\
Fe &  1.028910 & -3.901560 &  6.913100 \\
Fe & -0.048250 & -5.868010 &  5.514510 \\
C &  3.901820 & -4.717830 &  3.209480 \\
H &  3.857970 & -4.010190 &  2.369220 \\
H &  4.621720 & -4.362640 &  3.958760 \\
C &  4.355790 & -6.058390 &  2.685490 \\
C &  5.209360 &  0.259060 & -0.523650 \\
S & -0.106090 & -3.752410 &  4.983340 \\
C &  0.948620 & -2.639540 &  9.922430 \\
S & -1.839710 & -7.098820 &  5.019550 \\
N &  2.313860 & -2.384700 &  3.661300 \\
C &  1.420330 & -1.951150 &  2.564810 \\
O &  1.297840 & -4.318320 &  2.123330 \\
C &  1.879800 & -0.679540 &  1.831850 \\
S &  3.640700 &  0.737860 &  0.258770 \\
S &  0.607660 & -5.914020 &  7.700330 \\
S &  0.480570 & -2.145030 &  8.210370 \\
C &  4.627700 & -7.436670 &  0.732730 \\
C &  5.843750 & -6.488250 &  0.864430 \\
C & -1.365750 & -8.855870 &  5.301510 \\
O &  5.741360 & -5.888820 &  2.147320 \\
C &  9.228370 & -6.050520 & -1.922770 \\
S &  3.960690 & -7.633200 &  7.011810 \\
S &  1.721820 & -6.662750 &  4.352150 \\
C &  3.556020 & -6.675760 &  1.524830 \\
C &  1.123000 & -3.138460 &  1.615840 \\
N &  8.094760 & -6.011900 & -1.208580 \\
H &  4.856420 & -8.394390 &  1.225890 \\
S &  3.214900 & -3.990150 &  6.269220 \\
C &  3.205650 & -0.835340 &  1.086160 \\
H &  1.090430 & -0.408490 &  1.115410 \\
O &  4.253340 & -7.710980 & -0.604610 \\
C &  5.282180 & -6.886340 &  8.051800 \\
N &  7.138520 & -7.155350 &  0.745330 \\
O &  0.705410 & -2.915190 &  0.467330 \\
H &  2.777820 & -7.353950 &  1.910660 \\
H &  5.572670 &  1.139980 & -1.070640 \\
H &  5.955520 & -0.028920 &  0.230980 \\
N &  10.308050 & -6.837690 & -1.751150 \\
H &  5.057640 & -0.567730 & -1.232210 \\
N &  11.373000 & -8.539520 & -0.567110 \\
C &  8.115540 & -6.903030 & -0.202000 \\
C &  7.646800 & -8.147540 &  1.562760 \\
C &  9.164640 & -7.784280 &  0.107070 \\
O &  3.002540 & -5.729720 &  0.617090 \\
C &  10.301280 & -7.727220 & -0.730640 \\
N &  8.856060 & -8.553960 &  1.217830 \\
H &  2.265180 & -5.237380 &  1.096220 \\
H &  7.081500 & -8.521250 &  2.413300 \\
H &  4.429990 & -6.795530 &  3.502890 \\
H &  1.955350 &  0.140060 &  2.565450 \\
H &  9.289880 & -5.348020 & -2.760710 \\
H &  11.441230 & -9.121060 &  0.260960 \\
H &  5.825520 & -5.737780 &  0.055290 \\
H &  0.448060 & -1.730440 &  3.036360 \\
H &  2.144400 & -1.829480 &  4.504660 \\
H &  12.206930 & -8.368300 & -1.119540 \\
H &  3.297540 & -2.248150 &  3.409690 \\
H &  4.029900 & -1.090990 &  1.772010 \\
H &  3.648720 & -6.968580 & -0.840970 \\
H &  3.125380 & -1.624550 &  0.323990 \\
C &  4.855230 & -6.590180 &  9.493940 \\
H &  5.648450 & -5.964090 &  7.574660 \\
H &  6.109810 & -7.609260 &  8.079150 \\
N &  3.900700 & -5.472370 &  9.635590 \\
H &  3.785610 & -4.972080 &  8.744210 \\
H &  2.970420 & -5.802270 &  9.898630 \\
C &  6.089290 & -6.253970 &  10.372910 \\
H &  4.418530 & -7.504690 &  9.926490 \\
O &  7.083320 & -6.959430 &  10.462520 \\
O &  5.932690 & -5.107050 &  11.041380 \\
H &  4.982520 & -4.830130 &  10.702860 \\
C &  0.422940 & -1.638870 &  10.973690 \\
H &  2.045450 & -2.700570 &  10.002320 \\
H &  0.532870 & -3.635300 &  10.125240 \\
N &  1.069880 & -0.338970 &  10.848270 \\
H &  2.067300 & -0.435810 &  11.064680 \\
H &  0.692240 &  0.305010 &  11.548780 \\
C &  0.603670 & -2.275940 &  12.358580 \\
H & -0.659870 & -1.512980 &  10.812610 \\
O &  1.501760 & -2.019630 &  13.142660 \\
O & -0.364140 & -3.192040 &  12.625000 \\
H & -0.162050 & -3.586640 &  13.504790 \\
C & -2.318400 & -9.826960 &  4.573810 \\
H & -1.384150 & -9.073810 &  6.380790 \\
H & -0.335050 & -8.996500 &  4.947170 \\
N & -3.644990 & -9.811580 &  5.173730 \\
H & -3.583740 & -10.170930 &  6.131920 \\
H & -4.260080 & -10.457800 &  4.671620 \\
C & -1.671740 & -11.227180 &  4.548610 \\
H & -2.410210 & -9.491820 &  3.525300 \\
O & -2.012090 & -12.156950 &  5.254410 \\
O & -0.649160 & -11.403810 &  3.666420 \\
H & -0.492040 & -10.578690 &  3.157170 \\
\bottomrule
\end{longtable}
\end{center}

\begin{center}
\begin{longtable}{c r r r}
\caption{$\Omega$ intermediate, $\beta\alpha\alpha\beta$ conf.} \label{tab:baab epr 3024996} \\
\toprule
Element & x (\AA) & y (\AA) & z (\AA) \\
\midrule
\endfirsthead
\toprule
Element & x (\AA) & y (\AA) & z (\AA) \\
\midrule
\endhead
Fe &  1.812230 & -4.494670 &  4.354910 \\
Fe &  1.857010 & -6.182120 &  6.490000 \\
Fe &  0.279130 & -4.041990 &  6.726330 \\
Fe & -0.484440 & -6.106610 &  5.251000 \\
C &  3.729760 & -4.692240 &  3.525640 \\
H &  3.730670 & -3.978960 &  2.690190 \\
H &  4.370850 & -4.314660 &  4.333200 \\
C &  4.283430 & -6.006840 &  3.033200 \\
C &  5.299760 &  0.410320 &  0.041400 \\
S & -0.548000 & -3.942360 &  4.695810 \\
C &  0.770790 & -2.560070 &  9.586720 \\
S & -2.258220 & -7.224840 &  4.456130 \\
N &  2.020110 & -2.445270 &  3.800010 \\
C &  1.252200 & -2.005200 &  2.607530 \\
O &  1.269880 & -4.347550 &  2.054270 \\
C &  1.766120 & -0.691980 &  1.994490 \\
S &  3.656920 &  0.829730 &  0.695100 \\
S & -0.236950 & -6.161500 &  7.428500 \\
S & -0.320800 & -2.390530 &  8.114820 \\
C &  4.838730 & -7.327390 &  1.101620 \\
C &  5.995540 & -6.344530 &  1.400920 \\
C & -2.220260 & -8.823560 &  5.372820 \\
O &  5.719620 & -5.784740 &  2.677380 \\
C &  9.689880 & -5.602240 & -0.871880 \\
S &  2.935420 & -7.630210 &  7.872870 \\
S &  1.513640 & -6.775490 &  4.390210 \\
C &  3.652850 & -6.614540 &  1.768250 \\
C &  1.113180 & -3.159270 &  1.579380 \\
N &  8.470230 & -5.657380 & -0.318680 \\
H &  5.036750 & -8.289770 &  1.599220 \\
S &  2.539470 & -4.023730 &  6.473880 \\
C &  3.161400 & -0.794810 &  1.377260 \\
H &  1.048990 & -0.388720 &  1.217720 \\
O &  4.640100 & -7.578810 & -0.277060 \\
C &  4.674140 & -7.768730 &  7.280550 \\
N &  7.319920 & -6.959770 &  1.418410 \\
O &  0.824640 & -2.878030 &  0.401100 \\
H &  2.853000 & -7.322560 &  2.040840 \\
H &  5.708340 &  1.328830 & -0.402060 \\
H &  5.969130 &  0.070210 &  0.844780 \\
N &  10.775600 & -6.351260 & -0.598270 \\
H &  5.227710 & -0.363790 & -0.735990 \\
N &  11.757880 & -8.077270 &  0.622190 \\
C &  8.401050 & -6.608930 &  0.628360 \\
C &  7.763930 & -7.978500 &  2.241060 \\
C &  9.441880 & -7.463280 &  1.026840 \\
O &  3.193300 & -5.664280 &  0.816800 \\
C &  10.676950 & -7.303490 &  0.358970 \\
N &  9.026350 & -8.312830 &  2.040220 \\
H &  2.377830 & -5.200850 &  1.200870 \\
H &  7.111530 & -8.427000 &  2.986540 \\
H &  4.278910 & -6.767290 &  3.832010 \\
H &  1.756190 &  0.088370 &  2.773730 \\
H &  9.827520 & -4.846070 & -1.652090 \\
H &  11.746880 & -8.696070 &  1.425930 \\
H &  6.044740 & -5.571330 &  0.614710 \\
H &  0.224040 & -1.828690 &  2.963830 \\
H &  1.739470 & -1.892120 &  4.615760 \\
H &  12.650300 & -7.825170 &  0.209540 \\
H &  3.021820 & -2.274330 &  3.662530 \\
H &  3.918240 & -1.086730 &  2.123540 \\
H &  4.040650 & -6.850200 & -0.566100 \\
H &  3.164240 & -1.535270 &  0.564130 \\
C &  5.662500 & -6.997080 &  8.168970 \\
H &  4.750340 & -7.413880 &  6.241180 \\
H &  4.951830 & -8.831730 &  7.295940 \\
N &  5.479580 & -5.535810 &  8.080790 \\
H &  4.580570 & -5.301160 &  7.640550 \\
H &  5.473930 & -5.103890 &  9.006900 \\
C &  7.108830 & -7.312910 &  7.717190 \\
H &  5.560750 & -7.361820 &  9.203650 \\
O &  7.627110 & -8.416240 &  7.798900 \\
O &  7.727940 & -6.239690 &  7.211450 \\
H &  6.989120 & -5.519650 &  7.342630 \\
C &  0.460850 & -1.502670 &  10.667330 \\
H &  1.821210 & -2.461080 &  9.270160 \\
H &  0.634880 & -3.564250 &  10.010180 \\
N &  0.776980 & -0.156400 &  10.207060 \\
H &  1.786590 & -0.081630 &  10.045510 \\
H &  0.558750 &  0.521640 &  10.942340 \\
C &  1.218990 & -1.903870 &  11.939700 \\
H & -0.615150 & -1.554000 &  10.899440 \\
O &  2.281260 & -1.431200 &  12.308430 \\
O &  0.571960 & -2.880430 &  12.628060 \\
H &  1.134930 & -3.118740 &  13.400640 \\
C & -2.979880 & -9.935020 &  4.619420 \\
H & -2.675370 & -8.681490 &  6.365780 \\
H & -1.173770 & -9.120140 &  5.524300 \\
N & -4.403740 & -9.643340 &  4.541810 \\
H & -4.798510 & -9.637040 &  5.487990 \\
H & -4.884900 & -10.396600 &  4.042800 \\
C & -2.668440 & -11.287200 &  5.295090 \\
H & -2.581350 & -9.984770 &  3.590450 \\
O & -3.444040 & -11.896470 &  6.006140 \\
O & -1.432020 & -11.810090 &  5.063900 \\
H & -0.918140 & -11.222360 &  4.467980 \\
\bottomrule
\end{longtable}
\end{center}

\begin{center}
\begin{longtable}{c r r r}
\caption{$\Omega$ intermediate, $\beta\alpha\beta\alpha$ conf.} \label{tab:baba epr 3023111} \\
\toprule
Element & x (\AA) & y (\AA) & z (\AA) \\
\midrule
\endfirsthead
\toprule
Element & x (\AA) & y (\AA) & z (\AA) \\
\midrule
\endhead
Fe &  2.068910 & -4.506630 &  4.119610 \\
Fe &  2.489120 & -6.052810 &  6.336860 \\
Fe &  1.020450 & -3.921120 &  6.862430 \\
Fe & -0.044610 & -5.876170 &  5.437710 \\
C &  3.924220 & -4.716140 &  3.178690 \\
H &  3.888460 & -4.002450 &  2.343070 \\
H &  4.637320 & -4.366490 &  3.937110 \\
C &  4.382980 & -6.052770 &  2.648880 \\
C &  5.262480 &  0.304830 & -0.502870 \\
S & -0.096670 & -3.757120 &  4.922660 \\
C &  0.928360 & -2.678070 &  9.878190 \\
S & -1.840450 & -7.102880 &  4.947390 \\
N &  2.337560 & -2.385090 &  3.633310 \\
C &  1.452010 & -1.941270 &  2.534460 \\
O &  1.329070 & -4.304210 &  2.072460 \\
C &  1.915910 & -0.662600 &  1.816830 \\
S &  3.684280 &  0.771150 &  0.267720 \\
S &  0.587770 & -5.938980 &  7.629330 \\
S &  0.464410 & -2.173210 &  8.168040 \\
C &  4.673700 & -7.410940 &  0.684090 \\
C &  5.885990 & -6.460840 &  0.835410 \\
C & -1.400200 & -8.843200 &  5.357410 \\
O &  5.774210 & -5.879890 &  2.125990 \\
C &  9.288500 & -5.970250 & -1.921160 \\
S &  3.943170 & -7.661370 &  6.961000 \\
S &  1.734990 & -6.666010 &  4.287500 \\
C &  3.593960 & -6.660630 &  1.475620 \\
C &  1.161170 & -3.119900 &  1.572650 \\
N &  8.149680 & -5.946180 & -1.214690 \\
H &  4.901020 & -8.373240 &  1.168890 \\
S &  3.212470 & -4.009670 &  6.239150 \\
C &  3.246650 & -0.810190 &  1.078100 \\
H &  1.131000 & -0.384890 &  1.098020 \\
O &  4.310300 & -7.671930 & -0.658950 \\
C &  5.254900 & -6.928480 &  8.022850 \\
N &  7.183600 & -7.121820 &  0.715140 \\
O &  0.754440 & -2.886500 &  0.422380 \\
H &  2.814480 & -7.344810 &  1.848160 \\
H &  5.628990 &  1.192450 & -1.036740 \\
H &  6.001390 &  0.011270 &  0.256730 \\
N &  10.369740 & -6.756110 & -1.753460 \\
H &  5.121090 & -0.514950 & -1.221670 \\
N &  11.432220 & -8.471660 & -0.586990 \\
C &  8.166410 & -6.852080 & -0.221340 \\
C &  7.689680 & -8.124110 &  1.521540 \\
C &  9.216390 & -7.734120 &  0.082190 \\
O &  3.045510 & -5.707260 &  0.572490 \\
C &  10.358780 & -7.660710 & -0.746310 \\
N &  8.902780 & -8.521230 &  1.179220 \\
H &  2.304380 & -5.219010 &  1.050120 \\
H &  7.119620 & -8.512060 &  2.362470 \\
H &  4.449300 & -6.796230 &  3.461120 \\
H &  1.986360 &  0.149890 &  2.558780 \\
H &  9.353570 & -5.255150 & -2.748110 \\
H &  11.496350 & -9.065270 &  0.232820 \\
H &  5.869620 & -5.699550 &  0.036340 \\
H &  0.476440 & -1.725210 &  3.001380 \\
H &  2.164600 & -1.835160 &  4.479470 \\
H &  12.269580 & -8.289390 & -1.130650 \\
H &  3.323240 & -2.249920 &  3.388790 \\
H &  4.066980 & -1.070280 &  1.766940 \\
H &  3.704930 & -6.928880 & -0.891590 \\
H &  3.172280 & -1.592910 &  0.308730 \\
C &  4.812280 & -6.642330 &  9.462280 \\
H &  5.630940 & -6.003870 &  7.557970 \\
H &  6.079250 & -7.654980 &  8.054100 \\
N &  3.860860 & -5.521570 &  9.601520 \\
H &  3.757850 & -5.014080 &  8.712720 \\
H &  2.926290 & -5.849200 &  9.851970 \\
C &  6.037560 & -6.317780 &  10.357990 \\
H &  4.366840 & -7.558410 &  9.882280 \\
O &  7.027300 & -7.028350 &  10.454250 \\
O &  5.878220 & -5.174660 &  11.032280 \\
H &  4.932960 & -4.891510 &  10.684560 \\
C &  0.399680 & -1.684150 &  10.934320 \\
H &  2.025030 & -2.738910 &  9.960410 \\
H &  0.512750 & -3.675290 &  10.073900 \\
N &  1.044010 & -0.382240 &  10.816390 \\
H &  2.041930 & -0.478410 &  11.030960 \\
H &  0.665940 &  0.256470 &  11.521460 \\
C &  0.580310 & -2.328270 &  12.315960 \\
H & -0.683270 & -1.559580 &  10.773000 \\
O &  1.475740 & -2.072530 &  13.103340 \\
O & -0.384180 & -3.249840 &  12.575410 \\
H & -0.182390 & -3.648310 &  13.453540 \\
C & -2.414320 & -9.842780 &  4.763410 \\
H & -1.370140 & -8.965880 &  6.451320 \\
H & -0.393380 & -9.047530 &  4.967150 \\
N & -3.720010 & -9.706000 &  5.393800 \\
H & -3.640940 & -9.955500 &  6.385030 \\
H & -4.373660 & -10.380750 &  4.986860 \\
C & -1.827070 & -11.264490 &  4.881220 \\
H & -2.525480 & -9.618830 &  3.687710 \\
O & -2.155540 & -12.076090 &  5.724940 \\
O & -0.874920 & -11.606840 &  3.968950 \\
H & -0.720540 & -10.867040 &  3.341860 \\
\bottomrule
\end{longtable}
\end{center}

\begin{center}
\begin{longtable}{c r r r}
\caption{$\Omega$ intermediate, $\beta\beta\alpha\alpha$ conf.} \label{tab:bbaa epr 3024505} \\
\toprule
Element & x (\AA) & y (\AA) & z (\AA) \\
\midrule
\endfirsthead
\toprule
Element & x (\AA) & y (\AA) & z (\AA) \\
\midrule
\endhead
Fe &  1.597440 & -4.518160 &  4.563240 \\
Fe &  1.914010 & -5.995940 &  7.098010 \\
Fe &  0.459750 & -3.798550 &  7.066870 \\
Fe & -0.398960 & -6.046250 &  5.799140 \\
C &  3.537400 & -4.733830 &  3.697550 \\
H &  3.458730 & -4.144810 &  2.774430 \\
H &  4.190930 & -4.218750 &  4.413250 \\
C &  4.131190 & -6.078160 &  3.365560 \\
C &  4.892620 & -0.093390 & -0.377360 \\
S & -0.569830 & -3.931380 &  5.043840 \\
C &  0.923340 & -1.474260 &  9.328320 \\
S & -2.256040 & -7.175300 &  5.241060 \\
N &  1.810120 & -2.498470 &  3.818600 \\
C &  0.989290 & -2.238670 &  2.612070 \\
O &  1.008860 & -4.645270 &  2.383170 \\
C &  1.439690 & -1.010850 &  1.803180 \\
S &  3.268580 &  0.381750 &  0.285190 \\
S & -0.042650 & -5.805050 &  8.043390 \\
S & -0.328370 & -1.936090 &  8.055390 \\
C &  4.640950 & -7.626410 &  1.593160 \\
C &  5.748730 & -6.550140 &  1.671210 \\
C & -1.734650 & -8.902250 &  4.857840 \\
O &  5.535570 & -5.855220 &  2.891600 \\
C &  9.189890 & -5.897520 & -0.996000 \\
S &  3.327960 & -7.192180 &  8.352640 \\
S &  1.504360 & -6.778390 &  5.011530 \\
C &  3.459840 & -6.890820 &  2.243800 \\
C &  0.839280 & -3.526720 &  1.756520 \\
N &  8.025040 & -5.949590 & -0.334830 \\
H &  4.930610 & -8.499040 &  2.199460 \\
S &  2.602990 & -3.898660 &  6.663970 \\
C &  2.822010 & -1.158770 &  1.166480 \\
H &  0.693460 & -0.839510 &  1.013800 \\
O &  4.369000 & -8.077740 &  0.279250 \\
C &  4.988410 & -7.108360 &  7.561670 \\
N &  7.109670 & -7.078260 &  1.647810 \\
O &  0.527000 & -3.418130 &  0.557280 \\
H &  2.724480 & -7.594020 &  2.668700 \\
H &  5.267680 &  0.768520 & -0.946270 \\
H &  5.598320 & -0.323670 &  0.433900 \\
N &  10.343210 & -6.539810 & -0.727170 \\
H &  4.803720 & -0.958290 & -1.050210 \\
N &  11.535540 & -8.029200 &  0.612710 \\
C &  8.097040 & -6.770960 &  0.727240 \\
C &  7.687730 & -7.953700 &  2.548870 \\
C &  9.222840 & -7.503730 &  1.136180 \\
O &  2.880960 & -6.110830 &  1.205010 \\
C &  10.386740 & -7.362020 &  0.347540 \\
N &  8.950250 & -8.236510 &  2.280490 \\
H &  2.083300 & -5.615150 &  1.580170 \\
H &  7.129500 & -8.339100 &  3.398730 \\
H &  4.212290 & -6.710210 &  4.264940 \\
H &  1.424510 & -0.131100 &  2.468080 \\
H &  9.212140 & -5.244030 & -1.874610 \\
H &  11.626030 & -8.549070 &  1.478850 \\
H &  5.681500 & -5.878750 &  0.797300 \\
H & -0.030370 & -2.040560 &  2.982160 \\
H &  1.548810 & -1.855060 &  4.571910 \\
H &  12.371910 & -7.795350 &  0.087590 \\
H &  2.803180 & -2.338460 &  3.622130 \\
H &  3.600420 & -1.349520 &  1.923430 \\
H &  3.713670 & -7.425860 & -0.065460 \\
H &  2.820990 & -1.989770 &  0.446000 \\
C &  6.128840 & -7.265750 &  8.586730 \\
H &  5.099300 & -6.145300 &  7.038930 \\
H &  5.068320 & -7.913430 &  6.818970 \\
N &  6.162890 & -6.142150 &  9.506940 \\
H &  6.053390 & -5.225860 &  9.071370 \\
H &  6.922400 & -6.159380 &  10.186040 \\
C &  7.433540 & -7.533120 &  7.800900 \\
H &  5.950630 & -8.190340 &  9.161520 \\
O &  7.675050 & -8.579100 &  7.227040 \\
O &  8.323340 & -6.511010 &  7.772400 \\
H &  7.936300 & -5.789890 &  8.330510 \\
C &  0.315730 & -1.340040 &  10.741900 \\
H &  1.373150 & -0.509350 &  9.042570 \\
H &  1.718000 & -2.230850 &  9.338100 \\
N & -0.603350 & -0.209250 &  10.819620 \\
H & -0.070470 &  0.662100 &  10.728680 \\
H & -1.036220 & -0.178410 &  11.746870 \\
C &  1.477480 & -1.243510 &  11.737850 \\
H & -0.237700 & -2.265000 &  10.964680 \\
O &  1.946230 & -0.205650 &  12.177720 \\
O &  1.963040 & -2.466760 &  12.071770 \\
H &  2.736090 & -2.326290 &  12.666210 \\
C & -2.626710 & -9.547130 &  3.775890 \\
H & -1.790760 & -9.511140 &  5.774050 \\
H & -0.687770 & -8.885760 &  4.527300 \\
N & -3.983280 & -9.757800 &  4.264940 \\
H & -3.960440 & -10.451490 &  5.019770 \\
H & -4.556940 & -10.171590 &  3.524670 \\
C & -1.945430 & -10.844560 &  3.295250 \\
H & -2.672560 & -8.857480 &  2.914760 \\
O & -2.275930 & -11.963710 &  3.639040 \\
O & -0.899140 & -10.698040 &  2.436230 \\
H & -0.753920 & -9.747590 &  2.235020 \\
\bottomrule
\end{longtable}
\end{center}

\begin{center}
\begin{longtable}{c r r r}
\caption{Partially detached 5'-dAdo radical, high-spin configuration, $2S+1=20$} \label{tab:rad a5 opt 3022000} \\
\toprule
Element & x (\AA) & y (\AA) & z (\AA) \\
\midrule
\endfirsthead
\toprule
Element & x (\AA) & y (\AA) & z (\AA) \\
\midrule
\endhead
Fe & -4.520690 & -23.509440 &  21.567610 \\
Fe & -6.322300 & -24.038390 &  19.412030 \\
Fe & -6.151680 & -21.561170 &  20.469740 \\
Fe & -4.125320 & -22.578810 &  18.963700 \\
C & -3.313650 & -26.115430 &  22.050490 \\
H & -4.358320 & -25.950890 &  21.796740 \\
H & -2.837620 & -25.475440 &  22.792840 \\
C & -2.534020 & -27.356190 &  21.844890 \\
S & -6.910000 & -23.469580 &  21.485240 \\
S & -6.210440 & -22.007480 &  18.282180 \\
S & -3.886430 & -21.420780 &  20.860780 \\
S & -4.170250 & -24.786840 &  19.647830 \\
N & -4.626780 & -23.012180 &  23.739840 \\
C & -3.632190 & -23.680430 &  24.551460 \\
C & -2.461840 & -24.005980 &  23.638140 \\
O & -2.586660 & -23.949120 &  22.380900 \\
C & -4.197400 & -24.947480 &  25.239320 \\
C & -5.193640 & -25.735660 &  24.372430 \\
C & -5.797120 & -27.317700 &  22.182400 \\
O & -1.377460 & -24.329960 &  24.190590 \\
H & -4.360400 & -22.205250 &  23.562800 \\
H & -5.378730 & -22.980630 &  24.171900 \\
H & -3.338400 & -23.085400 &  25.215390 \\
H & -4.630320 & -24.689650 &  26.031540 \\
H & -3.480660 & -25.512760 &  25.459740 \\
O & -3.392000 & -28.419700 &  21.349390 \\
C & -1.350050 & -27.240990 &  20.817200 \\
C & -2.039830 & -27.745910 &  19.547470 \\
O & -1.119750 & -28.080690 &  18.532380 \\
C & -2.879780 & -28.914030 &  20.106870 \\
N & -1.969680 & -31.721500 &  15.870730 \\
C & -1.391080 & -31.720620 &  17.085630 \\
N & -1.669330 & -30.948240 &  18.147270 \\
C & -2.670890 & -30.099980 &  17.881410 \\
C & -3.372510 & -29.979120 &  16.672730 \\
C & -2.970910 & -30.848690 &  15.634810 \\
N & -3.541090 & -30.824840 &  14.402480 \\
N & -4.354360 & -29.003640 &  16.755070 \\
C & -4.244280 & -28.540480 &  17.982900 \\
N & -3.238550 & -29.150060 &  18.723100 \\
H & -3.295510 & -31.555180 &  13.745590 \\
H & -4.376670 & -30.275110 &  14.251330 \\
H & -1.615890 & -28.422540 &  17.769980 \\
H & -2.253540 & -29.810860 &  20.241500 \\
H & -2.736620 & -26.949990 &  19.218200 \\
H & -2.089110 & -27.698430 &  22.800850 \\
H & -4.848820 & -27.794090 &  18.360360 \\
H & -0.631180 & -32.406060 &  17.218900 \\
O & -0.804070 & -25.957060 &  20.686100 \\
H & -0.566320 & -27.995860 &  21.065870 \\
H & -0.333010 & -25.772760 &  21.539550 \\
S & -4.954290 & -27.458570 &  23.819530 \\
S & -7.294710 & -19.689030 &  21.168790 \\
H & -5.392200 & -25.180640 &  23.448620 \\
H & -6.165220 & -25.728100 &  24.894210 \\
H & -5.153940 & -27.643560 &  21.374510 \\
H & -6.166210 & -26.295070 &  22.039080 \\
H & -6.688420 & -27.965490 &  22.217390 \\
C & -6.913780 & -19.526750 &  22.958180 \\
H & -7.413260 & -18.628860 &  23.346320 \\
H & -5.831550 & -19.427160 &  23.114830 \\
H & -7.278890 & -20.402430 &  23.511230 \\
S & -2.550590 & -22.201130 &  17.331470 \\
C & -3.037840 & -23.296520 &  15.940390 \\
H & -2.326660 & -23.156280 &  15.115300 \\
H & -4.047680 & -23.047920 &  15.588890 \\
H & -3.016170 & -24.347630 &  16.255670 \\
S & -7.747800 & -25.481610 &  18.321460 \\
C & -7.161180 & -25.504310 &  16.580190 \\
H & -7.828820 & -26.148010 &  15.991930 \\
H & -6.139680 & -25.900870 &  16.518560 \\
H & -7.177500 & -24.490250 &  16.160140 \\
\bottomrule
\end{longtable}
\end{center}

\end{document}